\documentclass[preprint, astrosymb]{aastex631} %
\usepackage{CJK}
\usepackage{amsmath}
\usepackage{physics}
\hypersetup{linkcolor=blue,citecolor=blue,filecolor=blue,urlcolor=blue}

\begin{document}
\begin{CJK*}{UTF8}{gbsn}
\title{Multi-scale physical properties of NGC 6334 as revealed by local relative orientations between magnetic fields, density gradients, velocity gradients, and gravity}

\correspondingauthor{Junhao Liu}
\email{liujunhao42@outlook.com; j.liu@eaobservatory.org}

\author[0000-0002-4774-2998]{Junhao Liu (刘峻豪)}
%\altaffiliation{EAO Fellow}
\affiliation{East Asian Observatory, 660 N. A`oh\={o}k\={u} Place, University Park, Hilo, HI 96720, USA}

\author[0000-0003-2384-6589]{Qizhou Zhang}
\affiliation{Center for Astrophysics $\vert$ Harvard \& Smithsonian, 60 Garden Street, Cambridge, MA 02138, USA}

\author[0000-0003-2777-5861]{Patrick M. Koch}
\affiliation{Institute of Astronomy and Astrophysics, Academia Sinica, 11F of Astronomy-Mathematics Building, AS/NTU No.1, Sec. 4, Roosevelt Rd, Taipei 10617, Taiwan, Republic of China}

\author[0000-0003-2300-2626]{Hauyu Baobab Liu}
\affiliation{Institute of Astronomy and Astrophysics, Academia Sinica, 11F of Astronomy-Mathematics Building, AS/NTU No.1, Sec. 4, Roosevelt Rd, Taipei 10617, Taiwan, Republic of China} 

\author[0000-0002-7402-6487]{Zhi-Yun Li}
\affiliation{Astronomy Department, University of Virginia, Charlottesville, VA 22904-4325, USA}

\author[0000-0003-1275-5251]{Shanghuo Li}
\affiliation{Max Planck Institute for Astronomy, Konigstuhl 17, D-69117 Heidelberg, Germany}

\author[0000-0002-3829-5591]{Josep Miquel Girart}
\affil{Institut de Ci$\grave{e}$ncies de l'Espai (ICE, CSIC), Can Magrans s/n, E-08193 Cerdanyola del Vall$\grave{e}$s, Catalonia, Spain}
\affil{Institut d'Estudis Espacials de de Catalunya (IEEC), E-08034 Barcelona, Catalonia, Spain}

\author[0000-0002-9774-1846]{Huei-Ru Vivien Chen}
\affiliation{Institute of Astronomy and Astrophysics, Academia Sinica, 11F of Astronomy-Mathematics Building, AS/NTU No.1, Sec. 4, Roosevelt Rd, Taipei 10617, Taiwan, Republic of China}

\author[0000-0001-8516-2532]{Tao-Chung Ching}
\affiliation{National Radio Astronomy Observatory, P.O. Box O, Socorro, NM 87801, USA}

\author[0000-0002-3412-4306]{Paul T. P. Ho}
\affiliation{East Asian Observatory, 660 N. A`oh\={o}k\={u} Place, University Park, Hilo, HI 96720, USA}
\affiliation{Institute of Astronomy and Astrophysics, Academia Sinica, 11F of Astronomy-Mathematics Building, AS/NTU No.1, Sec. 4, Roosevelt Rd, Taipei 10617, Taiwan, Republic of China}

\author[0000-0001-5522-486X]{Shih-Ping Lai}
\affiliation{Institute of Astronomy and Department of Physics, National Tsing Hua University, Hsinchu 30013, Taiwan, Republic of China}
\affiliation{Institute of Astronomy and Astrophysics, Academia Sinica, 11F of Astronomy-Mathematics Building, AS/NTU No.1, Sec. 4, Roosevelt Rd, Taipei 10617, Taiwan, Republic of China}

\author[0000-0002-5093-5088]{Keping Qiu}
\affiliation{School of Astronomy and Space Science, Nanjing University, 163 Xianlin Avenue, Nanjing 210023, Jiangsu, People's Republic of China}
\affiliation{Key Laboratory of Modern Astronomy and Astrophysics (Nanjing University), Ministry of Education, Nanjing 210023, Jiangsu, People's Republic of China}

\author[0000-0002-1407-7944]{Ramprasad Rao}
\affiliation{Institute of Astronomy and Astrophysics, Academia Sinica, 11F of Astronomy-Mathematics Building, AS/NTU No.1, Sec. 4, Roosevelt Rd, Taipei 10617, Taiwan, Republic of China}
\affiliation{Center for Astrophysics $\vert$ Harvard \& Smithsonian, 60 Garden Street, Cambridge, MA 02138, USA}

\author[0000-0002-0675-276X]{Ya-wen Tang}
\affiliation{Institute of Astronomy and Astrophysics, Academia Sinica, 11F of Astronomy-Mathematics Building, AS/NTU No.1, Sec. 4, Roosevelt Rd, Taipei 10617, Taiwan, Republic of China}



\begin{abstract}

We present ALMA dust polarization and molecular line observations toward 4 clumps (I(N), I, IV, and V) in the massive star-forming region NGC 6334. In conjunction with large-scale dust polarization and molecular line data from JCMT, Planck, and NANTEN2, we make a synergistic analysis of relative orientations between magnetic fields ($\theta_{\mathrm{B}}$), column density gradients ($\theta_{\mathrm{NG}}$), local gravity ($\theta_{\mathrm{LG}}$), and velocity gradients ($\theta_{\mathrm{VG}}$) to investigate the multi-scale (from $\sim$30 pc to 0.003 pc) physical properties in NGC 6334. We find that the relative orientation between $\theta_{\mathrm{B}}$ and $\theta_{\mathrm{NG}}$ changes from statistically more perpendicular to parallel as column density ($N_{\mathrm{H_2}}$) increases, which is a signature of trans-to-sub-Alfv\'{e}nic turbulence at complex/cloud scales as revealed by previous numerical studies. Because $\theta_{\mathrm{NG}}$ and $\theta_{\mathrm{LG}}$ are preferentially aligned within the NGC 6334 cloud, we suggest that the more parallel alignment between $\theta_{\mathrm{B}}$ and $\theta_{\mathrm{NG}}$ at higher $N_{\mathrm{H_2}}$ is because the magnetic field line is dragged by gravity. At even higher $N_{\mathrm{H_2}}$, the angle between $\theta_{\mathrm{B}}$ and $\theta_{\mathrm{NG}}$ or $\theta_{\mathrm{LG}}$ transits back to having no preferred orientation or statistically slightly more perpendicular, suggesting that the magnetic field structure is impacted by star formation activities. A statistically more perpendicular alignment is found between $\theta_{\mathrm{B}}$ and $\theta_{\mathrm{VG}}$ throughout our studied $N_{\mathrm{H_2}}$ range, which indicates a trans-to-sub-Alfv\'{e}nic state at small scales as well and signifies an important role of magnetic field in the star formation process in NGC 6334. The normalised mass-to-flux ratio derived from the polarization-intensity gradient (KTH) method increases with $N_{\mathrm{H_2}}$, but the KTH method may fail at high $N_{\mathrm{H_2}}$ due to the impact of star formation feedback. 

\end{abstract}



\section{Introduction} \label{sec:intro}
Turbulence and magnetic fields are the two major forces that compete with gravity within self-gravitating molecular clouds. The balance among these forces controls the star formation process \citep{2007ARA&A..45..565M}. The role of magnetic fields in star formation is less understood than turbulence due to relatively fewer observations. Understanding the interactions between magnetic fields and the other two forces have been a key topic in the study of star formation \citep{2012ARA&A..50...29C}. 

Assuming that the shortest axis of a fraction of irregular dust grains is aligned with the magnetic field, the plane-of-sky (POS) magnetic field orientation can be traced by rotating the observed position angle of linearly polarized dust emission by 90$\degr$ \citep{1949PhRv...75.1605D, 2007JQSRT.106..225L, 2007MNRAS.378..910L, 2015ARA&A..53..501A}. There has been an increasing number of dust polarization observations that reveal the POS magnetic field orientation in star-forming molecular clouds \citep{2019FrASS...6....3H, 2019FrASS...6...15P}. The Davis-Chandrasekhar-Fermi (DCF) method \citep{1951PhRv...81..890D, 1953ApJ...118..113C} and its modified forms have been the most widely used method to indirectly derive the magnetic field strength with statistics of field orientations. The compilation of previous DCF estimations suggests that magnetically trans-to-super-critical and averagely trans-to-super-Alfv\'{e}nic clumps/cores form in sub-critical clouds \citep{2022ApJ...925...30L}. However, the breakdown of the DCF assumptions, such as energy equipartition \citep{2021A&A...647A.186S} or turbulence isotropy \citep{2022ApJ...935...77L}, in specific physical conditions (e.g., in non-self-gravitating media) might bring some uncertainties to the DCF estimations \citep[see a review of the DCF method in][]{ 2022FrASS...9.3556L}. Thus, it is essential to study the magnetic field properties with other statistical methods as well. 
% and trans-to-sub-Alfv\'{e}nic

Well-ordered magnetic field structures (e.g., hourglass or toroidal shapes) are seldom observed in star-forming regions \citep[see a review of observed hourglass-shaped magnetic fields in][]{2019FrASS...6....3H}. Many star formation regions show complex magnetic field structures, which brings difficulties in interpreting the field topology. The development of statistical techniques has made it possible to infer the physical properties of star formation regions by comparing the magnetic field orientation with other orientations (e.g., the column density gradient/column density contour/intensity gradient\footnote{The column density gradient is perpendicular to the column density contour. The column density gradient is parallel to the intensity gradient if the physical parameters of gas and dust are constant or their variations are along the density gradient.}, the direction of local gravity, and the velocity gradient) that can be obtained through astronomical observations. For instance, the Histogram of relative orientation analysis \citep[HRO, ][]{2013ApJ...774..128S} measures the statistical relation between magnetic fields and density structures and can be used to link the physical properties of observations and simulations. The observational HRO studies reveal that the magnetic field and column density contour changes from a preferential parallel alignment to a perpendicular alignment with increasing column densities \citep[e.g.,][]{2016AA...586A.138P}, and may transit back to a random alignment at higher column densities \citep[e.g.,][]{2020ApJ...904..168B, 2022ApJ...926..163K}. The observational trends suggest that star formation is ongoing in trans-to-sub-Alfv\'{e}nic clouds and the magnetic field is likely affected by star formation activities in high-density regions, but the exact reason for the different alignment at different column densities is still under debate \citep[see a review of the HRO analysis in ][]{2022FrASS...9.3556L}. On the other hand, the polarization-intensity gradient method \citep[Koch-Tang-Ho or KTH method,][]{2012ApJ...747...79K} proposes to determine the local magnetic field strength as well as the local normalized mass-to-flux ratio ($\lambda_{\mathrm{KTH}}$) under the assumption of ideal magnetohydrodynamics (MHD) by comparing the local orientations of magnetic fields, intensity gradients, and local gravity. Several observational KTH studies found that the magnetic field strength or $\lambda_{\mathrm{KTH}}$ estimated with the KTH method are not far from the DCF estimations \citep[e.g.,][]{2013ApJ...769L..15S, 2013ApJ...772...69G, 2020A&A...644A..52A}. A review of all the observational KTH studies can be found in \citet{2022FrASS...9.3556L}. Moreover, the velocity gradients from molecular line observations are expected to be perpendicular to the local magnetic field orientation in the absence of gravity due to the intrinsic property of MHD turbulence, where the correlation between the magnetic field and velocity gradient should be weaker for larger Alfv\'{e}nic Mach numbers \citep{2017ApJ...835...41G, 2018ApJ...853...96L}. In high-density regions, the infalling gas due to strong self-gravity may drag the magnetic field lines and align magnetic fields with velocity gradients, where the alignment may be used to identify self-gravitating regions \citep{2017arXiv170303026Y}. A synergistic analysis with these techniques using relative orientations between different angles will be advantageous in revealing the physical conditions in star-forming molecular clouds. 
%, which, in conjunction with simulations

The massive ($>10^5 M_{\odot}$) star-forming complex NGC 6334 at a distance of $\sim$1.3 kpc \citep{2014ApJ...784..114C, 2014A&A...566A..17W} is one of the nearest massive star-forming regions from the sun and has been extensively studied at various wavelengths \citep[see a review by][]{2008hsf2.book..456P}. The predominant structure in the NGC 6334 complex is a 10 pc-long filamentary cloud (hereafter the NGC 6334 cloud or NGC 6334 filament) elongated along the direction of the galactic plane. The NGC 6334 filament harbors six massive star-forming molecular clumps (N6334I-V and N6334I(N)) that were identified with far-infrared/sub-mm/mm observations \citep[e.g., ][]{1978ApJ...226L.149C, 1979ApJ...232L.183M, 1982ApJ...259L..29G}. The high-luminosity ($>10^4 L_{\odot}$), large gas reservoir, presence of H$_2$O, OH, and CH$_3$OH (class I and II) masers, and detections of compact and ultra-compact HII regions, outflows, young stellar objects, and massive stars within or in the vicinity of these clumps suggest that these clumps are undergoing active intermediate- to high-mass star formation \citep[e.g.,][]{1982ApJ...255..103R, 1986ApJ...303..629L, 2007ApJ...668..906M, 2008hsf2.book..456P, 2012A&A...538A.142R, 2013ApJ...778...96W, 2016A&A...592A..54A}. 
%(N6334I, I(N), IV, and V) .Outflows are detected toward N6334I, I(N), IV, and V \citep{2014ApJ...792..116Z}, suggesting active massive star formation. 

The magnetic field structure of the NGC 6334 region at different scales has been previously studied with dust polarization observations \citep{2014ApJ...792..116Z, 2015Natur.520..518L, 2017ApJ...844...44J, 2021ApJ...912..159P, 2021A&A...647A..78A, 2021ApJ...923..204C}. Specifically, the multi-scale magnetic field study by \citet{2015Natur.520..518L} revealed that the orientation of magnetic fields does not change much from cloud scales to clump and core scales in N6334I and I(N), where the area-averaged magnetic field is perpendicular to the area-averaged elongation of density structures at each scale. This signifies a dynamically important role played by magnetic fields on guiding gravitational collapse which leads to a self-similar fragmentation across various scales. Different trends have been found in some sub-regions (N6334IV and V) where the magnetic fields might be affected by stellar feedback or converging flows \citep{2015Natur.520..518L, 2017ApJ...844...44J}. In addition, \citet{2021A&A...647A..78A} found that the magnetic field changes from being perpendicular or randomly aligned with the outer part of the sub-filaments to being parallel to the inter part of the sub-filaments that merges into the main filament, which may indicate infalling gas flows from sub-filaments to the main filament. 

In this paper, we present high-resolution ($\sim$900 AU) Atacama Large Millimeter/submillimeter Array (ALMA) 1.3mm dust polarization and molecular line observations toward clumps N6334I(N), I, IV, and V in NGC 6334 to study the physical properties within molecular dense cores. We also collect the historical dust polarization data and molecular line data at coarser resolutions for the study of the large-scale physical conditions at cloud and clump scales. In Section \ref{sec:observation}, we describe the observational data. In Section \ref{sec:results}, we present maps of the density structure, magnetic field structure, and velocity structure. In Section \ref{sec:RON}, we study the relative orientations between magnetic fields, column density gradients, local gravity, and velocity gradients at different column densities and discuss their implication on the physical properties in NGC 6334. A summary of this paper is provided in Section \ref{sec:summary}. We only focus on the statistical properties of physical conditions at different scales in this paper and will present detailed analyses of ALMA observations toward individual clumps in future papers. 
%, calculate the angular dispersion function (ADF) and velocity centroid dispersion function (VDF)

\section{Observation} \label{sec:observation}

\subsection{ALMA dust polarization and molecular line observations}\label{sec:obsalma}
\begin{deluxetable}{cccc}[t!]
\tablecaption{Source coordinates of ALMA observations \label{tab:sources}}
\tablecolumns{4}
\tablewidth{0pt}
\tablehead{
\colhead{Source} &
\colhead{Field} &
\colhead{$\alpha_{\mathrm{J2000}}$} & 
\colhead{$\delta_{\mathrm{J2000}}$}
}
\startdata
N6334I & NGC6334I & $17^{\mathrm{h}}20^{\mathrm{m}}53^{\mathrm{s}}.41$ & $-35\degr46\arcmin57\arcsec.8$ \\\hline
N6334I(N) & NGC6334In.1 & $17^{\mathrm{h}}20^{\mathrm{m}}54^{\mathrm{s}}.97$ & $-35\degr45\arcmin05\arcsec.6$ \\
& NGC6334In.2 & $17^{\mathrm{h}}20^{\mathrm{m}}54^{\mathrm{s}}.53$ & $-35\degr45\arcmin18\arcsec.8$ \\
& NGC6334In.3 & $17^{\mathrm{h}}20^{\mathrm{m}}56^{\mathrm{s}}.00$ & $-35\degr45\arcmin27\arcsec.5$ \\\hline
N6334IV & NGC6334IV.1\tablenotemark{a} & $17^{\mathrm{h}}20^{\mathrm{m}}19^{\mathrm{s}}.72$ & $-35\degr54\arcmin38\arcsec.0$ \\
& NGC6334IV.2\tablenotemark{a} & $17^{\mathrm{h}}20^{\mathrm{m}}18^{\mathrm{s}}.24$ & $-35\degr54\arcmin42\arcsec.7$ \\
& NGC6334IV.3\tablenotemark{a} & $17^{\mathrm{h}}20^{\mathrm{m}}18^{\mathrm{s}}.19$ & $-35\degr54\arcmin52\arcsec.7$ \\\hline
N6334V & NGC6334V & $17^{\mathrm{h}}19^{\mathrm{m}}57^{\mathrm{s}}.55$ & $-35\degr57\arcmin50\arcsec.8$ \\\hline
\enddata
\tablenotetext{a}{There is a typo in the ALMA data archive. Field NGC6334VI in the archive should be NGC6334IV.}
\end{deluxetable}

\begin{deluxetable*}{ccccccc}[t!]
\tablecaption{Parameters of ALMA observations \label{tab:observation}}
\tablecolumns{8}
\tablewidth{0pt}
\tablehead{
\colhead{Date} &
\colhead{Configuration} &
\colhead{$N_{ant}$\tablenotemark{a}} & 
%\colhead{PWV} & 
\colhead{Bandpass} & 
\colhead{Gain} &  
\colhead{Flux} &  
\colhead{Polarization} \\
\colhead{} & \colhead{} &
\colhead{} & 
%\colhead{(mm)} & 
\colhead{calibrator} & \colhead{calibrator} & \colhead{calibrator} & \colhead{calibrator}
}
\startdata
2018 Jun 28 & C43-1 & 47  &J1751+0939 & J1851+0035 & J1751+0939 & J1924-2914 \\
2018 Sep 02 & C43-4 & 44   &J1924-2914 & J1733-3722 & J1924-2914 & J1924-2914 \\
%& 1.4& 1.7& 2.1 & 1.3
\enddata
\tablenotetext{a}{Number of antennas.}

\end{deluxetable*}

Four clumps (N6334I(N), I, IV, and V) in the massive cloud NGC 6334 were observed with ALMA on 2018 June 28 (in C43-1 configuration) and 2018 September 02 (in C43-4 configuration) under the project 2017.1.00793.S (PI: Qizhou Zhang). Tables \ref{tab:sources} and \ref{tab:observation} list the detailed information of the observations. The correlator was configured in the full polarization mode in ALMA band 6 with 3 spectral windows to cover the dust continuum at $\sim$215.5–219.5 GHz and $\sim$232.5–234.5 GHz, and 4 spectral windows to cover the $^{12}$CO (2-1), OCS (19-18), $^{13}$CS (5-4), and N$_2$D$^+$ (3-2) lines. The 3 spectral windows covering the dust continuum have a total bandwidth of 5.6 GHz (three basebands, with 1.875 GHz effective bandwidth each). The line spectral windows have a channel width of 122 kHz (0.16 km s$^{-1}$) over a bandwidth of 58.6 MHz ($\sim$76 km s$^{-1}$). 
%The spectral setups are the same as the ALMA data reported in \citet{2020ApJ...895..142L}.

The data were calibrated by the ALMA supporting staff with Common Astronomy Software Applications \citep[CASA, ][]{2007ASPC..376..127M}. We performed two rounds of phase-only self-calibration on the manually extracted line-free channels of the Stokes $I$ data for the dust continuum using CASA. We imaged the molecular line cubes and Stokes $I$, $Q$, and $U$ maps of dust continuum using the CASA task \textit{TCLEAN} with a Briggs weighting parameter of robust = 0.5. The maps for N6334I(N) and N6334IV are each constructed from three-pointing mosaics. The synthesized beam of the combined (C43-1 plus C43-4) images is $\sim0.7\arcsec\times0.5\arcsec$ ($\sim$0.004-0.003 pc or $\sim$900-700 AU at a distance of 1.3 kpc). The maximum recoverable scale\footnote{https://almascience.eso.org/observing/observing-configuration-schedule/prior-cycle-observing-and-configuration-schedule} is $\sim$13$\arcsec$ ($\sim$0.08 pc at 1.3 kpc). Before primary beam correction, the 1$\sigma$ root-mean-square (RMS) noises are $\sim$0.8, 3.8, 0.6 and 0.8 mJy beam$^{-1}$ for the Stokes $I$ dust continuum maps and $\sim$0.08, 0.09, 0.05, and 0.06 mJy beam$^{-1}$ for the Stokes $Q$ or $U$ dust continuum maps of N6334I(N), I, IV, and V, respectively. The debiased polarized intensity $PI$ and its corresponding uncertainty $\sigma_{PI}$ are calculated as 
$PI = \sqrt{Q^2 + U^2 - \sigma_{QU}^2} $ \citep{2006PASP..118.1340V} and $\sigma_{PI} \sim \sqrt{2}\sigma_{QU}$, where $\sigma_{QU}$ is the 1$\sigma$ rms noise on the background region ($Q \sim U \sim 0$) of the $Q$ or $U$ maps. The polarization position angle $\theta_{\mathrm{p}}$ is estimated with $\theta_{\mathrm{p}} = 0.5 \arctan(U/Q)$. The uncertainty on the polarization position angle \citep{1993A&A...274..968N} is given by  $\delta \theta = 0.5 \sqrt{\sigma_{QU}^2/(Q^2 + U^2}) \sim 20\degr.26 (\sigma_{PI}/PI) \sim 28\degr.65 (\sigma_{QU}/PI)$, where we assume $\sigma_{QU}$ is not far from the observational error of $Q$ and $U$ (i.e., $\delta Q$ and $\delta U$). The RMS noises of the spectral line cubes (before primary beam correction) with a velocity channel width of 0.16 km s$^{-1}$ are $\sim$3.8, 8.7, 3.0, and 5.2 mJy beam$^{-1}$ for N6334I(N), I, IV, and V, respectively. We also imaged several CH$_3$OH lines in the low-resolution continuum spectral windows to derive the gas temperature with the rotation diagram analysis (see Appendix \ref{sec:Tgas}). All the ALMA images shown in this paper are before primary beam correction. The continuum fluxes used for the column density estimation in Appendix \ref{sec:Tgas} are after primary beam correction. 
% The debiased polarized intensity $PI$, debiased polarization percentage $P$, polarization position angle $\theta_{\mathrm{p}}$, and their associated uncertainties are derived with the same procedure as in \citet{2020ApJ...895..142L}.
%For the molecular line observations, we only include the OCS data for the studies of kinematics in this paper. 0.5 km s$^{-1}$ are $\sim$2.2, 5.0, 1.7, and 3.0 mJy beam$^{-1}$
% After primary beam correction, the 1$\sigma$ root-mean-square (RMS) noises are $\sim$1.4, 5.2, 1.2 and 1.0 mJy beam$^{-1}$ for the Stokes $I$ dust continuum maps and $\sim$0.14, 0.15, 0.09, and 0.10 mJy beam$^{-1}$ for the Stokes $Q$ or $U$ dust continuum maps of N6334I(N), I, IV, and V, respectively.

\subsection{JCMT 850 $\mu$m dust polarization and $^{13}$CO (3-2) data}
We adopt the 850 $\mu$m ($\sim$353 GHz) $I$, $Q$, and $U$ images observed with SCUBA-2/POL-2 \citep{2013MNRAS.430.2513H, 2016SPIE.9914E..03F} on the James Clerk Maxwell Telescope (JCMT) at a resolution of $\sim$14$\arcsec$ ($\sim$0.09 pc) toward the whole NGC 6334 filamentary cloud. The JCMT data (program code: M17BL011) were previously published by \citet{2021A&A...647A..78A} as part of the JCMT large program B-field In STar-forming Region Observations \citep[BISTRO, ][]{2017ApJ...842...66W}. For area with signal-to-noise ratio (SNR) greater than 25 for the $I$ map, the mean values for the observational error of $I$, $Q$, and $U$ (i.e., $\delta I$, $\delta Q$, and $\delta U$) are $\sim$1.6, 1.13, and 1.6 mJy beam$^{-1}$, respectively \citep{2021A&A...647A..78A}. The debiased polarized intensity $PI$ and its corresponding error\footnote{It should be noted that the propagated observational error $\delta PI$ for the JCMT and Planck polarization map is position-dependent and is a different type of uncertainty from the RMS noise $\sigma_{PI}$ for the ALMA polarization map. There is $\delta PI \sim \delta Q \sim \delta U$ but $\sigma_{PI} \sim \sqrt{2}\sigma_{QU} \sim \sqrt{2}\delta Q \sim \sqrt{2}\delta U$. } $\delta PI$ are calculated as $PI = \sqrt{Q^2 + U^2 - 0.5(\delta Q^2 + \delta U^2)}$ and $ \delta PI = (Q \delta Q +  U \delta U)/\sqrt{Q^2 + U^2} \sim \sqrt{(Q^2 \delta Q^2 +  U^2 \delta U^2)/(Q^2 + U^2)}$, respectively. The polarization position angle $\theta_{\mathrm{p}}$ and its uncertainty $\delta \theta$ \citep{1993A&A...274..968N} are estimated with $\theta_{\mathrm{p}} = 0.5 \arctan(U/Q) $ and $\delta \theta = 0.5 \sqrt{(Q^2 \delta U^2 +  U^2 \delta Q^2)/(Q^2 + U^2)^2} \sim 28\degr.65 (\delta PI/PI)$, respectively, where we assume $\delta PI \sim \delta Q \sim \delta U$.
%The debiased $PI$, debiased $P$, $\theta$, and their associated uncertainties are derived with the same procedure as in \citet{2021A&A...647A..78A}. 

Additionally, we include in our analysis the $^{13}$CO (3-2) line cubes toward N6334I(N) and N6334I taken with the Heterodyne Array Receiver Program and Auto-Correlation Spectrometer
and Imaging System \citep[HARP and ACSIS, ][]{2009MNRAS.399.1026B} from the JCMT data archive (program code: M11BN07). The spatial and spectral resolutions of the $^{13}$CO (3-2) data are $\sim$14$\arcsec$ and 0.055 km s$^{-1}$, respectively. The map size is $2\arcmin \times 2\arcmin$ ($\sim$0.76 pc $\times$ 0.76 pc) for each field. The RMS noises of N6334I(N) and N6334I are 0.90 and 0.45 K per channel, respectively, in corrected antenna temperature ($T^\ast _{\mathrm{A}}$). The pipeline-produced data cubes in the barycentric velocity frame are converted to the kinematic local standard of rest (LSRK) radio velocity frame with Starlink \citep{2014ASPC..485..391C}. Because the two clumps are larger than the beam, we estimate the antenna radiation temperature ($T_{\mathrm{R}}^\ast$) from $T^\ast _{\mathrm{A}}$ adopting a forward efficiency\footnote{https://www.eaobservatory.org/jcmt/instrumentation/heterodyne/harp/} of $\eta_{\mathrm{fss}}=$0.75.

\subsection{Planck 353 GHz dust polarization data}
Planck maps towards the NGC 6334 region and its surrounding area observed with the High Frequency Instrument \citep[HFI, ][]{2010A&A...520A...9L} at 353 GHz are included in our analysis to study the global-scale density structure and magnetic fields. We adopt the 353 GHz Stokes $Q$ and $U$ maps of the thermal dust emission \citep[version R3.00, ][]{2020A&A...641A...4P} constructed with the Generalized Needlet Internal Linear Combination method \citep[GNILC, ][]{2011MNRAS.418..467R} and the earlier released dust optical depth ($\tau_{353}$) and temperature maps \citep[version R1.02, ][]{2014A&A...571A..11P}. The Planck maps are at a resolution of 5$\arcmin$ ($\sim$1.9 pc). Within our considered map area, the mean values for the uncertainties of $Q$ and $U$ (i.e., $\delta Q$ and $\delta U$) are $\sim$3 and 4 $\mu$K$_{\mathrm{CMB}}$, respectively. The debiased polarized intensity $PI$ and its corresponding uncertainty $\delta PI$ are calculated as $PI = \sqrt{Q^2 + U^2 - 0.5(\delta Q^2 + \delta U^2)}$ and $ \delta PI \sim \sqrt{(Q^2 \delta Q^2 +  U^2 \delta U^2)/(Q^2 + U^2)}$, respectively. The adopted Planck $Q$ and $U$ maps downloaded from the Planck Legacy Archive\footnote{http://pla.esac.esa.int/} are in galactic coordinates.  We estimate the polarization position angle in equatorial coordinates with $\theta_{\mathrm{p}} = 0.5 \arctan(U/Q) - \Delta\theta_{\mathrm{p}}^{\mathrm{g-e}}$, where
\begin{equation}
    \Delta\theta_{\mathrm{p}}^{\mathrm{g-e}} =  \arctan( \frac{ \cos(l-32.9\degr) }{ \cos b \cot 62.9\degr - \sin b  \sin(l-32.9\degr)} )
\end{equation}
is the angle between the galactic and equatorial reference directions \citep{1998MNRAS.297..617C}. For NGC 6334 at $l=351.33\degr$ and $b=0.68\degr$, we adopt $\Delta\theta_{\mathrm{p}}^{\mathrm{g-e}} \approx 55.22\degr$. Similar to the JCMT data, the uncertainty on the Planck polarization position angle is given by $\delta \theta \sim 28\degr.65 (\delta PI/PI)$.
% (map size $2\deg \times 2 \deg$ or $\sim$45 pc $\times$ 45 pc)

\subsection{NANTEN2 $^{12}$CO (1-0) data}
We also include in our analysis the $^{12}$CO (1-0) data from \citet{2018PASJ...70S..41F} to study the global-scale velocity fields. The data were obtained with NANTEN2, which is a 4 m millimeter/sub-millimeter radio telescope in Chile. The spatial and spectral resolutions of the  $^{12}$CO (1-0) cubes are $\sim$3$\arcmin$ ($\sim$1.1 pc) and 0.16 km s$^{-1}$, respectively. The typical RMS noise level is $\sim$1.2 K per channel. In this study, the NANTEN2  $^{12}$CO (1-0) data is convolved to a beam size of 5$\arcmin$ to match the Planck resolution. 

\section{Results} \label{sec:results}
\subsection{Dust continuum and magnetic fields}
In this subsection, we briefly overview the multi-scale magnetic field structures in the NGC 6334 region traced by Planck, JCMT, and ALMA dust polarization observations. Assuming that the observed linear dust polarization is due to dust grain alignment, the dust polarization position angle is rotated by 90$\degr$ to reveal the magnetic field orientation. It is possible that the observed polarization of the ALMA dust emission peaks is affected by other possible dust polarization mechanisms \citep[e.g., disk self-scattering or dichroic extinction,][]{2018ApJ...856L..27G, 2021ApJ...914...25L}. But these mechanisms that are predominant at scales smaller than 100-200 AU should not be significant for our ALMA observations with a resolution of $\sim$900 AU. 

\begin{figure}[!htbp]
 \gridline{\fig{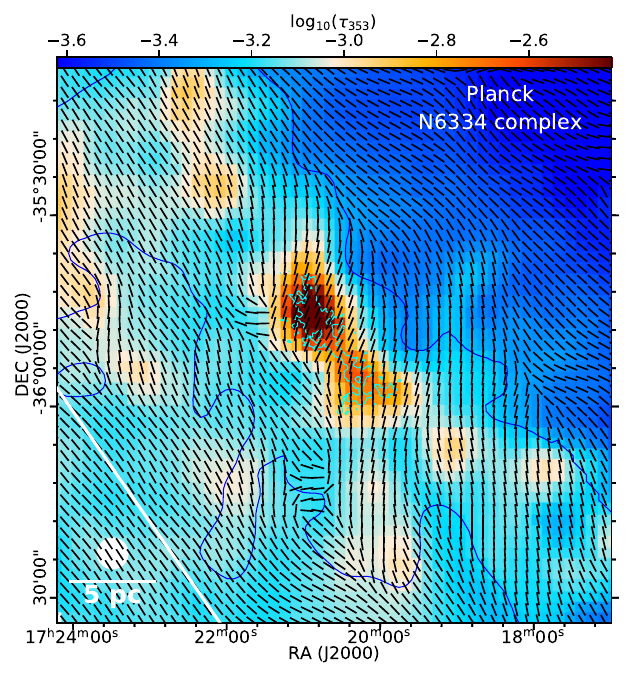}{0.48\textwidth}{(a)}
  \fig{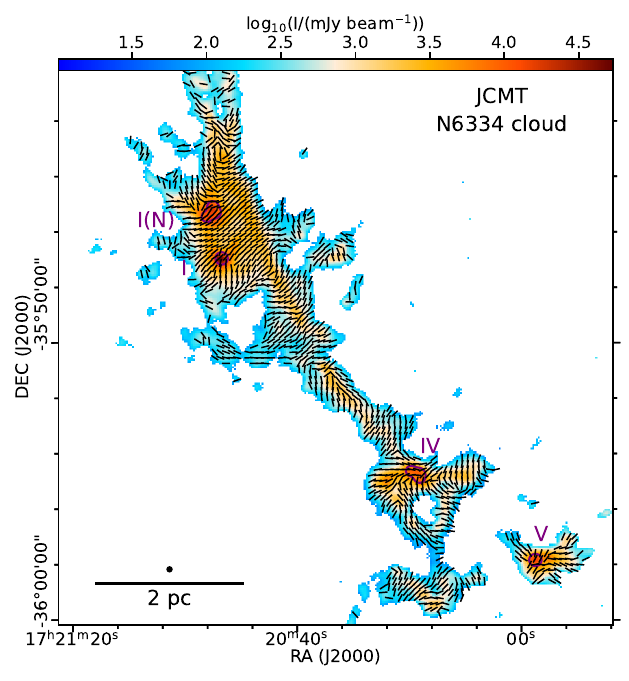}{0.48\textwidth}{(b)}
 }
\caption{(a) Planck magnetic field orientations (black line segments) overlaid on the dust optical depth map (colorscales) toward the NGC 6334 complex. Line segments are of arbitrary length. All Planck polarization detections have signal-to-noise ratios (SNR) greater than 3. The 5$\arcmin$  ($\sim$1.9 pc) beam (white circle), a scale bar of 5 pc, and the galactic plane (white line) are indicated in the lower left corner. The 10 pc-long NGC 6334 filament is elongated along the direction of the galactic plane. The cyan contour indicates the region with SNR($I$)=25 for the JCMT observation. The blue contour indicates the region with NANTEN2 integrated  $^{12}$CO (1-0) intensity greater than 25 K km s$^{-1}$ within which we perform the relative orientation analysis. (b) JCMT magnetic field orientations (black line segments) overlaid on the total intensity map of the dust emission (colorscales) toward the NGC 6334 filament from BISTRO observations \citep{2021A&A...647A..78A}. Line segments are of arbitrary length. Only line segments with SNR($I$)$>$25 and SNR($PI$)$>$3 are shown. The 14$\arcsec$ ($\sim$0.09 pc) beam (black circle) and a scale bar of 2 pc are indicated in the lower left corner. Purple contours indicate the Full Width at Half-Maximum (FWHM) field of view of our ALMA observations. \label{fig:N6334_large}}
\end{figure}
% White dashed circles show regions with radii of 70$\arcsec$ ($\sim$0.44 pc) surrounding each clump.

Figure \ref{fig:N6334_large}(a) shows the magnetic field orientation of the NGC 6334 complex traced by Planck dust polarization observations, which has been briefly reported by \citet{2021A&A...647A..78A}. The well-ordered magnetic fields in the diffuse region surrounding NGC 6334 mostly show a northeast-southwest orientation, which tends to be parallel to the galactic plane and the NGC 6334 filament. Toward the 10-pc long predominant NGC 6334 filament, the magnetic field in the northern part (containing N6334I(N) and N6634I) of the filament changes to be mostly perpendicular to the main filament and is pinched toward the northern end. The magnetic field in the southern part (containing N6334IV and N6634V) of the filament also deviates from the global-scale magnetic field and changes to a north-south orientation.
%\citet{2021A&A...647A..78A} has reported Planck and JCMT polarization observations toward the NGC 6334 region. 

Figure \ref{fig:N6334_large}(b) shows the JCMT BISTRO observations of the magnetic field orientation in the NGC 6334 filament \citep{2021A&A...647A..78A}. The magnetic field near N6334I and N6334I(N) is mostly perpendicular to the filament and shows dragged-in structures toward the north of N6334I(N) and the south of N6334I, which agrees with the large-scale magnetic field revealed by Planck. The magnetic field in N6334IV and N6334V shows complex structures without a prevailing orientation. 

\begin{figure*}[!htbp]
 \gridline{\fig{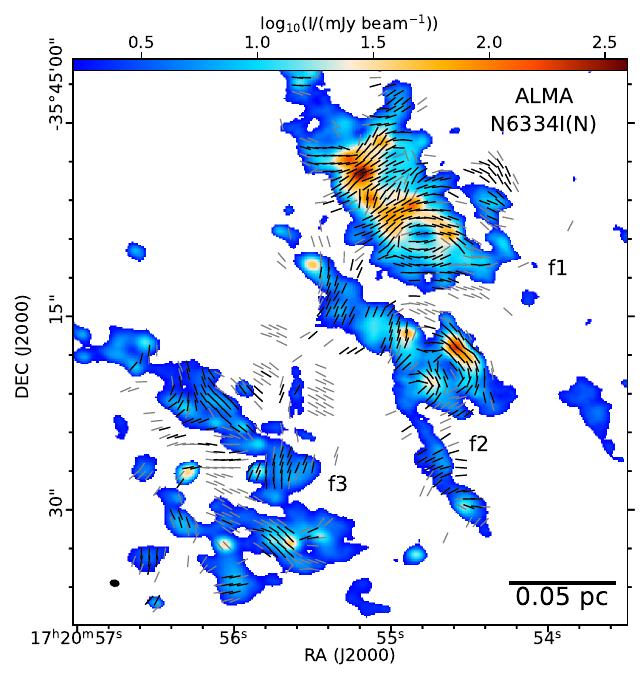}{0.48\textwidth}{}
 \fig{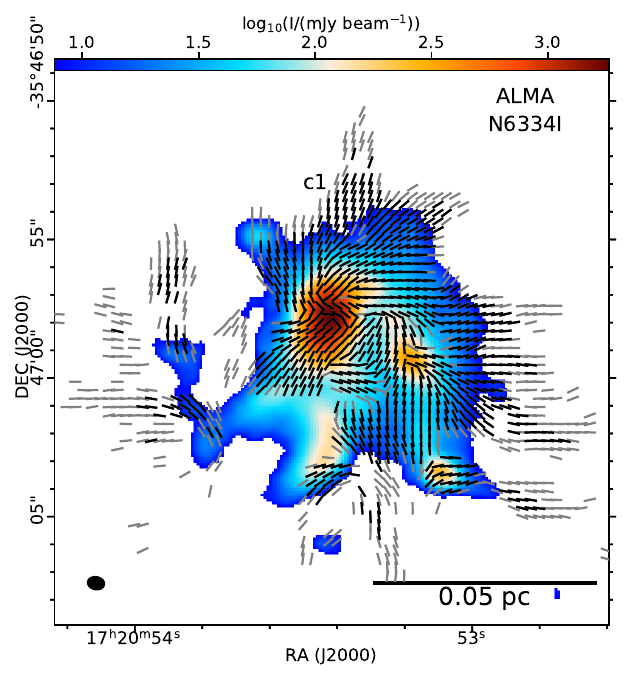}{0.48\textwidth}{}
 }
  \gridline{\fig{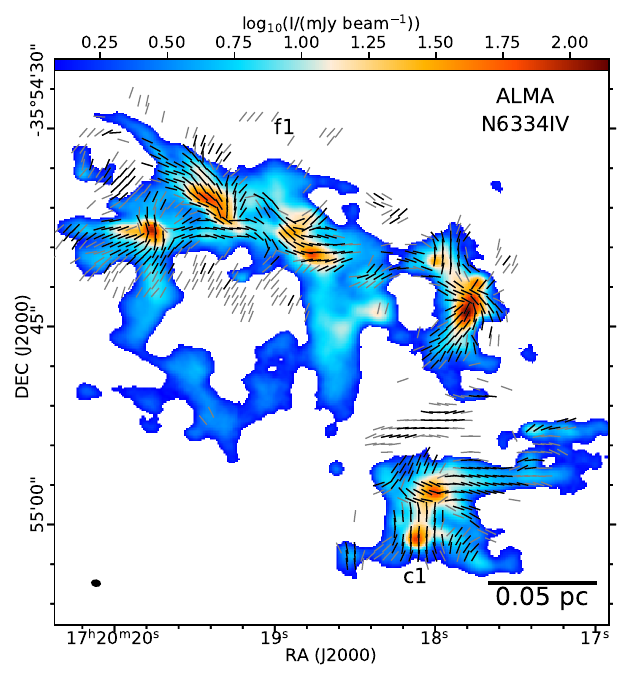}{0.48\textwidth}{}
 \fig{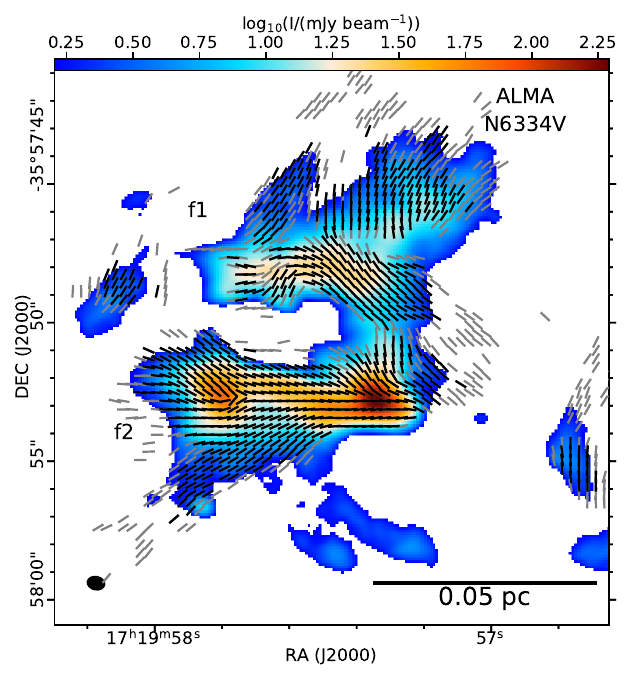}{0.48\textwidth}{}
 }
\caption{ALMA observations (C43-1 and C43-4 combined) toward the massive clumps N6334I(N), I, IV, and V. Magnetic field orientations (black and grey line segments indicate SNR($PI$)$>$3 and 2$<$SNR($PI$)$<$3, respectively) are overlaid on the total intensity map (SNR($I$)$>$2) of the dust emission (colorscales). Line segments are of arbitrary length. The $\sim0.7\arcsec\times0.5\arcsec$ ($\sim$0.004-0.003 pc or $\sim$900-700 AU) synthesized beam (black ellipse) and a scale bar of 0.05 pc are indicated in the lower left and right corner of each panel, respectively.\label{fig:N6334_alma_B}}
\end{figure*}
% The map edges are noisy due to primary beam correction.  Grey dashed contours correspond to the FWHM field of view of the ALMA observations 

Our ALMA polarization observations have revealed the continuum emission structures and magnetic field structures in the four massive clumps (N6334I(N), I, IV, and V) down to a resolution of $<1000$ AU. Figure \ref{fig:N6334_alma_B} shows the ALMA observational results of the magnetic field orientation in the four clumps. Clump N6334I(N) is resolved into three parallel 0.1-0.2 pc long elongated filamentary structures (hereafter I(N)-f1, f2, and f3 from north to south) that follow the direction of the N6334 main filament and are perpendicular to the clump-scale magnetic field revealed by JCMT. Each 0.1 pc-scale filament fragments into a group of compact sources. The magnetic field in the predominant I(N)-f1 is overall perpendicular to the direction of the elongation and shows pinched (or hourglass) field morphology, which agrees with the previous ALMA polarization observations with poorer sensitivity \citep{2021ApJ...923..204C}. The magnetic field in I(N)-f2 and I(N)-f3 are more complex, which might be affected by turbulence or (proto-)stellar feedback. In N6334I, the dominating 0.1 pc core (I-c1) fragments into a cluster of compact sources. The magnetic field in the core shows radial patterns in the outer region, pinched structures near fainter emission peaks, and spiral-like structures near the brightest emission peak, which deviates from the coherent and approximately straight field at cloud and clump scales \citep{2014ApJ...792..116Z, 2015Natur.520..518L, 2021A&A...647A..78A}. The magnetic field pattern in core I-c1 seems to indicate that the field is dragged by gravity and influenced by core/disk rotation. The radial field patterns in the outer region may be related to accretion streamers that are connected to the central core. Clump N6334IV is resolved into an elongated filamentary structure (IV-f1) containing several compact sources in the north and a core (IV-c1) in the south. In the central part of IV-f1, the magnetic field is mostly along the elongation of IV-f1, which agrees with previous SMA polarization observations \citep{2014ApJ...792..116Z}. Around the two compact sources in the east and west ends of IV-f1, the magnetic field orientation shows bimodal distributions, which may suggest a magnetized collapse \citep{2018MNRAS.477.2760M}. The northern part of core IV-c1 shows a prevailing east-west magnetic field orientation, while the southern part shows a prevailing north-south orientation. Clump N6334V is resolved into two nearly parallel elongated structures (V-f1 and V-f2 from north to south) connected with each other in the western part. The magnetic field in V-f2 shows a dominant east-west orientation, which follows the elongation of V-f2 and agrees with previous SMA observations \citep{2017ApJ...844...44J}. The magnetic field also shows a radial pattern in the eastern end of V-f2 and signs of pinched structure in the western end. The magnetic field in V-f1 is overall complex but shows well ordered and consistent field in some sub-regions. 

\subsection{Molecular lines and velocity fields}\label{sec:linem1}

\begin{figure}[!htbp]
 \gridline{
 \fig{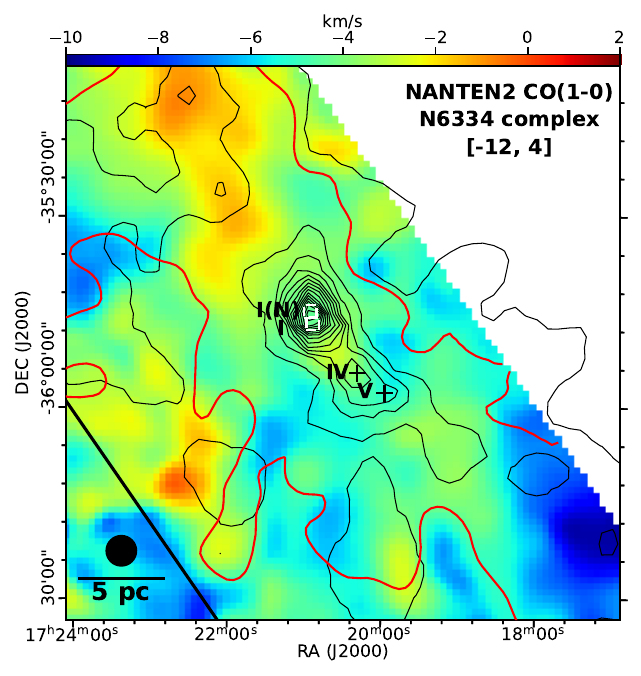}{0.33\textwidth}{(a)}
 \fig{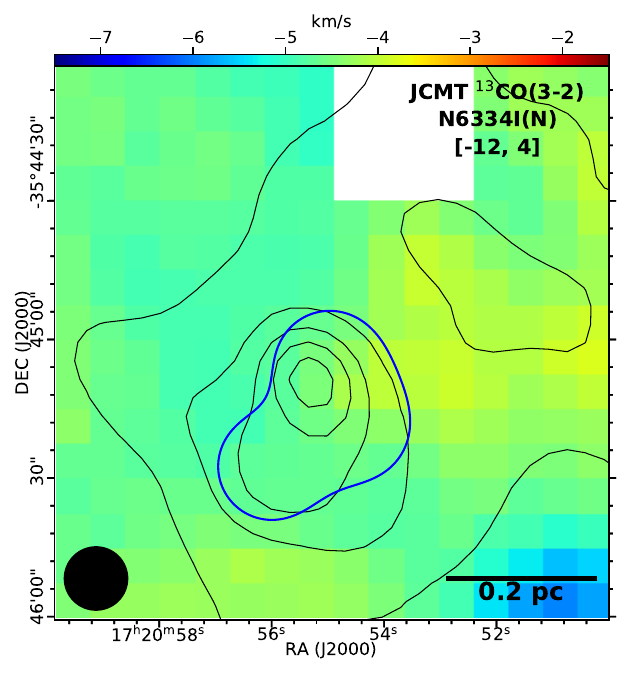}{0.33\textwidth}{(b)}
 \fig{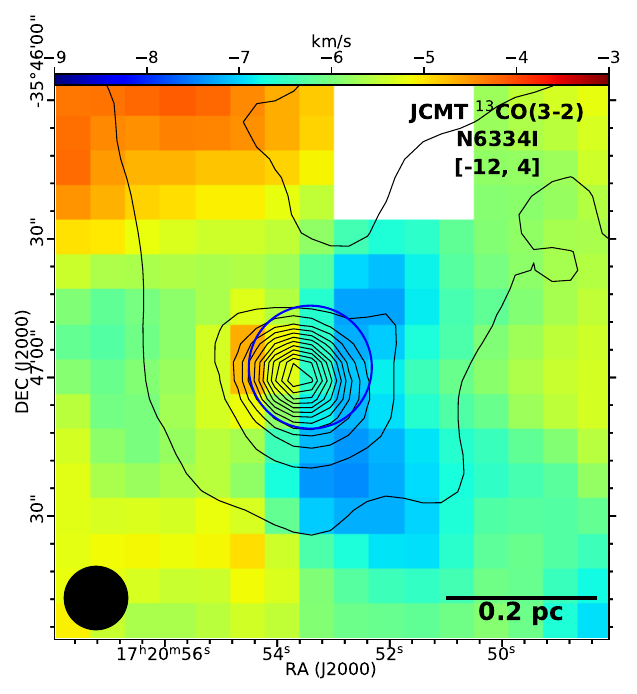}{0.33\textwidth}{(c)}
 }
\caption{(a) Velocity centroid map of NANTEN2  $^{12}$CO (1-0) line emission toward NGC 6334 complex \citep{2018PASJ...70S..41F}. The line data is convolved to a beam size of 5$\arcmin$. The black contour levels correspond to the Planck 353 GHz optical depth ($\tau_{353}$) map. The contour starts at 0.0004 and continues with an interval of 0.0004. The white rectangles indicate the map area of the JCMT fields toward N6334I(N) and I in (b) and (c). Black crosses indicate the positions of N6334IV and V. The red contour indicates the region with NANTEN2 integrated $^{12}$CO (1-0) intensity greater than 25 K km s$^{-1}$ within which we perform the relative orientation analysis. (b)-(c) Velocity centroid maps of JCMT $^{13}$CO (3-2) line emission toward N6334I(N) and N6334I. The black contour levels correspond to the JCMT 850 $\mu$m dust continuum map. Contour starts at 2 Jy beam$^{-1}$ and continues with an interval of 4 Jy beam$^{-1}$. Blue contours show the FWHM field of view of our ALMA observations. \label{fig:N6334_large_line_m1}}
\end{figure}

\begin{figure*}[!htbp]
 \gridline{\fig{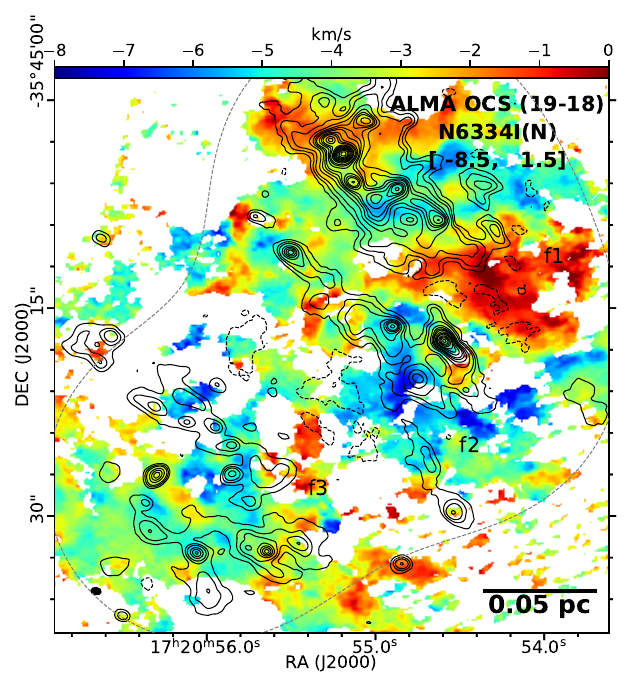}{0.33\textwidth}{(a)}
 \fig{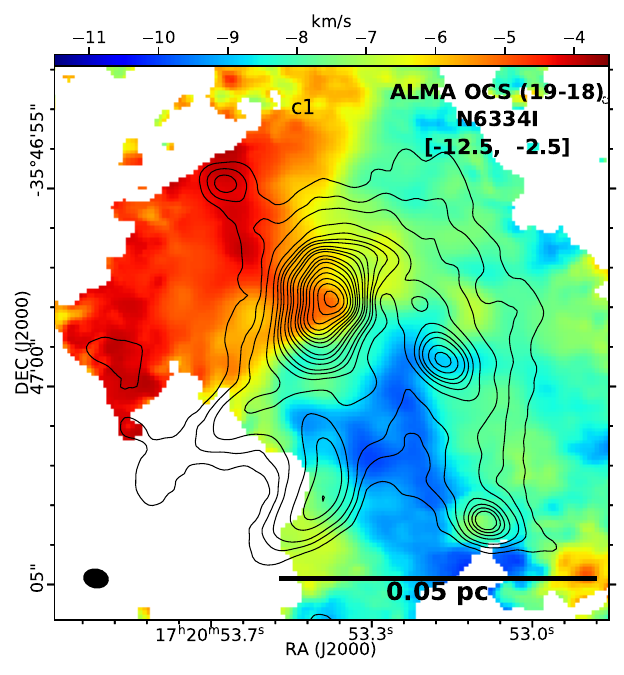}{0.33\textwidth}{(b)}
\fig{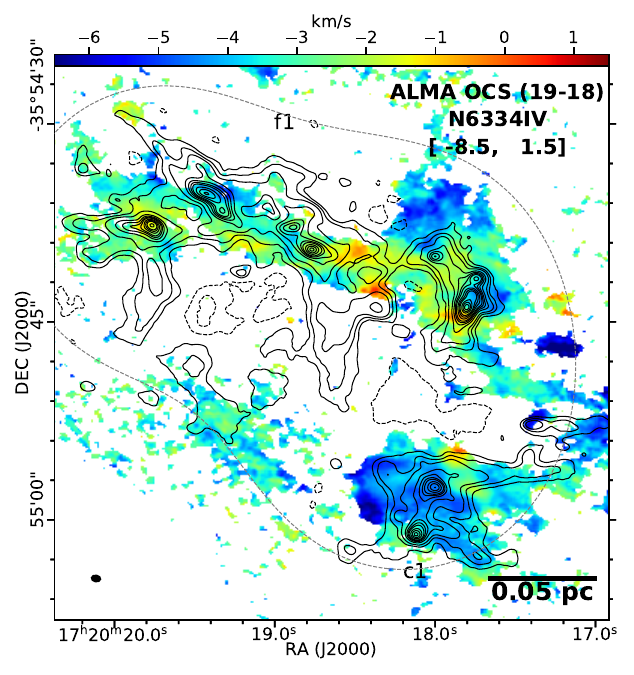}{0.33\textwidth}{(c)}
 }
 \gridline{\fig{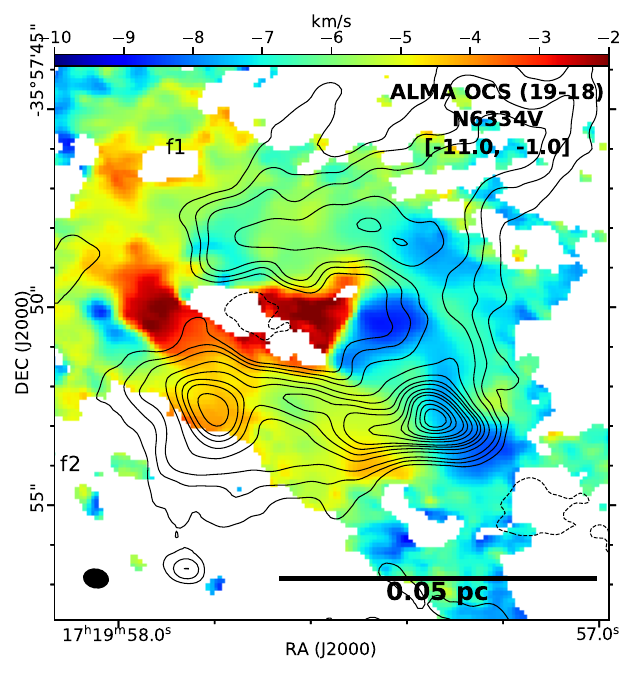}{0.33\textwidth}{(d)}
 \fig{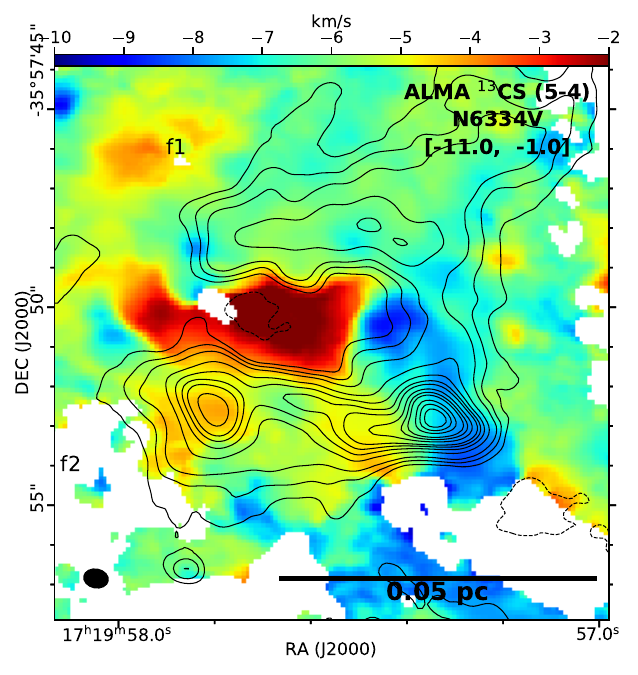}{0.33\textwidth}{(e)}
 }
\gridline{\fig{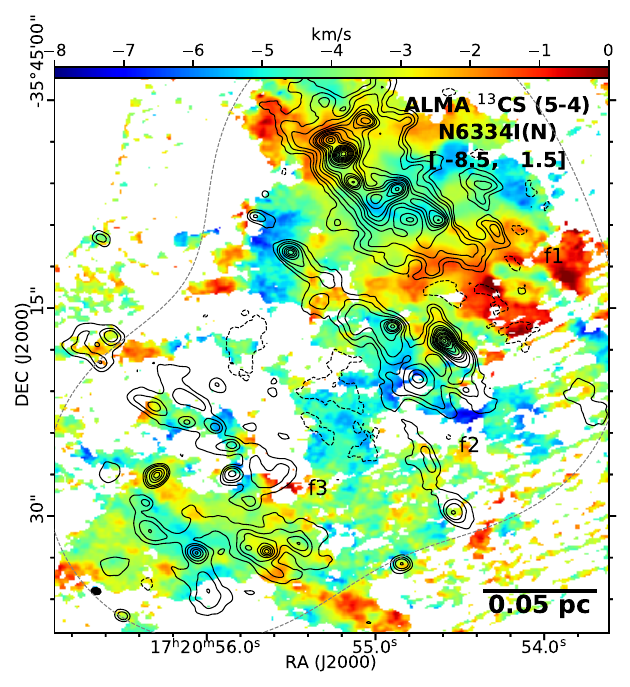}{0.33\textwidth}{(f)}
 \fig{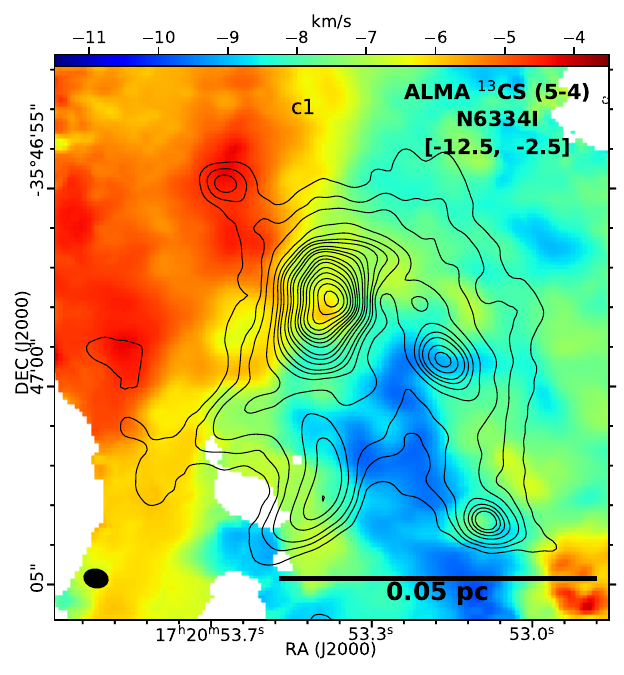}{0.33\textwidth}{(g)}
 \fig{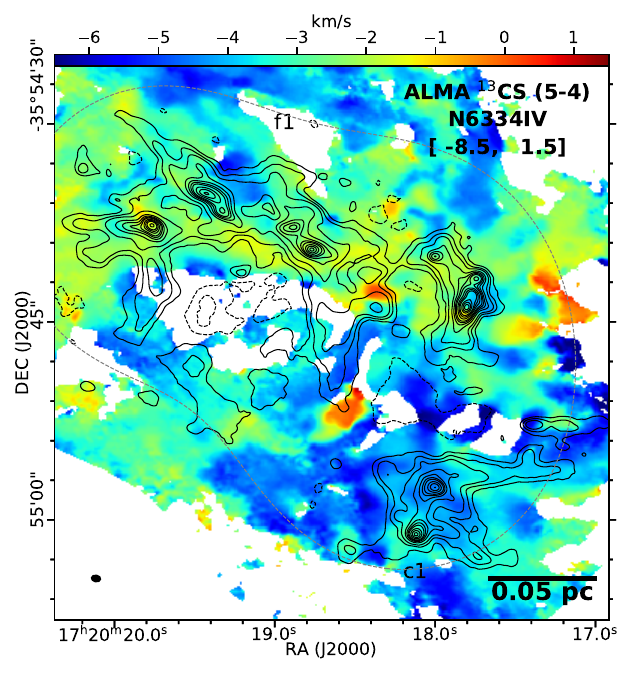}{0.33\textwidth}{(h)}
 }
\caption{Velocity centroid maps of ALMA OCS ((a)-(d)) and $^{13}$CS ((e)-(h)) observations. The black contour levels correspond to the ALMA dust continuum map. Contour levels are ($\pm$3, 6, 10, 20, 30, 40, 50, 70, 90, 110, 150, 180, 210, 250, 290, 340, 390, 450) $\times \sigma_{I}$, where $\sigma_{I}$ is the RMS noise of the Stokes $I$ maps (see Section \ref{sec:obsalma}). Grey dashed contours correspond to the FWHM field of view of the ALMA observations. \label{fig:N6334_alma_line_m1}}
\end{figure*}

\begin{figure*}[!htbp]
 \gridline{\fig{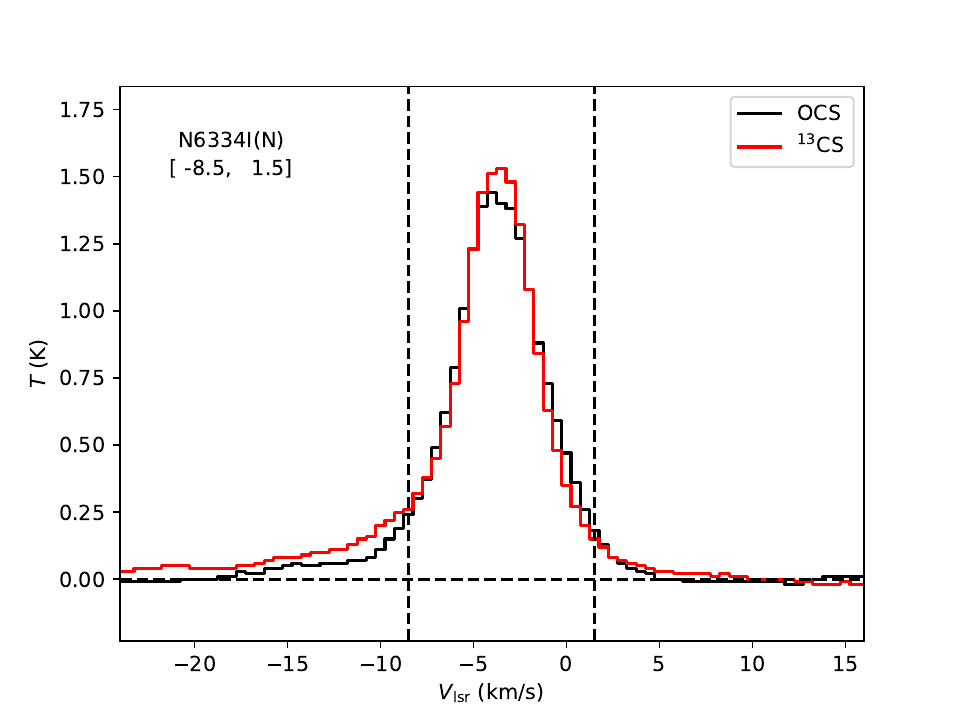}{0.4\textwidth}{}
 \fig{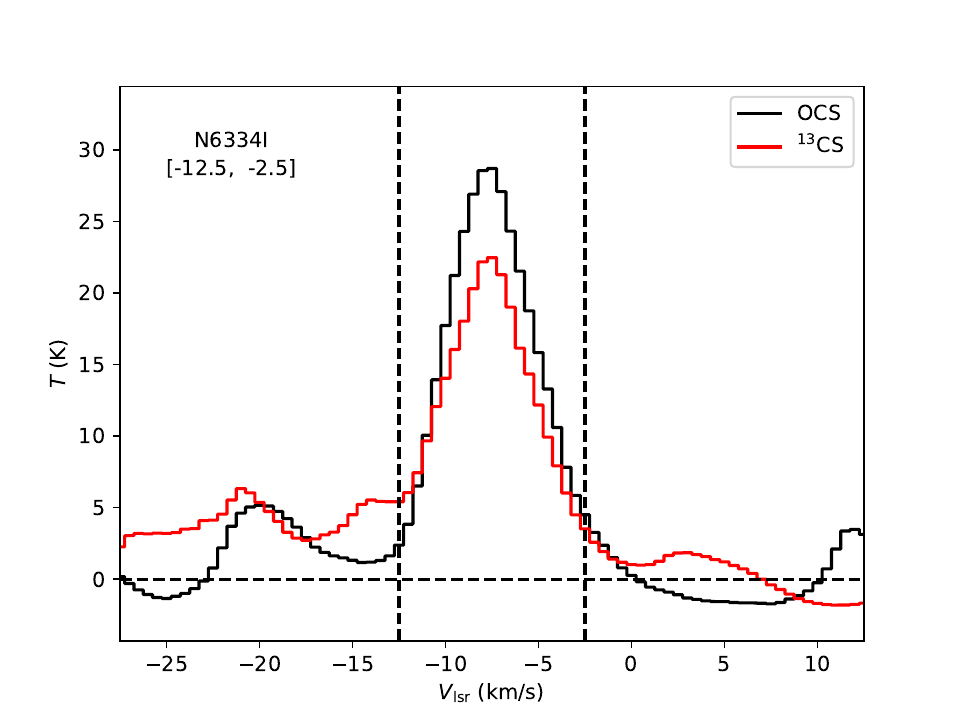}{0.4\textwidth}{}
 }
  \gridline{\fig{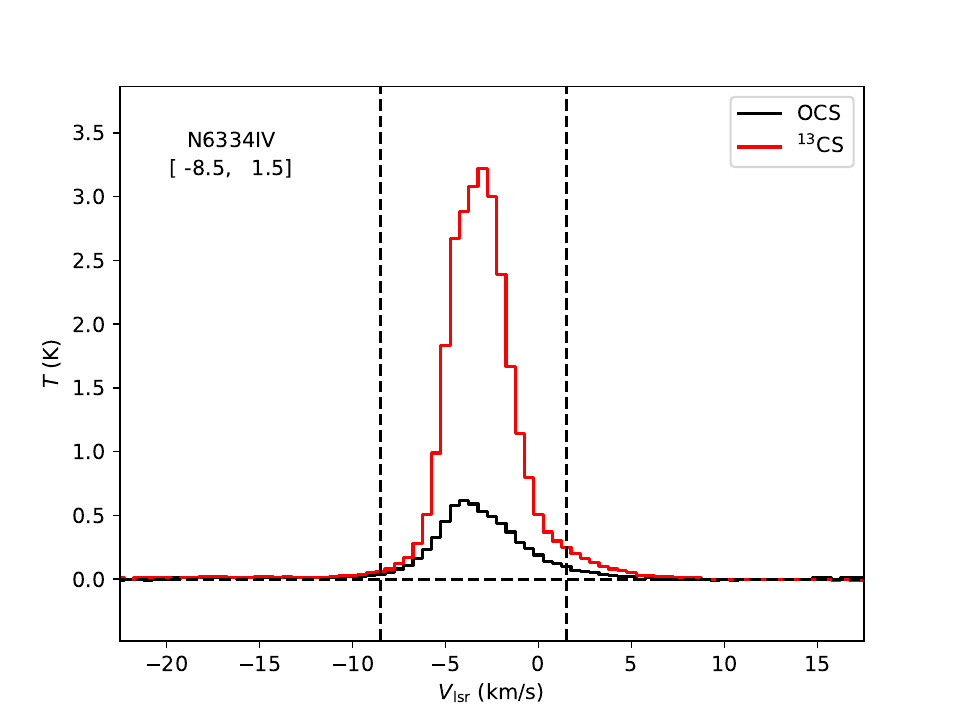}{0.4\textwidth}{}
 \fig{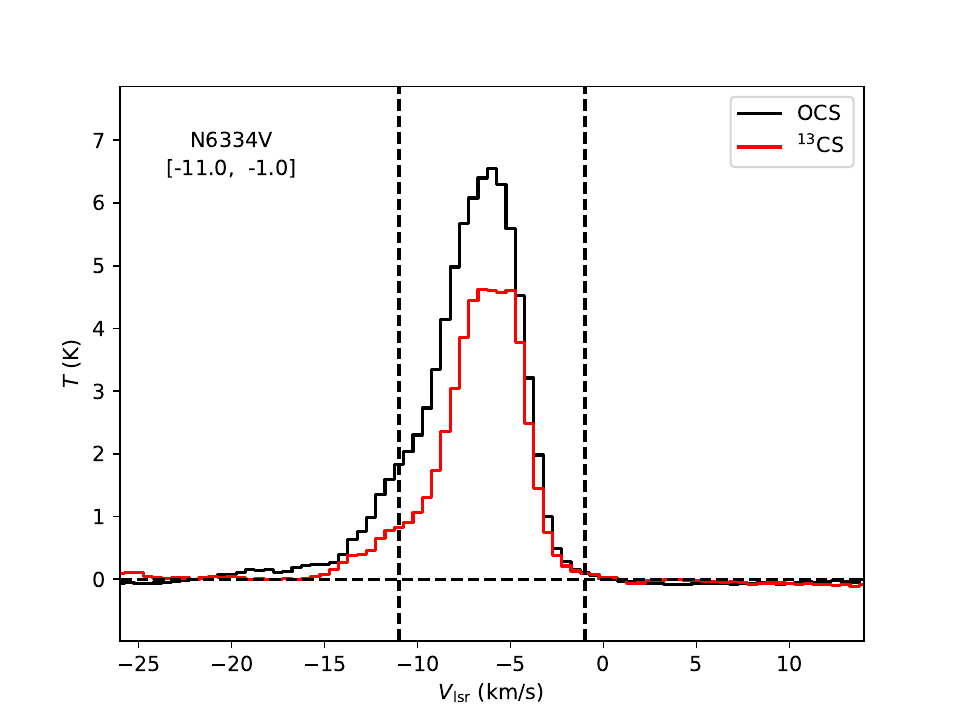}{0.4\textwidth}{}
 }
\caption{The averaged ALMA OCS (black histogram) and $^{13}$CS (red histogram) spectra of the considered area for each clump within which we perform the relative orientation analysis. The vertical dashed lines indicate the velocity range within which we calculate the integrated intensity and velocity centroid.  \label{fig:N6334_alma_spec}}
\end{figure*}

We use NANTEN2  $^{12}$CO (1-0), JCMT $^{13}$CO (3-2), and ALMA OCS and $^{13}$CS data to study the kinematics at different scales. In this subsection, we briefly overview the multi-scale intensity-weighted velocity (moment 1 or velocity centroid $V_c$) structures (Figures \ref{fig:N6334_large_line_m1} and \ref{fig:N6334_alma_line_m1}) in the NGC 6334 region. The integrated intensity (moment 0) maps of these lines are shown in Appendix \ref{sec:linem0}. The velocity centroid $V_c(\boldsymbol{x})$ at position $\boldsymbol{x}$ is calculated with 
\begin{equation}
V_c(\boldsymbol{x}) = \frac{\Sigma_i^{N_{\mathrm{ch}}} I_i(\boldsymbol{x}) v_i \Delta v_{\mathrm{ch}}}{\Sigma_i^{N_{\mathrm{ch}}} I_i(\boldsymbol{x}) \Delta v_{\mathrm{ch}}},
\end{equation}
where $I_i(\boldsymbol{x})$, $v_i$, $\Delta v_{\mathrm{ch}}$, and $N_{\mathrm{ch}}$ are the 
line intensity, line-of-sight velocity, channel width, and number of integrated channels, respectively. The propagated uncertainty of the calculated velocity centroid is given by \citep{1985ApJ...295..479D, 2019RNAAS...3...74T}
\begin{equation}
\delta V_c(\boldsymbol{x}) = \frac{\sigma_{ch} \Delta v_{\mathrm{ch}}\sqrt{\Sigma_i^{N_{\mathrm{ch}}} (v_i - V_c(\boldsymbol{x}))^2}}{\Sigma_i^{N_{\mathrm{ch}}} I_i(\boldsymbol{x}) \Delta v_{\mathrm{ch}}},
\end{equation}
where $\sigma_{ch}$ is the noise of one spectral channel (reported in Section \ref{sec:observation}).  For the NANTEN2  $^{12}$CO (1-0) and JCMT $^{13}$CO (3-2) observations, we only consider the line emission from -12 to 4 km s$^{-1}$ since most of the large-scale line emission in the NGC 6334 region is within this velocity range \citep{2022A&A...660A..56A}. A second and fainter velocity component in the NGC 6334 region from -20 to -12 km s$^{-1}$ has been previously reported \citep{2018PASJ...70S..41F} but is not considered in this work. At small scales and near young stellar objects, the outflow usually dominates at $\gtrsim$5 km s$^{-1}$ \citep[e.g.,][]{2009ApJ...696...66Q, 2018ApJ...860..106L} with respect to the local-standard-of-rest (LSR) velocity ($V_{\mathrm{lsr}}$) of the central source within massive star formation regions. The low-velocity ($<$5 km s$^{-1}$) outflowing gas is usually indistinguishable from the clump bulk gas. Thus, we only consider velocities within $\sim$5 km s$^{-1}$ with respect to the $V_{\mathrm{lsr}}$ of each clump for the ALMA OCS and $^{13}$CS lines. The LSR velocities are $\sim$-3.5, -7.5, -3.5, and -6 km s$^{-1}$ for N6334I(N), I, IV, and V, respectively. In Figure \ref{fig:N6334_alma_spec}, we indicate the considered velocity ranges for each clump on the averaged ALMA OCS and $^{13}$CS spectra. 

% The white dashed circles shows regions with radii of 70$\arcsec$ ($\sim$0.44 pc) surrounding each clump.

Figure \ref{fig:N6334_large_line_m1}(a) shows the velocity centroid map of the NGC 6334 complex traced by NANTEN2  $^{12}$CO (1-0) observations \citep{2018PASJ...70S..41F}. The velocity structures of NGC 6334 and its surrounding material are coherent and there is a global velocity gradient of 0.1 km s$^{-1}$ pc$^{-1}$ from northeast to southwest along the direction of the galactic plane, but the origin of this global velocity gradient is still unclear \citep{2022A&A...660A..56A}. 
%The velocity centroid variation within the NGC 6334 filament is relatively small.

Figures \ref{fig:N6334_large_line_m1}(b) and (c) show the velocity centroid map of N6334I(N) and I traced by JCMT $^{13}$CO (3-2) observations. The velocity centroid variation is small in N6334I(N), which might be because this clump is at an early star formation stage \citep{2008hsf2.book..456P}. There is a large-scale velocity gradient from northeast to southwest in N6334I, which agrees with the global velocity gradient seen in Figure \ref{fig:N6334_large_line_m1}(a). 

Figure \ref{fig:N6334_alma_line_m1} shows the velocity centroid map of N6334I(N), I, IV, and V traced by ALMA OCS and $^{13}$CS observations. The velocity centroid maps of the two lines are very similar. In N6334I(N)-f1, there is a clear gradient from northeast to southwest, which should have a local origin at core scales since this gradient is not seen in JCMT observations at clump scales (Figures \ref{fig:N6334_large_line_m1}(b)). The gradient is reversed near the southwest edge of N6334I(N)-f1, which may indicate local converging flows. In N6334I, the dominating northeast-southwest velocity gradient agrees with the large-scale and global-scale velocity gradients seen by JCMT and NANTEN2 (Figure \ref{fig:N6334_large_line_m1}). In N6334IV, the velocity centroid variation is relatively small compared to the other 3 clumps and there are no clear signs of ordered velocity gradients. In N6334V-f2, there is a clear east-west gradient, and the gradient is reversed in the west edge, which agrees with previous SMA H$^{13}$CO$^+$ and CH$_3$OH observations at a resolution of 2$\arcsec$ \citep{2017ApJ...844...44J}. \citet{2017ApJ...844...44J} interpreted this velocity structure as converging flows. In N6334V-f1, we do not find the east-west gradient previously reported by \citet{2017ApJ...844...44J}, which may be due to the difference in the beam resolution, filtering scale, or line excitation condition between our and their observations. 
%In N6334V-f1, there are no clear signs of ordered velocity gradients, which is in contrast to \citet{2017ApJ...844...44J} that has reported an east-west gradient. 

\section{Relative orientation analysis and discussion} \label{sec:RON}
%between magnetic fields, density gradient, gravity direction, and velocity gradient
The relative orientation between magnetic field ($\theta_{\mathrm{B}}$), gas column density gradient ($\theta_{\mathrm{NG}}$), local gravity ($\theta_{\mathrm{LG}}$), and velocity centroid gradient ($\theta_{\mathrm{VG}}$) and their varying trend with column density are informative of the physical conditions of star-forming regions \citep{2012ApJ...747...79K, 2013ApJ...774..128S, 2017ApJ...835...41G}. In NGC 6334, the information on the magnetic field orientation and its uncertainty is available from the dust polarization observation. We implement a $3\times3$ Sobel kernel \citep[e.g., ][]{2013ApJ...774..128S} on the column density maps (see Appendix \ref{sec:Tgas}) and line moment 1 maps (see Section \ref{sec:linem1}) to derive the column density gradient ($\theta_{\mathrm{NG}}$) and velocity centroid gradient\footnote{It should be noted that the term ``velocity gradient'' in our analysis refers to the local velocity gradient and is different from the subblock-averaged velocity gradient in the VGT.}($\theta_{\mathrm{VG}}$) at different positions. The uncertainties of the column density and velocity centroid gradients are calculated following \citet{2016AA...586A.138P}. The calculation of the gradients and their uncertainties is described in detail in Appendix \ref{sec:uncer}. Considering the significant SNR and the rather uniformly distributed observational error for the dust continuum emission observations, the uncertainty on the orientation of the column density gradient should be negligible \citep[e.g.,][]{2016AA...586A.138P}. Taking into account the gas mass of pixels with SNR($I$)$>$3 (see Appendix \ref{sec:Tgas}), we calculate the map-wise 2D direction of local gravitational force ($\theta_{\mathrm{LG}}$) with the standard formula of gravitation \citep[e.g.,][]{2012ApJ...747...79K, 2020ApJ...895..142L}. Calculating the uncertainty on the local gravity direction is time-consuming and we are unable to do so due to our limited computer resources.

Combining the approaches of the KTH method \citep{2012ApJ...747...79K} and the HRO analysis \citep[][]{2013ApJ...774..128S}, we calculate and study the angular difference among these orientations. We use the alignment measure ($AM$) parameter introduced by the Velocity Gradient Technique \citep[VGT, ][]{2017ApJ...835...41G, 2018ApJ...853...96L} to characterize the alignment between different orientations. The $AM$ is given by
\begin{equation} \label{eq:am}
AM = \langle \cos (2\phi_{o1}^{o2}) \rangle,
\end{equation}
where $\phi_{\mathrm{o1}}^{\mathrm{o2}} = \vert \theta_{\mathrm{o1}} - \theta_{\mathrm{o2}} \vert$ is the angle between orientation 1 ($\theta_{\mathrm{o1}}$) and orientation 2 ($\theta_{\mathrm{o2}}$) and is in the range of 0 to 90$\degr$. In the calculation of $AM$ within each column density bin, different pixels are weighted equally. The $AM$ is in the range of -1 (perpendicular) to 1 (parallel). $AM>0$ (i.e., approximately $\overline{\phi_{\mathrm{o1}}^{\mathrm{o2}}}<45\degr$) indicates two orientations are statistically more parallel than perpendicular in the considered region and vice versa. The uncertainty of $\phi_{\mathrm{o1}}^{\mathrm{o2}}$ is given by $\delta\phi_{\mathrm{o1}}^{\mathrm{o2}} = \sqrt{\delta\theta_{\mathrm{o1}}^2+\delta\theta_{\mathrm{o2}}^2}$. As discussed above, we adopt $\delta\theta=0$ for the column density gradient and gravity direction. We exclude data points with $\delta\phi>10\degr$ in our analysis.
The uncertainty of $AM$ is given by (see Appendix \ref{sec:uncer})
\begin{equation} \label{eq:dam}
\delta AM = \sqrt{(\langle (\cos (2\phi_{o1}^{o2}))^2 \rangle - AM^2 + \Sigma_i^{n'} (2\sin (2\phi_i) \delta \phi_i)^2)/n'},
\end{equation}
where $n'$ is the number of data points considered.

We calculate the $AM$ for different relative orientations at different column densities. A NANTEN2 integrated  $^{12}$CO (1-0) intensity of 25 K km s$^{-1}$ approximately separates the emission from the NGC 6334 complex and the background galactic plane emission \citep{2018PASJ...70S..41F}, thus we exclude positions with NANTEN2 integrated  $^{12}$CO (1-0) intensity smaller than 25 K km s$^{-1}$ for the Planck and NANTEN2 maps in our analysis. We consider every pixel with SNR$>$3 detection in the JCMT and ALMA maps. Similar to \citet{2016AA...586A.138P}, we calculate $AM$ in different $N_{\mathrm{H_2}}$ bins containing approximately equal number of pixels for each instrument (10, 15, and 15 $N_{\mathrm{H_2}}$ bins for Planck/NANTEN2, JCMT, and ALMA, respectively). The typical number of pixels per bin\footnote{The number of pixels per bin varies for different relative orientations due to the different detection area for the total dust emission, polarized dust emission, and molecular line emission.} is $\sim$120-170, $\sim$200-900, and $\sim$200-1000 for Planck/NANTEN2, JCMT, and ALMA, respectively. We test and find that varying the number of $N_{\mathrm{H_2}}$ bins by a factor of 2 does not significantly affect the general trend on the relative orientation-column density (RO-N) relations. For the JCMT observation, we only derive the RO-N relation for $\phi_{VG}^{B}$, $\phi_{VG}^{NG}$, and $\phi_{VG}^{LG}$ in N6334I(N) and N6334I in 5 $N_{\mathrm{H_2}}$ bins with $\sim$20-50 pixels per bin. Figures \ref{fig:N6334_NG_LG}, \ref{fig:N6334_B_NG}, \ref{fig:N6334_B_LG}, \ref{fig:N6334_VG_B}, \ref{fig:N6334_VG_NG}, and \ref{fig:N6334_VG_LG} show the relative orientation between $\theta_{\mathrm{B}}$, $\theta_{\mathrm{NG}}$, $\theta_{\mathrm{LG}}$, and $\theta_{\mathrm{VG}}$ characterized by $AM$ as functions of column density. Because the atmospheric emission as well as the extended emission outside of the SNR-based masks (ASTMASK and PCAMASK) in the data reduction process are filtered out for POL-2 observations and the ALMA observation filters the extended spatial emission limited by the minimal separation of antenna pairs, the JCMT and ALMA observations can underestimate the actual column density. For NGC 6334, the JCMT observation filters out the large-scale emission corresponding to $N_{\mathrm{H_2}} \sim 3 \times 10^{22}$ cm$^{-2}$ \citep{2021A&A...647A..78A}. Our ALMA observation filters out large-scale emissions at scales of $>$0.08 pc, but the filtered column density at this scale is unclear. Thus, the $AM$ at the similar $N_{\mathrm{H_2}}$ but from different instruments are not comparable. On the other hand, the highest $N_{\mathrm{H_2}}$ bin of Planck/NANTEN2 observations contains the area of the NGC 6334 filament covered by the JCMT observation and the highest $N_{\mathrm{H_2}}$ bin of JCMT observations contains the area of N6334I(N), I, IV, and V covered by the ALMA observation. Thus, we should regard the Planck/NANTEN2, JCMT, and ALMA observations as tracing low, intermediate, and high column densities, respectively. 
%, we also derive the relative orientation-column density relation (RO-N relation) for $\phi_{LG}^{NG}$, $\phi_{B}^{NG}$, and $\phi_{B}^{LG}$ for individual clumps ($r\sim0.44$ pc region), and
%$N_{\mathrm{H_2}}$ smaller than the median $N_{\mathrm{H_2}}$ ($\sim 2 \times 10^{22}$ cm$^{-2}$) of our considered region
%we should not compare the $AM$ at low $N_{\mathrm{H_2}}$ bins for JCMT observations with the $AM$ at the same $N_{\mathrm{H_2}}$ bins for Planck observations. Similarly, we should not compare the $AM$ at low $N_{\mathrm{H_2}}$ bins for ALMA observations with the $AM$ at the same $N_{\mathrm{H_2}}$ bins for JCMT observations due to the minimal separation of antenna pairs of ALMA observations. 
%We look into each RO-N relation in detail below. 
%NG to IG. 

\subsection{Column density gradient versus local gravity}\label{sec:NG_LG}

%Different symbols represent different regions.
The relative orientation between column density gradient and local gravity ($\phi_{LG}^{NG}$) may indicate how effectively gravity can shape the density structure. 

\begin{figure*}[!htbp]
 \gridline{\fig{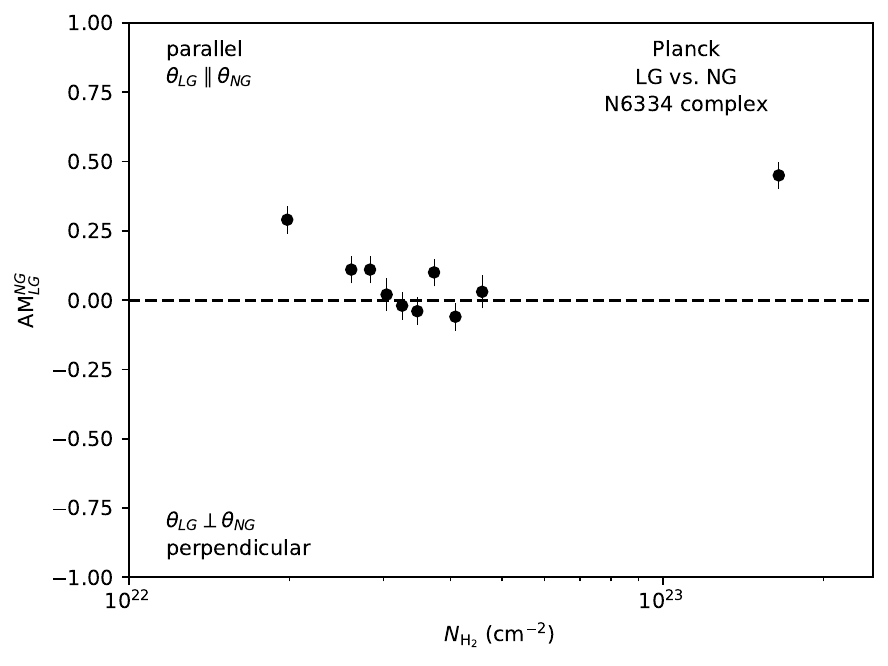}{0.33\textwidth}{}
 \fig{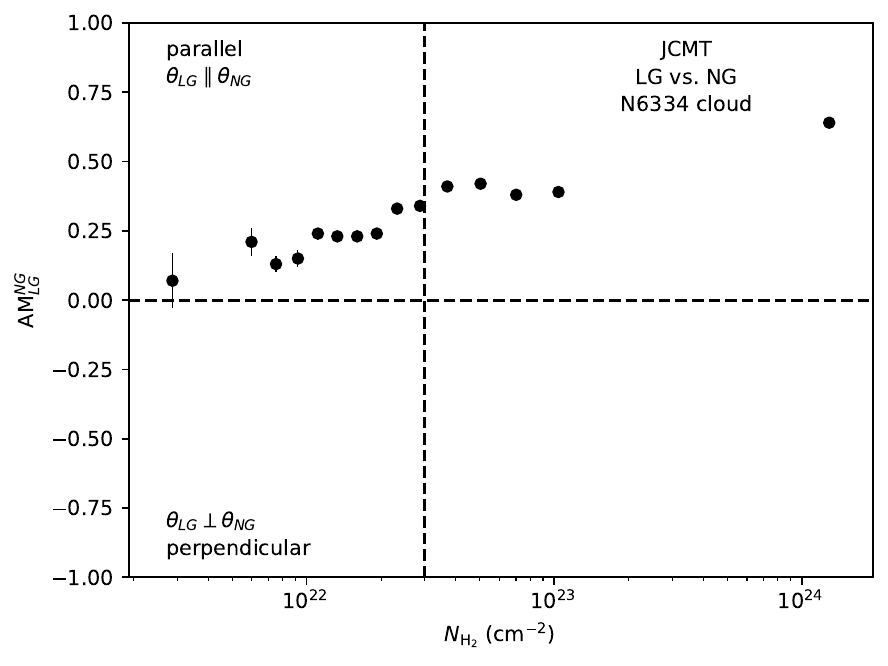}{0.33\textwidth}{}
 \fig{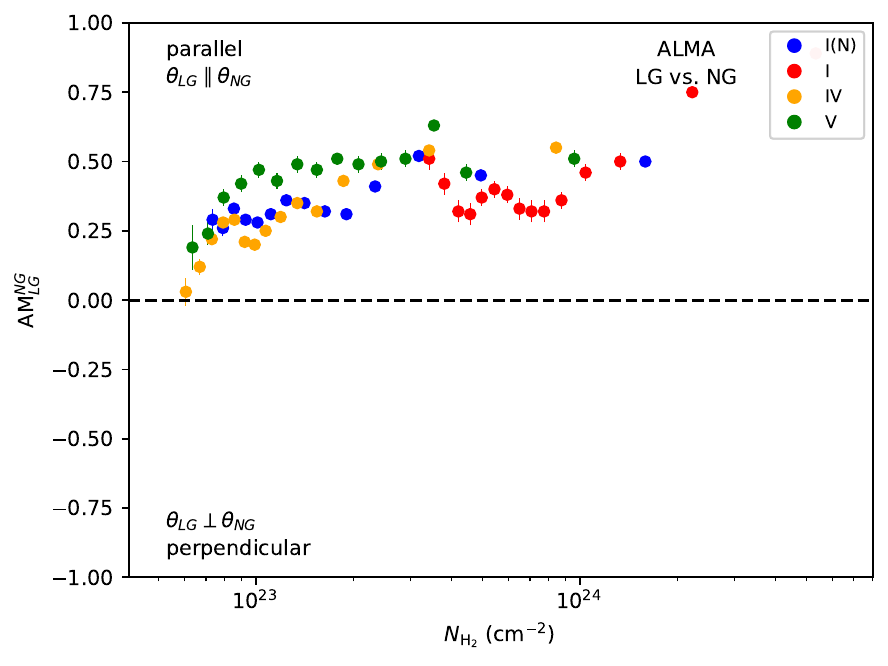}{0.33\textwidth}{}
 }
\caption{Relative orientations (characterized by $AM$. See Equation \ref{eq:am}) between column density gradient ($\theta_{\mathrm{NG}}$) and local gravity ($\theta_{\mathrm{LG}}$) as a function of column density for Planck (left), JCMT (middle), and ALMA (right) observations. Different colors indicate different clumps. The JCMT observation filters out the large-scale emission corresponds to $N_{\mathrm{H_2}} \sim 3 \times 10^{22}$ cm$^{-2}$ \citep{2021A&A...647A..78A} which is indicated by the vertical dashed line. The ALMA observation filters out the large-scale emission at scales $>$0.08 pc. The highest $N_{\mathrm{H_2}}$ bin of Planck observations contains the area of the NGC 6334 filament covered by the JCMT observation. The highest $N_{\mathrm{H_2}}$ bin of JCMT observations contains the area of the N6334I(N), I, IV, and V covered by the ALMA observation. The absolute column densities from different instruments are not comparable. Planck, JCMT, and ALMA observations trace low, intermediate, and high column densities, respectively. $AM>0$ and $AM<0$ indicate a statistically more parallel and perpendicular alignment, respectively. \label{fig:N6334_NG_LG}}
\end{figure*}

Figure \ref{fig:N6334_NG_LG} shows the RO-N relation for $\phi_{LG}^{NG}$. For the Planck observation, there is $AM\sim0$ in most $N_{\mathrm{H_2}}$ bins, but the value of $AM$ is clearly positive in the lowest and highest $N_{\mathrm{H_2}}$ bins. While the highest $N_{\mathrm{H_2}}$ bin corresponds to the NGC 6334 main filament, the $AM>0$ at the lowest $N_{\mathrm{H_2}}$ bin may just be a coincidence of specific geometry since the gravity is not expected to be significant in diffuse regions. 
At higher density revealed by JCMT and ALMA, the two angles are always statistically more parallel than perpendicular ($AM>0$). For the JCMT observation toward the whole filament, we see that $AM$ increases with $N_{\mathrm{H_2}}$. For ALMA observations toward individual clumps, the trend of increasing $AM$ with $N_{\mathrm{H_2}}$ persists. Thus, we suggest that gravity plays an increasingly important role in shaping the density structure at higher densities. 
% It should be noted that $AM\sim0$ does not necessarily indicate an uniform distribution of $\phi_{LG}^{NG}$, but there could be local regions in each $N_{\mathrm{H_2}}$ bin where $\theta_{\mathrm{NG}}$ and $\theta_{\mathrm{LG}}$ are more parallel or more perpendicular. 
%be due to a directional bias of our NANTEN2  $^{12}$CO (1-0) mask for the NGC 6334 complex
% except in N6334I, where $AM$ does not show a clear relation with $N_{\mathrm{H_2}}$. Toward the same clump, the $AM$ for ALMA observations tends to be lower than that for JCMT observations, which may be because of the accretion or rotation activities near the emission center \citep{2012ApJ...747...79K} or because the complex fragmentary density structure at higher column density is smoothed by the coarser resolution of JCMT. 

\subsection{Magnetic field versus column density gradient}\label{sec:B_NG}

The angle between the magnetic field and column density gradient ($\phi_{B}^{NG}$) is complementary to the angle between the magnetic field and column density contour ($\phi_{B}^{N}$) that has been extensively studied by the HRO analysis\footnote{The alignment measure parameter ($AM_{B}^{NG}$) for the magnetic field and column density gradient adopted by this work should not be confused with the HRO shape parameter \citep[$\xi$, ][]{2016AA...586A.138P} for the magnetic field and column density contour adopted by the HRO analysis. $AM_{B}^{NG}<0$ approximately corresponds to $\xi>0$, and vice versa.} both observationally \citep[e.g.,][]{2016AA...586A.138P, 2022ApJ...926..163K, 2020ApJ...904..168B} and numerically \citep[e.g.,][]{2013ApJ...774..128S, 2017A&A...603A..64S, 2020MNRAS.497.4196S, 2021MNRAS.507.5641G}. A detailed review of the observational and numerical HRO studies can be found in \citet{2022FrASS...9.3556L}.

\begin{figure*}[!htbp]
 \gridline{\fig{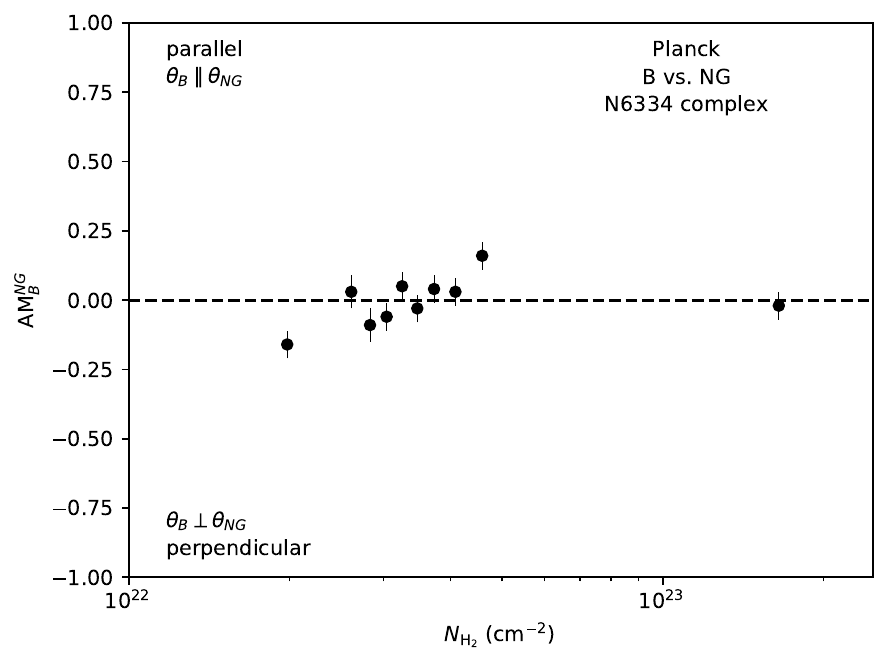}{0.33\textwidth}{}
 \fig{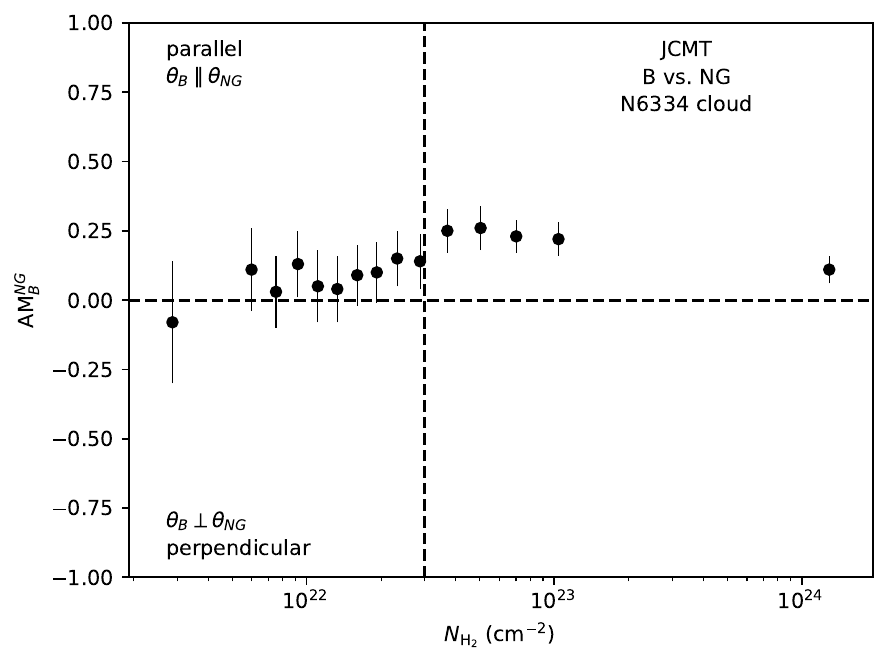}{0.33\textwidth}{}
 \fig{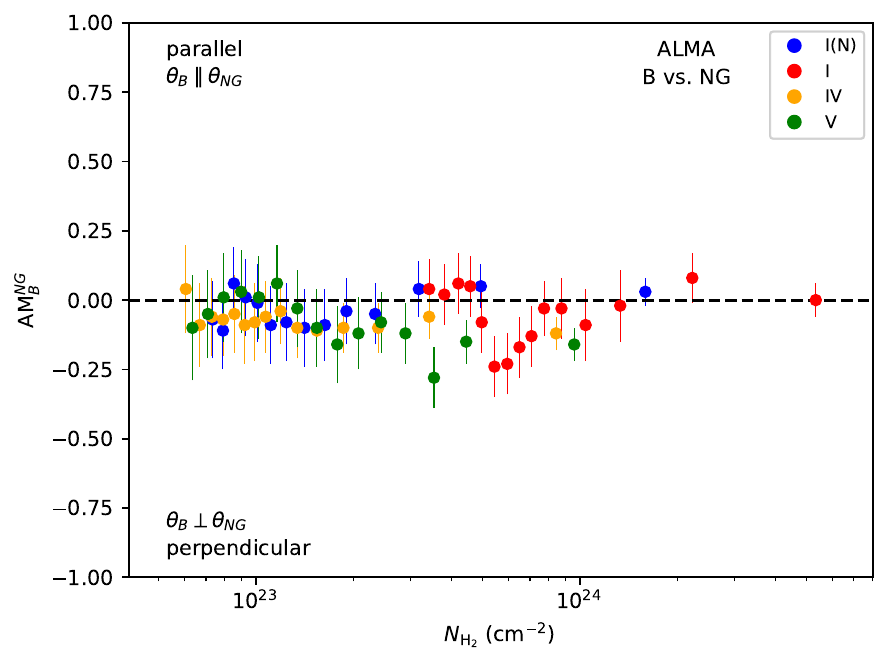}{0.33\textwidth}{}
 }
\caption{Same as Figure \ref{fig:N6334_NG_LG} but for the relative orientation between magnetic field ($\theta_{\mathrm{B}}$) and column density gradient ($\theta_{\mathrm{NG}}$). \label{fig:N6334_B_NG}}
\end{figure*}

Figure \ref{fig:N6334_B_NG} shows the RO-N relation for $\phi_{B}^{NG}$ from the Planck, JCMT, and ALMA observations. For the Planck observation, the overall statistical trend is that the magnetic field and column density gradient change from a statistically slightly more perpendicular alignment ($AM_{B}^{NG} \lesssim 0$) at lower column densities to a slightly more parallel alignment ($AM_{B}^{NG} \gtrsim 0$) at higher column densities. At the highest $N_{\mathrm{H_2}}$ bin, the alignment measure of two angles transits back to $AM_{B}^{NG}\sim0$ (i.e., no preferred orientation), which might be due to insufficient resolution. The transition from $AM_{B}^{NG}<0$ to $AM_{B}^{NG}>0$ is in agreement with trans-to-sub-Alfv\'{e}nic simulations in numerical HRO studies \citep[see a review in][]{2022FrASS...9.3556L}, which suggests the NGC 6334 is trans-to-sub-Alfv\'{e}nic at complex/cloud scale. Similar trans-to-sub-Alfv\'{e}nic states have been reported in the Gould Belt clouds from previous observational HRO and VGT studies \citep{2016AA...586A.138P, 2019NatAs...3..776H}. The statistically more perpendicular alignment between the magnetic field and column density gradient (i.e., more parallel alignment between the magnetic field and column density contour) at low column densities may be due to the stretch of an initially super-Alfv\'{e}nic turbulence or due to the intrinsic property of a large-scale sub-Alfv\'{e}nic turbulence \citep[see][ and references therein]{2022FrASS...9.3556L}. The direct reason for the transition from $AM_{B}^{NG}<0$ to $AM_{B}^{NG}>0$ is still under debate \citep[][]{2022FrASS...9.3556L}. We refrain from deriving the transition column density for $AM_{B}^{NG}\sim0$ due to the uncertainty of our estimated column densities (see Appendix \ref{sec:Tgas}) and the inconsistency of absolute column densities between different instruments. 
At intermediate column densities revealed by JCMT, the two angles are mostly statistically more parallel ($AM>0$). The value of $AM_{B}^{NG}$ increases with $N_{\mathrm{H_2}}$ and then decreases with $N_{\mathrm{H_2}}$. Because the NGC 6334 region also tends to be trans-to-sub-Alfv\'{e}nic at intermediate and high column densities (see discussions in Section \ref{sec:VG_B} below), the more parallel alignment at this $N_{\mathrm{H_2}}$ range cannot be due to a local super-Alfv\'{e}nic turbulence as proposed by some numerical studies \citep[e.g.,][]{2016ApJ...829...84C}, but is more likely due to the interaction between the magnetic field and local gravity \citep[see discussions in Section \ref{sec:B_LG} and][]{2021MNRAS.507.5641G}. 
At even higher $N_{\mathrm{H_2}}$ bins revealed by ALMA, the alignment measure of two angles transits back to $AM_{B}^{NG}\lesssim0$ (i.e., no preferred orientation or slightly more perpendicular). The reason for the reverse transition is also unclear, but may be related to the impact of converging gas flows, outflows, disk rotation, and/or the projection effect \citep{2022FrASS...9.3556L}. It should be noted that our results do not conflict with \citet{2015Natur.520..518L} which have found that the area-averaged magnetic field orientation and density structure orientation are perpendicular to each other at different scales in NGC 6334. This is because the global average statistics in \citet{2015Natur.520..518L} and the local statistics in our work trace different physics. i.e., While the global ordered magnetic field could guide gravitational collapse and lead to self-similar fragmentation \citep{2015Natur.520..518L}, the local field orientation can be distorted by gravity or affected by star formation activities.
% revealed by Planck 
% has been found in previous observational HRO studies and
% \citep[e.g., N6334V][]{2017ApJ...844...44J}  \citep[e.g., G327.3][]{2020ApJ...904..168B}
%The column density gradient and the column density contours are complementary to each other. Because the Planck data has $AM_{B}^{NG}<0$ for most $N_{\mathrm{H_2}}$ bins and the JCMT data filters out the large-scale emission, we refrain from determining the transition column density for $AM_{B}^{NG}\sim0$. 
% Because the JCMT data has $AM_{B}^{NG}>0$ at the filtering column density ($N_{\mathrm{H_2}} \sim 3 \times 10^{22}$ cm$^{-2}$) and the Planck data has $AM_{B}^{NG}\lesssim0$ at this column density, we suggest that the transition column density for $AM_{B}^{NG}\sim0$ is not far from $N_{\mathrm{H_2}} \sim 3 \times 10^{22}$ cm$^{-2}$.

\subsection{Magnetic field versus local gravity}\label{sec:B_LG}

The relative orientation between the magnetic field and local gravity ($\phi_{B}^{LG}$) may indicate how effectively gravity can shape the magnetic field structure and how effectively the magnetic field can resist gravitational collapse \citep{2012ApJ...747...79K}. 

\begin{figure*}[!htbp]
 \gridline{\fig{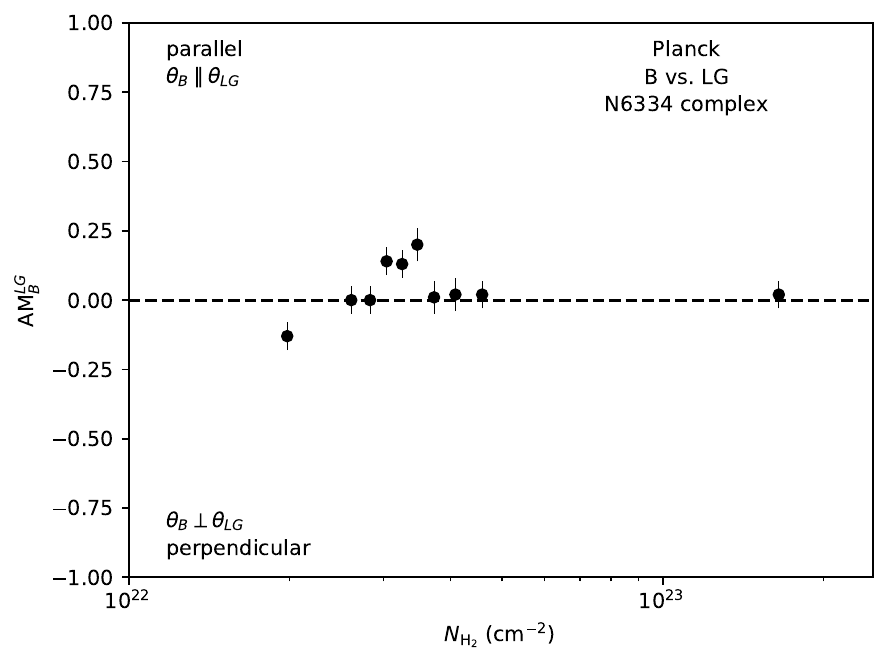}{0.33\textwidth}{}
 \fig{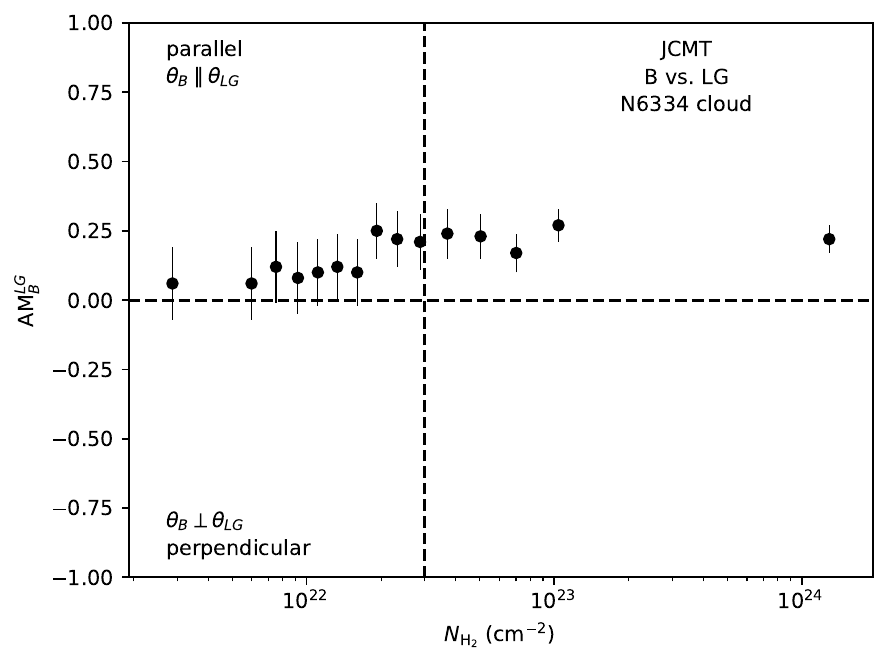}{0.33\textwidth}{}
 \fig{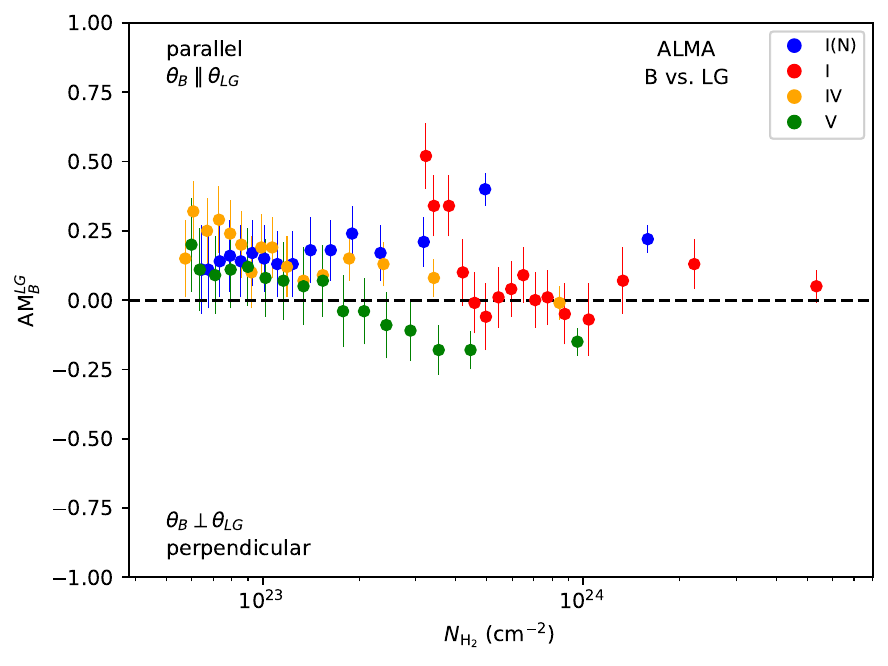}{0.33\textwidth}{}
 }
\caption{Same as Figure \ref{fig:N6334_NG_LG} but for the relative orientation between magnetic field ($\theta_{\mathrm{B}}$) and local gravity ($\theta_{\mathrm{LG}}$). \label{fig:N6334_B_LG}}
\end{figure*}

Figure \ref{fig:N6334_B_LG} shows the RO-N relation for $\phi_{B}^{LG}$ from the Planck, JCMT, and ALMA observations. For the Planck observation, the magnetic field and local gravity change from a statistically slightly more perpendicular alignment ($AM_{B}^{LG}\lesssim0$) to a slightly more parallel alignment ($AM_{B}^{LG}\gtrsim0$), then change to a random alignment as $N_{\mathrm{H_2}}$ increases. Because the gravity is not expected to actively interact with the magnetic field in the diffuse region, the increasing $AM$-$N$ trend at the lowest several $N_{\mathrm{H_2}}$ bins of the Planck data may be attributed to specific geometries where the direction of gravity coincidently correlates with the complex-scale magnetic field in low-density regions within our direction-biased NANTEN2  $^{12}$CO (1-0) mask.  
For the JCMT observation toward the NGC 6334 filament, $AM_{B}^{LG}$ is always positive and increases with $N_{\mathrm{H_2}}$, which indicates an increasingly important role of gravity in shaping the magnetic field structure at higher densities. The similarity between the RO-N relation for $\phi_{B}^{NG}$ and $\phi_{B}^{LG}$ suggests that the direct reason for the transition from $AM_{B}^{NG}<0$ to $AM_{B}^{NG}>0$ (see Section \ref{sec:B_NG} and Figure \ref{fig:N6334_B_NG}) may be related to the interplay between magnetic field and gravity. The statistically more parallel alignment between magnetic field, local gravity, and column density gradient at intermediate column densities probed by JCMT in NGC 6334 can be naturally explained by the scenario of a magnetized gravitational collapse \citep{1976ApJ...206..753M, 1976ApJ...207..141M}. 
At even higher $N_{\mathrm{H_2}}$ bins revealed by ALMA observations toward individual clumps, the $AM_{B}^{LG}$ shows a prevailing decreasing trend with increasing $N_{\mathrm{H_2}}$ and transits back to $AM_{B}^{LG}\sim0$ (in N6334I and IV) or even $AM_{B}^{LG}<0$ (in N6334V) at the highest several $N_{\mathrm{H_2}}$ bins, except that the value of $AM_{B}^{LG}$ in N6334I(N) stays positive across the $N_{\mathrm{H_2}}$ range. This may suggest that the magnetic field structure in high-density regions is not only shaped by gravity, but also affected by star formation activities (e.g., converging flows, accretion, outflows, rotation, and et al.). The distinct $AM$-$N$ relations in different clumps may indicate their different star formation activities. 
For the ALMA observation, the magnetic field is better aligned with the local gravity than with the column density gradient (see Figures \ref{fig:N6334_B_NG} and \ref{fig:N6334_B_LG}), suggesting that $\phi_{B}^{LG}$ is better than $\phi_{B}^{NG}$ in studying the interaction between magnetic fields and gravity.
The spatial distribution of $\phi_{B}^{LG}$ shows some patterns (see Appendix \ref{sec:B_LG_map}), where local regions with small and large $\phi_{B}^{LG}$ values indicate weak and strong magnetic resistance against gravity \citep{2018ApJ...855...39K}, respectively. But more detailed analytical explanations for the spatial $\phi_{B}^{LG}$ distribution are yet to be established.
%to the directional bias of our NANTEN2  $^{12}$CO (1-0) mask

\subsection{Velocity gradient versus magnetic field}\label{sec:VG_B}

The relative orientation between velocity gradient and magnetic field ($\phi_{VG}^{B}$) can be used as an indicator of the property of Alfv\'{e}nic turbulence\citep{2017ApJ...835...41G, 2018ApJ...853...96L, 2018ApJ...865...46L} due to its intrinsic anisotropic nature \citep{1995ApJ...438..763G} in the absence of gravity. The degree of turbulence anisotropy increases as the Alfv\'{e}nic Mach number decreases (i.e., stronger magnetic field and weaker turbulence). 

\begin{figure*}[!htbp]
 \gridline{\fig{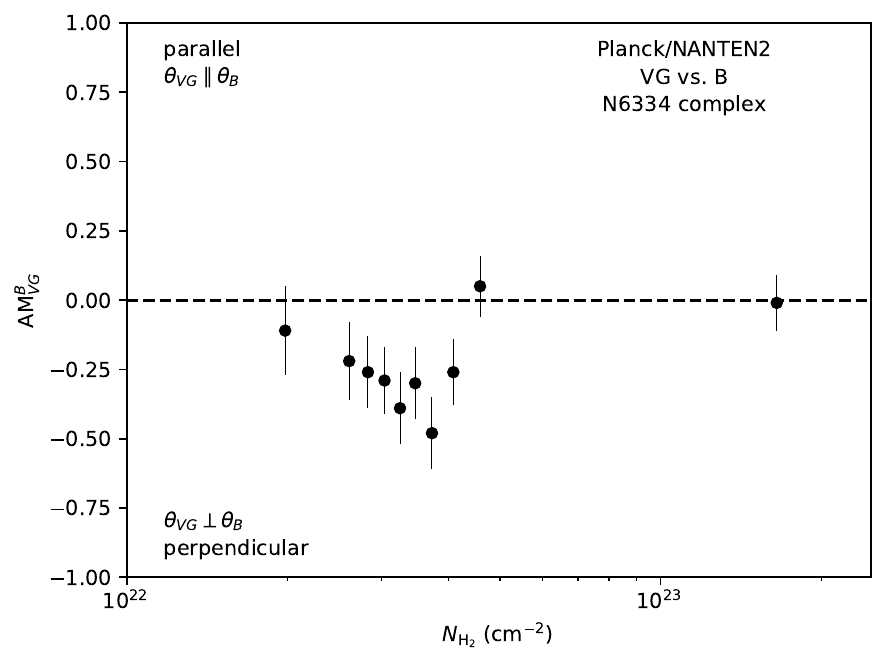}{0.33\textwidth}{}
 \fig{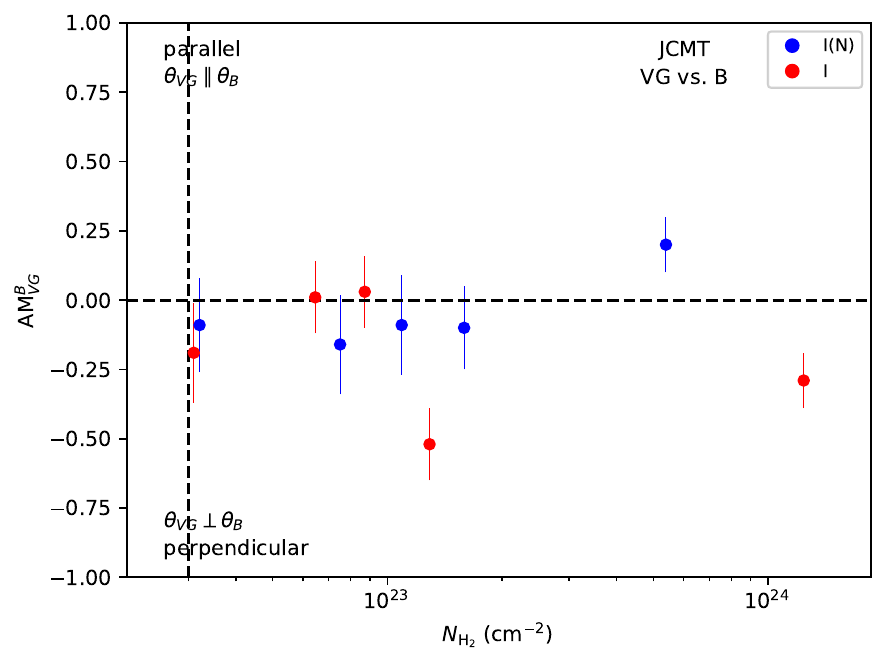}{0.33\textwidth}{}
 \fig{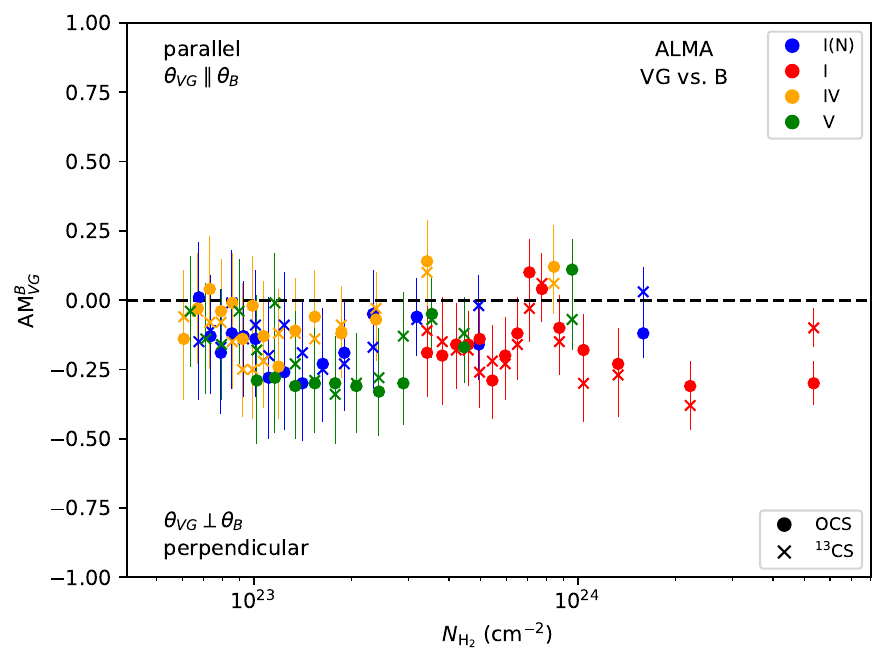}{0.33\textwidth}{}
 }
\caption{Same as Figure \ref{fig:N6334_NG_LG} but for the relative orientation between velocity gradient ($\theta_{\mathrm{VG}}$) and magnetic field ($\theta_{\mathrm{B}}$). \label{fig:N6334_VG_B}}
\end{figure*}

Figure \ref{fig:N6334_VG_B} shows the RO-N relation for $\phi_{VG}^{B}$ from the Planck, NANTEN2, JCMT, and ALMA observations. A clear trend is that the magnetic field and velocity gradient are statistically more perpendicular ($AM_{VG}^{B}<0$) to each other at different column densities across several orders of magnitude. The more perpendicular alignment at low column densities is as expected from previous numerical studies and is consistent with previous observations \citep[e.g., ][]{2017ApJ...835...41G, 2017ApJ...837L..24Y}. However, the more perpendicular alignment at high column densities is different from previous numerical and observational studies which have found that the magnetic field and velocity gradient transits to statistically more parallel in high-density regions due to the magnetized gravitational collapse \citep[e.g.,][]{2017arXiv170303026Y, 2019ApJ...878...10T}. Note that the perpendicular alignment itself does not necessarily indicate sub-Alfv\'{e}nic turbulence because velocity anisotropy is also expected in super-Alfv\'{e}nic turbulence 
at $<L_{inj}M_A^{-3}$ scales for a continuous turbulence cascade \citep{2006ApJ...645L..25L}, where $L_{inj}$ is the turbulence injection scale and $M_A$ is the Alfv\'{e}nic Mach number. On the other hand, the statistical turbulence anisotropy level characterized by $AM_{VG}^{B}$ does not vary too much at different scales ($AM_{VG}^{B} \sim$-0.20, -0.12, and -0.15 for Planck/NANTEN2, JCMT, and ALMA observations, respectively) despite a few outliers and some scatters, so we tentatively suggest that the average Alfv\'{e}nic Mach number at each scale should be similar. i.e., the high-density clumps/cores/condensations in NGC 6334 should also be trans-to-sub-Alfv\'{e}nic since NGC 6334 is trans-to-sub-Alfv\'{e}nic at complex/cloud scales (see discussions in Section \ref{sec:B_NG}). To our knowledge, our results may have provided one of the first observational evidences for a statistically more perpendicular local alignment between the magnetic field and velocity gradient in high-density regions with significant self-gravity. This suggests that even if the magnetic field is distorted by gravity (see Section \ref{sec:B_LG} and Appendix \ref{sec:B_LG_map}) or impacted by star formation activities (see Sections \ref{sec:B_NG} and \ref{sec:B_LG}), a strong magnetic field can still create anisotropic MHD turbulence locally. The trans-to-sub-Alfv\'{e}nic state across scales of several orders of magnitude implies a significant role of the magnetic field in the star formation process in NGC 6334, which can explain the self-similar fragmentation at different scales as reported by \citet{2015Natur.520..518L}. It should be noted that the trans-to-sub-Alfv\'{e}nic state at clump/core/condensations scales in NGC 6334 does not conflict with the previous DCF estimations \citep{2022ApJ...925...30L, 2022FrASS...9.3556L, 2022arXiv220311179P} because individual sources could still be sub-Alfv\'{e}nic while the average state for a large sample of cloud substructures is trans-to-super-Alfv\'{e}nic. On the other hand, the local magnetic field and velocity gradient are only weakly correlated (i.e., small $\vert AM_{VG}^{B} \vert$ values), so the local velocity gradient cannot be directly used as a tracer of the magnetic field orientation\footnote{It should be noted that our approach in the comparison between the magnetic field and velocity gradient is different from that of the VGT which requires subblock-averaging \citep{2017ApJ...837L..24Y} for the velocity gradient. Thus, our results are not against the validity of the VGT.} and the slightly anisotropic turbulence should not significantly affect the traditional DCF analysis that requires an assumption of isotropic turbulence. The exact relation between the $AM_{VG}^{B}$ and the Alfv\'{e}nic Mach number is unclear and is worth future numerical studies. 
%Tang 2019, aligned. 

\subsection{Velocity gradient versus column density gradient}\label{sec:VG_NG}

MHD turbulence can affect the scaling relation and anisotropy of the density structure \citep{2003MNRAS.345..325C, 2005ApJ...624L..93B, 2007ARA&A..45..565M}, where the column density gradient should be perpendicular to the magnetic field and parallel to the velocity gradient for sub-Alfv\'{e}nic turbulence in the absence of gravity. Although several numerical studies found that the line intensity gradient tends to be parallel to the subblock-averaged velocity gradient in self-gravitating regions \citep[e.g.,][]{2017arXiv170303026Y}, there is a lack of numerical studies on how the local velocity gradient and column density gradient should be correlated when gravity is significant. 

\begin{figure*}[!htbp]
 \gridline{\fig{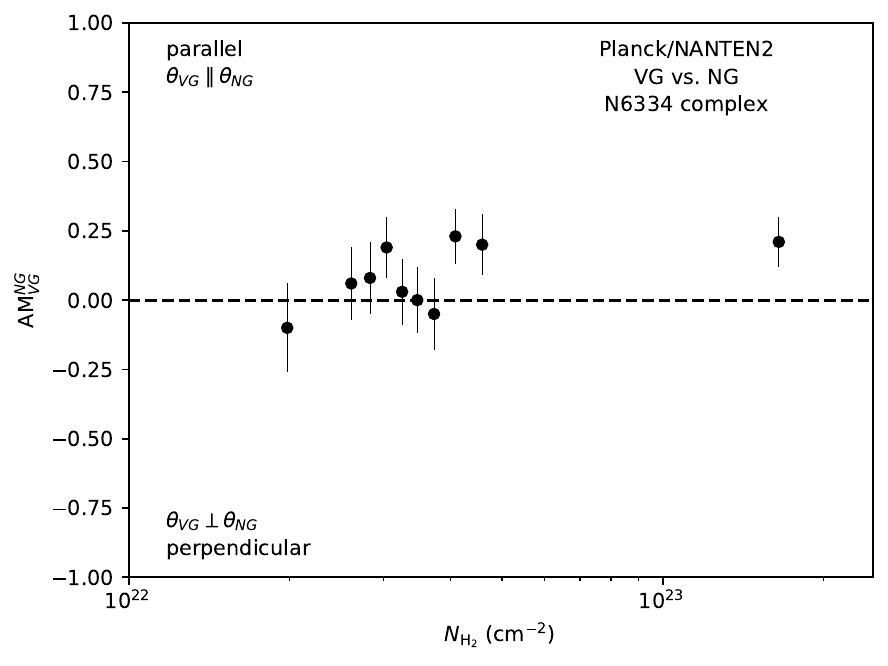}{0.33\textwidth}{}
 \fig{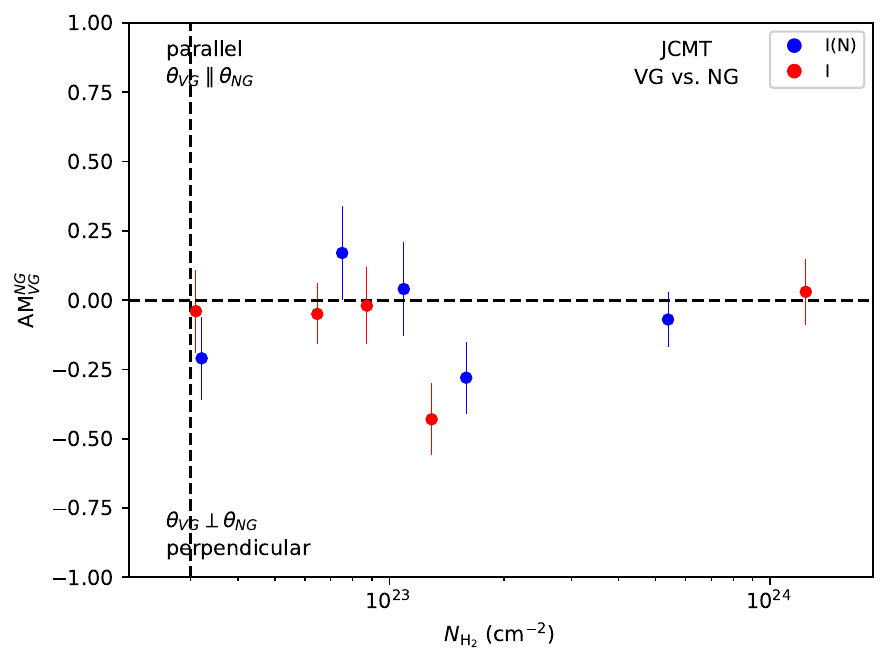}{0.33\textwidth}{}
 \fig{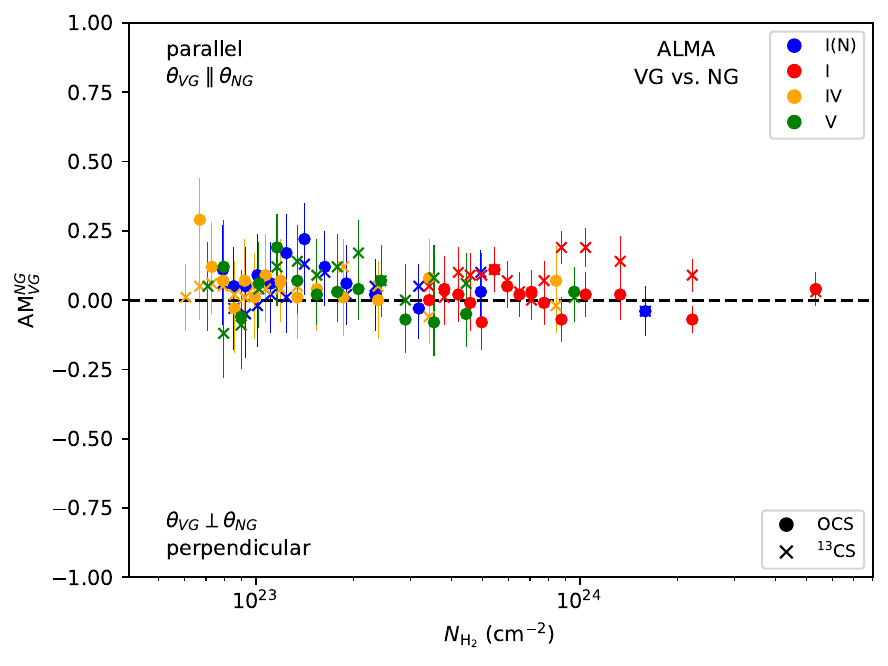}{0.33\textwidth}{}
 }
\caption{Same as Figure \ref{fig:N6334_NG_LG} but for the relative orientation between velocity gradient ($\theta_{\mathrm{VG}}$) and column density gradient ($\theta_{\mathrm{NG}}$). \label{fig:N6334_VG_NG}}
\end{figure*}

Figure \ref{fig:N6334_VG_NG} shows the RO-N relation for $\phi_{VG}^{NG}$ from the Planck, NANTEN2, JCMT, and ALMA observations. The velocity gradient and column density gradient tend to be statistically slightly more parallel at lower column densities revealed by Planck and NANTEN2 observations, which agrees with the theoretical predictions for sub-Alfv\'{e}nic turbulence. For JCMT and ALMA observations, there is no strong statistical relationship between the velocity gradient and column density gradient ($AM_{VG}^{NG}\sim0$). We suggest that the local alignment between the two gradients does not provide too much information on the property of MHD turbulence or gravitational collapse in the self-gravitating region. 
%Sub-Alfv\'{e}nic

\subsection{Velocity gradient versus local gravity}\label{sec:VG_LG}

It is expected that the gas motion will follow the direction of gravity during gravitational collapse. Thus, one may expect the velocity gradient to be aligned with the gravity direction. 

\begin{figure*}[!htbp]
 \gridline{\fig{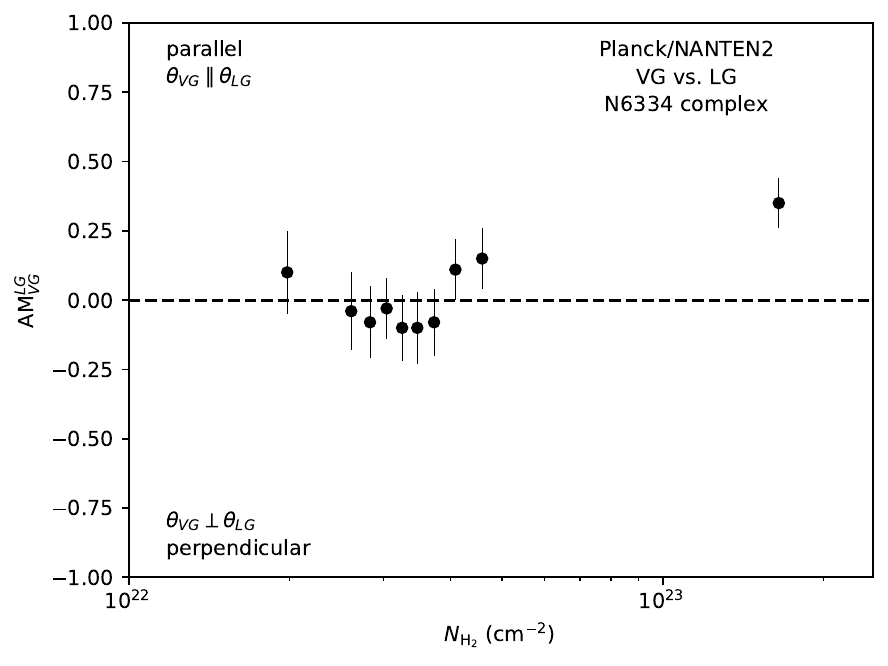}{0.33\textwidth}{}
 \fig{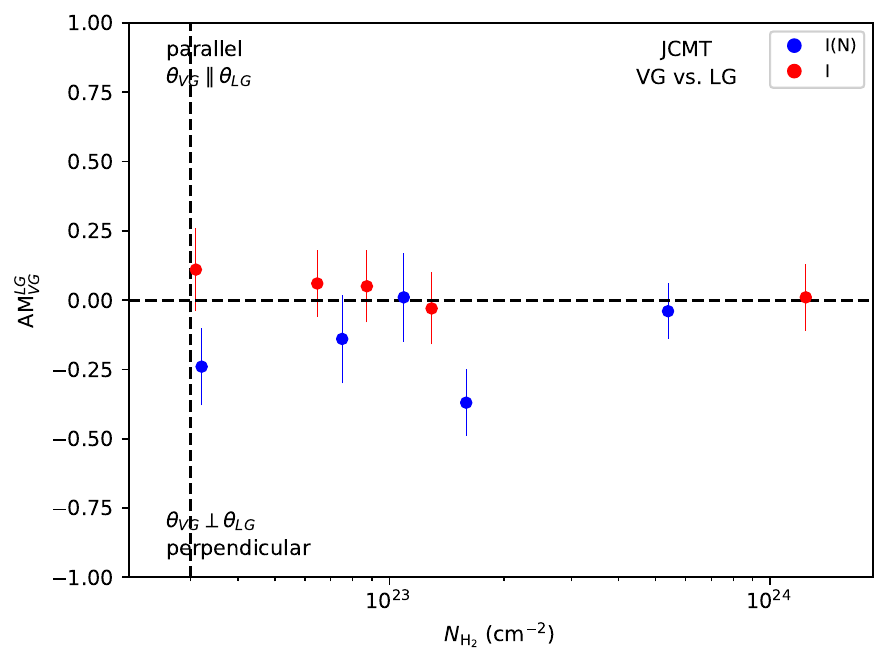}{0.33\textwidth}{}
 \fig{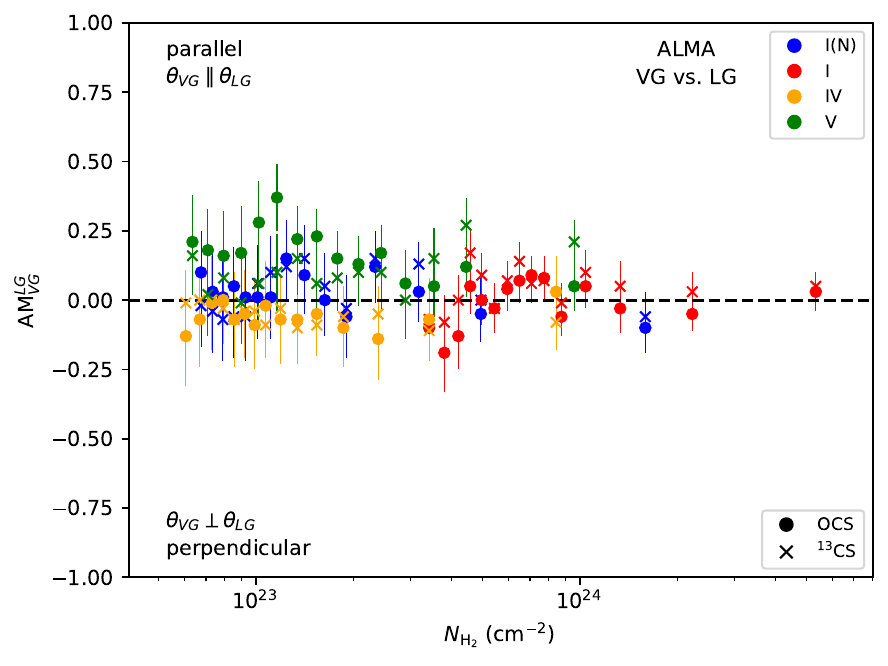}{0.33\textwidth}{}
 }
\caption{Same as Figure \ref{fig:N6334_NG_LG} but for the relative orientation between velocity gradient ($\theta_{\mathrm{VG}}$)  and local gravity ($\theta_{\mathrm{LG}}$). \label{fig:N6334_VG_LG}}
\end{figure*}

Figure \ref{fig:N6334_VG_LG} shows the RO-N relation for $\phi_{VG}^{LG}$ from the Planck, NANTEN2, JCMT, and ALMA observations. No strong statistical relation is found between the local velocity gradient and gravity, except that the two angles tend to be slightly statistically more perpendicular to each other for the JCMT observation toward N6334I(N), where the reason for this perpendicular alignment is unclear. There could be several possible reasons for the general statistical uncorrelation between local velocity gradient and gravity: (1) despite there being large-scale velocity gradients in the NGC 6334 region across scales of several orders of magnitude (see Section \ref{sec:linem1}), the small-scale local velocity gradient could be more reflecting the property of anisotropic MHD turbulence (see Section \ref{sec:VG_B}) instead of the large-scale ordered velocity field; (2) the velocity gradient is just an approximation of the POS velocity, but it does not perfectly trace the POS velocity; (2) several lines tend to be slightly optically thick \citep[e.g., see Figure \ref{fig:N6334_alma_spec} and][]{2022A&A...660A..56A} and do not trace the densest part of the gas that is more gravity dominant; (4) the OCS and $^{13}$CS line could be affected by specific star formation activities (shocks, outflows, rotation, et al.) and chemical processes in each clump. Thus, it is not surprising that the local velocity gradient and gravity are statistically not correlated with each other. 
%the $^{13}$CO (3-2) line tends to be optically thick in N6334I(N) and I \citep{2022A&A...660A..56A}, while the $^{13}$CS line also tends to be optically thick (broad lines and flatter line centers) in N6334I(N) and I. 
% due to the projection effect.  (broad line profile and flat line center)

\subsection{Normalised mass-to-flux ratio}\label{sec:lambda_KTH}

Based on ideal MHD equations, \citet{2012ApJ...747...79K} proposed that the local ratio between the magnetic field force ($F_{B}$) and the gravitational force ($F_{G}$) can be measured with
\begin{equation}\label{eq:sigmaB}
    \Sigma_B = \frac{\sin \phi_{LG}^{IG}}{\sin (90\degr - \phi_{B}^{IG})} = \frac{F_B}{\vert F_G\vert},
\end{equation}
if the hydrostatic gas pressure is negligible, where ``IG'' stands for intensity gradient. Later, \citet{2012ApJ...747...80K} further suggested that the mass-to-flux ratio normalized to the critical value within a specific region is given by
\begin{equation}\label{eq:lambdaKTH}
    \lambda_{\mathrm{KTH}} = \langle \Sigma_B^{-1/2} \rangle \pi^{-1/2}.
\end{equation}
$\lambda_{\mathrm{KTH}} > 1$ indicates that gravity dominates the magnetic field (i.e., magnetically super-critical), and vice versa. We calculate $\lambda_{\mathrm{KTH}}$ at different $N_{\mathrm{H_2}}$ bins. The basic assumption of the KTH method is that the dust emission intensity gradient traces the transport of matter as a result of the MHD force equation. Because the matter distribution is reflected by the column density map rather than the dust intensity map, we use the column density gradient instead of the intensity gradient in the calculation.

\begin{figure}[!htbp]
 %\gridline{\fig{./A_N6334_lambda.pdf}{0.48\textwidth}{}
 %}
  \gridline{\fig{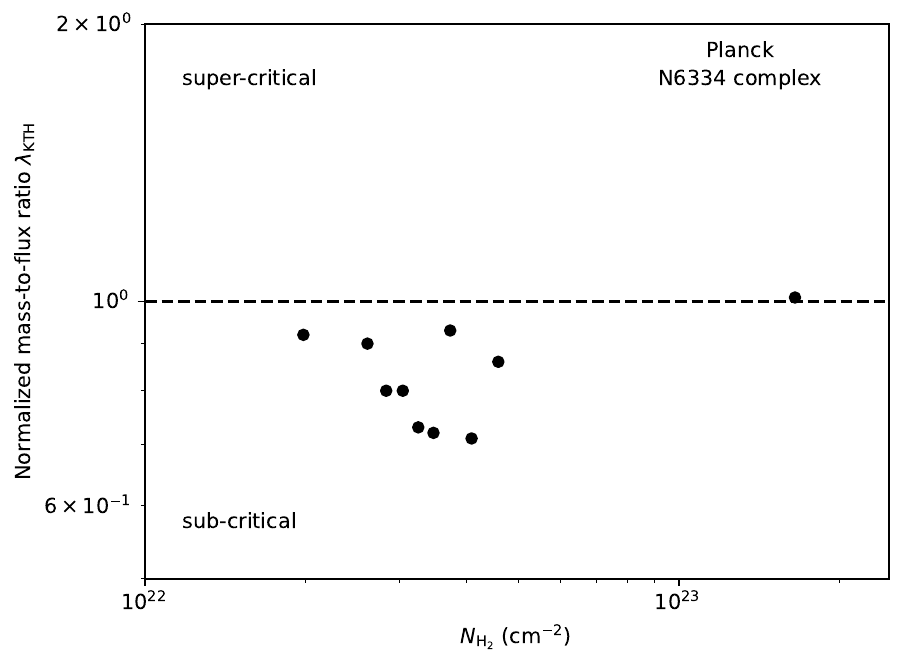}{0.33\textwidth}{}
 \fig{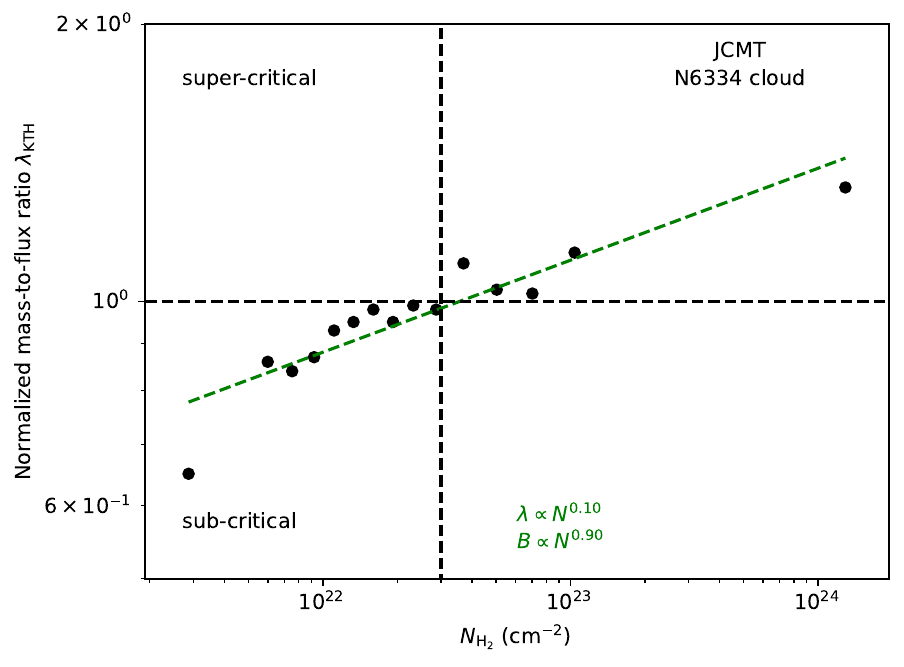}{0.33\textwidth}{}
 \fig{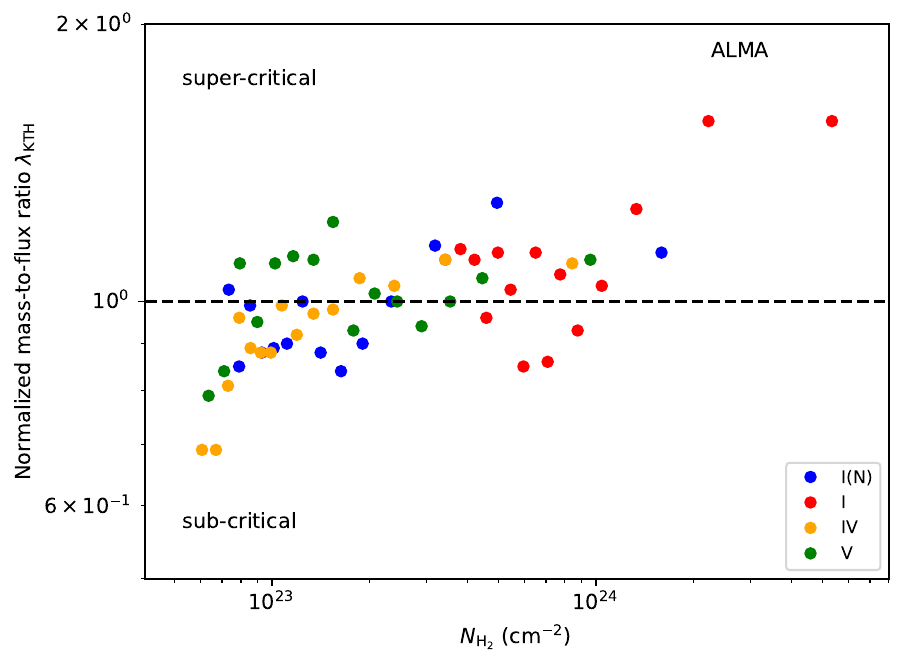}{0.33\textwidth}{}
 }
\caption{Normalised mass-to-flux ratio derived from the KTH method as a function of column density for Planck (left), JCMT (middle), and ALMA (right) observations. The JCMT observation filters out the large-scale emission corresponds to $N_{\mathrm{H_2}} \sim 3 \times 10^{22}$ cm$^{-2}$ \citep{2021A&A...647A..78A} which is indicated by the vertical dashed line. The ALMA observation filters out the large-scale emission at scales $>$0.08 pc. The highest $N_{\mathrm{H_2}}$ bin of Planck observations contains the area of the NGC 6334 filament covered by the JCMT observation. The highest $N_{\mathrm{H_2}}$ bin of JCMT observations contains the area of the N6334I(N), I, IV, and V covered by the ALMA observation. The absolute column densities from different instruments are not comparable. Planck, JCMT, and ALMA observations trace low, intermediate, and high column densities, respectively. \label{fig:N6334_lambda_kth}}
\end{figure}

Figure \ref{fig:N6334_lambda_kth} shows $\lambda_{\mathrm{KTH}}$ as a function of $N_{\mathrm{H_2}}$ from the Planck, JCMT, and ALMA observations. For the majority of Planck observations, there is $\lambda_{\mathrm{KTH}}<1$. Only at the highest $N_{\mathrm{H_2}}$ bin we see $\lambda_{\mathrm{KTH}}\sim1$. 
For the JCMT observation of the whole NGC 6334 filament, the $\lambda_{\mathrm{KTH}}$ increases with increasing $N_{\mathrm{H_2}}$ and transits from $\lambda_{\mathrm{KTH}}<1$ to $\lambda_{\mathrm{KTH}}>1$. 
The ALMA observations toward individual clumps show similar trends of increasing $\lambda_{\mathrm{KTH}}$ with $N_{\mathrm{H_2}}$. 
It should be noted that the magnetic field could be affected by star formation feedback (e.g., outflow, HII regions, et al.) in the vicinity of young stellar objects at high column densities revealed by ALMA, which can violate the assumption of the KTH method and make the estimated $\lambda_{\mathrm{KTH}}$ unreliable. If we only look at the Planck and JCMT observations, the prevailing increasing trend of $\lambda_{\mathrm{KTH}}$ with $N_{\mathrm{H_2}}$ is consistent with previous DCF estimations \citep{2022ApJ...925...30L, 2022FrASS...9.3556L}. Assuming uncertainties of a factor of 2 for both $\lambda_{\mathrm{KTH}}$ and $N_{\mathrm{H_2}}$, we perform a simple least-square fit for the power-law relation between $\lambda_{\mathrm{KTH}}$ and $N_{\mathrm{H_2}}$ for the JCMT observation. We obtain $\lambda\propto N^{0.10}$, which transfers to the relation between the magnetic field and column density as $B\propto N^{0.90}$ adopting $\lambda\propto N/B$ \citep[e.g.,][]{2004ApJ...600..279C}. The power-law index of 0.90 for the $B$-$N$ relation is larger than the value of 0.72 previously reported for the compilation of DCF estimations \citep{2022ApJ...925...30L}. Note that the uncertainty of the $\lambda_{\mathrm{KTH}}$ estimated from the KTH method is unknown due to the lack of direct numerical tests. Moreover, the uncertainty on the absolute column density of the Planck observation and JCMT observation (in the extended region) is also unclear (see Appendix \ref{sec:Tgas}). Thus, we stop at discussions of the $\lambda_{\mathrm{KTH}}-N_{\mathrm{H_2}}$ trend and refrain from determining the transition column density for $\lambda_{\mathrm{KTH}}\sim1$. 
%, but $\lambda_{\mathrm{KTH}}$ from ALMA observations are lower than the $\lambda_{\mathrm{KTH}}$ from JCMT observations toward the same clump. 
%, but using the column density gradient or intensity gradient does not significantly affect the estimated $\lambda_{\mathrm{KTH}}$

%Different symbols represent different regions. Different colors indicate different instruments. 

%\section{Discussion} \label{sec:discussion}

%\subsection{Formation of molecular clouds and substructures}

%Li Hua-Bai 2011 (M33), 2014. Feed by cloud-cloud collision? 

%\subsection{Origin of non-thermal velocity gradients in NGC 6334}
%Turbulence cascade, power law index. outflow driven. Li Shanghuo 2020 N6334IR. VG doesn't perfectly trace POS V. DV depend on density as well (Heyer 2009, Yuen 2021). Multi-point VDF prevents us from deriving the fraction of turbulent and non-turbulent components in non-thermal motions and also. Gravity driven NT DV density dependent (Heyer 2009, Balle.. 2011). massive SF region higher N, larger dv. 

\section{Summary} \label{sec:summary}
With ALMA dust polarization and molecular line observations toward 4 massive clumps (NGC6334I(N), I, IV, and V) in the massive star-forming region NGC 6334, in conjunction with the large-scale dust polarization and molecular line data from Planck, NANTEN2, and JCMT, we reveal the relative orientations between magnetic fields, gas column density gradients, local gravity, and velocity gradients and study their varying trend with column density in NGC 6334. We suggest that a synergistic study of local relative orientations between different angles is powerful at revealing the physical condition of molecular clouds at different scales. The major findings and conclusions are:
\begin{enumerate}

\item The column density gradient and local gravity do not have a preferred relative orientation in the diffuse region surrounding the NGC 6334 filament, suggesting that the density structure of the low-density region is not significantly affected by gravity. Within the NGC 6334 filament, the two angles are closely aligned with each other, suggesting an important role of gravity in shaping the density structure in self-gravitating star formation clouds. 

\item As the column density increases, the alignment between magnetic fields and column density gradients transits from statistically more perpendicular to parallel, which agrees with trans-to-sub-Alfv\'{e}nic simulations of previous numerical studies and suggests NGC 6334 is trans-to-sub-Alfv\'{e}nic at complex/cloud scale. At low column densities, the more perpendicular alignment may be due to the interaction between magnetic fields and turbulence. At intermediate column densities, the alignment between magnetic fields and local gravity shows a similar density-varying trend to the relative orientation between magnetic fields and column density gradients, which suggests the magnetic field is entrained by gravity and the statistically more parallel alignment between magnetic fields and column density gradients is most likely due to a magnetized gravitational collapse. At even higher column densities, the magnetic field and column density gradient/local gravity transits back to having no preferred orientation or statistically slightly more perpendicular, which may suggest the magnetic field structure is impacted by star formation activities. Our results in conjunction with the results in \citet{2015Natur.520..518L} suggest that the magnetic field can guide gravitational collapse and self-similar fragmentation globally but is distorted by gravity and affected by star formation activities locally.

\item The local velocity gradient tends to be statistically more perpendicular to the local magnetic field orientation across our considered spatial scales. The degree of alignment does not change too much at different column densities, which may suggest that the NGC 6334 region remains trans-to-sub-Alfv\'{e}nic at small scales. This signifies an important role of magnetic fields in the star formation process in NGC 6334 despite being dragged by gravity and impacted by star formation activities in intermediate- to high-density regions. 

\item No clear general statistical relation is found between the velocity gradient and column density gradient/local gravity.

\item The normalized mass-to-flux ratio derived from the KTH method tends to increase with column density, which agrees with previous DCF estimations. But the KTH method may fail at high column densities due to the breakdown of the underlying assumptions. 

\end{enumerate}

%% Also note that the akcnowlodgment environment does not support long amounts of text. If you have a lot of people and institutions to acknowledge, do not use this command. Instead, create a new \section{Acknowledgments}.
\begin{acknowledgments}
We thank the anonymous referee for the constructive comments. We thank Dr. Doris Arzoumanian for sharing the JCMT dust polarization maps and the Herschel temperature maps. We thank Dr. Yasuo Fukui and Dr. Mikito Kohno for sharing the NANTEN2 data. J.L. thanks Dr. Daniel Seifried for helpful comments.

J.L. acknowledges the support from the EAO Fellowship Program under the umbrella of the East Asia Core Observatories Association. K.Q. is supported by National Key R\&D Program of China grant No. 2017YFA0402600. K.Q. acknowledges the support from National Natural Science Foundation of China (NSFC) through grant Nos. U1731237, 11590781, and 11629302. H.B.L. is supported by the Ministry of Science and Technology (MoST) of Taiwan (Grant Nos. 108-2112-M-001-002-MY3, 108-2923-M-001-006-MY3, 111-2112-M-001-089-MY3). Z.Y.L is supported in part by NSF AST-1815784 and NASA 20NSSC18K1095. This work was also partially supported by the program Unidad de Excelencia Maria de Maeztu CEX2020-001058-M. J.M.G also acknowledges support by the grant PID2020-117710GB-I00 (MCI-AEI-FEDER, UE).

This paper makes use of the following ALMA data: ADS/JAO.ALMA\#2017.1.00793.S. ALMA is a partnership of the ESO (representing its member states), NSF (USA) and NINS (Japan), together with NRC (Canada), MOST and ASIAA (Taiwan), and KASI (Republic of Korea), in cooperation with the Republic of Chile. The Joint ALMA Observatory is operated by ESO, AUI/NRAO, and NAOJ. The National Radio Astronomy Observatory is a facility of the National Science Foundation operated under cooperative agreement by Associated Universities, Inc.
The JCMT is operated by the EAO on behalf of NAOJ; ASIAA; KASI; CAMS as well as the National Key R\&D Program of China (No. 2017YFA0402700). Additional funding support is provided by the STFC and participating universities in the UK and Canada. Additional funds for the construction of SCUBA-2 were provided by the Canada Foundation for Innovation.
This work is based on observations obtained with Planck (http://www.esa.int/Planck), an ESA science mission with instruments and contributions directly funded by ESA Member States, NASA, and Canada. 
The present study has also made use of NANTEN2 data. NANTEN2 is an international collaboration of ten universities: Nagoya University, Osaka Prefecture University, University of Cologne, University of Bonn, Seoul National University, University of Chile, University of New SouthWales, Macquarie University, University of Sydney, and Zurich Technical University. 
\end{acknowledgments}


\vspace{5mm}
\facilities{Planck(HFI), NANTEN2, JCMT(HARP, SCUBA-2, POL-2), ALMA}


\software{Astropy \citep{2013A&A...558A..33A,2018AJ....156..123A},  
Matplotlib \citep{2007CSE.....9...90H}.
          }



\appendix
\section{Integrated line intensity maps}\label{sec:linem0}
\begin{figure}[!htbp]
 \gridline{\fig{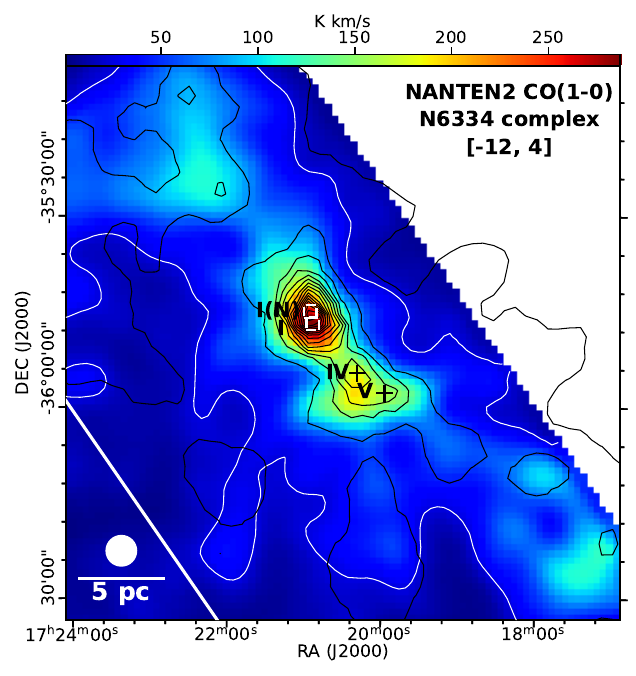}{0.33\textwidth}{(a)}
 \fig{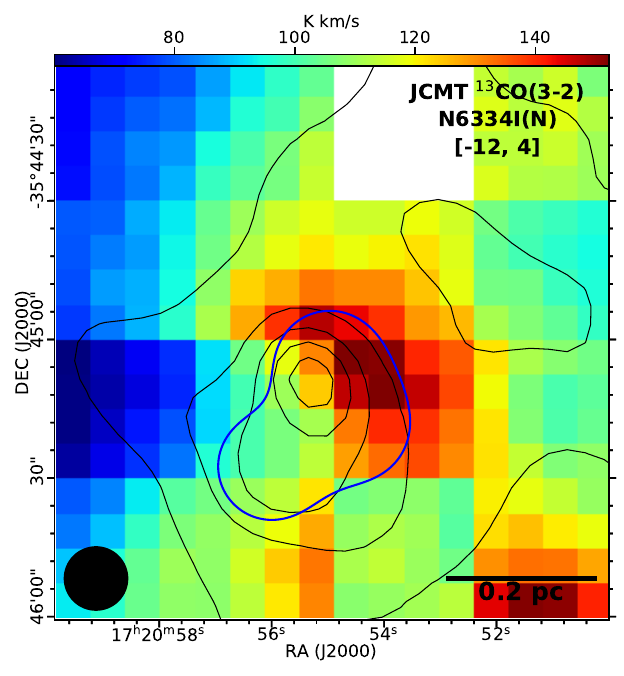}{0.33\textwidth}{(b)}
 \fig{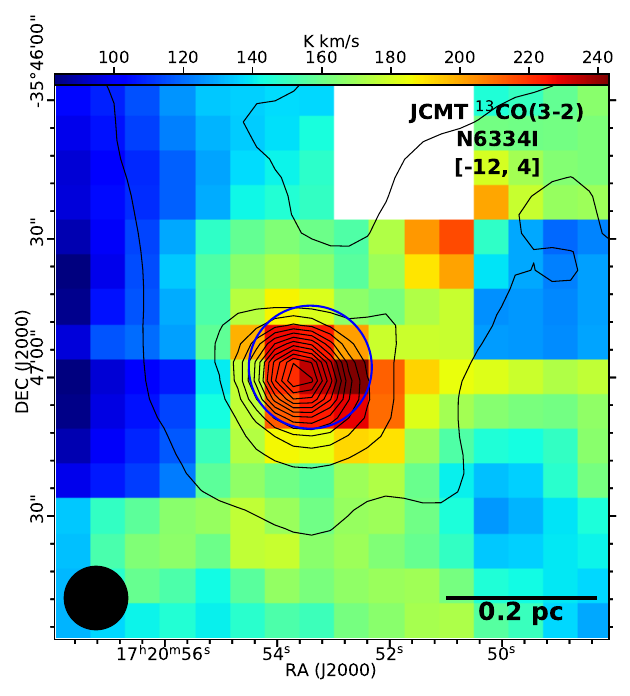}{0.33\textwidth}{(c)}
 }
\caption{(a). Moment 0 map (colorscale) of NANTEN2  $^{12}$CO (1-0) line emission toward NGC 6334 complex \citep{2018PASJ...70S..41F} from -12 to 4 km s$^{-1}$. The line data is convolved to a beam size of 5$\arcmin$. The black contour levels correspond to the Planck $\tau_{353}$ map. Contour starts at 0.0004 and continues with an interval of 0.0004. The white rectangles indicate the map area of the JCMT fields toward N6334I(N) and I in (b) and (c). Black crosses indicate the positions of N6334IV and V. The white contour indicates the region with NANTEN2 integrated  $^{12}$CO (1-0) intensity greater than 25 K km s$^{-1}$ within which we perform the relative orientation analysis. (b)-(c). Moment 0 maps (colorscale) of JCMT $^{13}$CO (3-2) line emission from -12 to 4 km s$^{-1}$ toward N6334I(N) and N6334I. The black contour levels correspond to the JCMT 850 $\mu$m dust continuum map. Contour starts at 2 Jy beam$^{-1}$ and continues with an interval of 4 Jy beam$^{-1}$. Blue contours show the FWHM field of view of our ALMA observations. \label{fig:N6334_large_line_m0}}
\end{figure}

\begin{figure*}[!htbp]
 \gridline{\fig{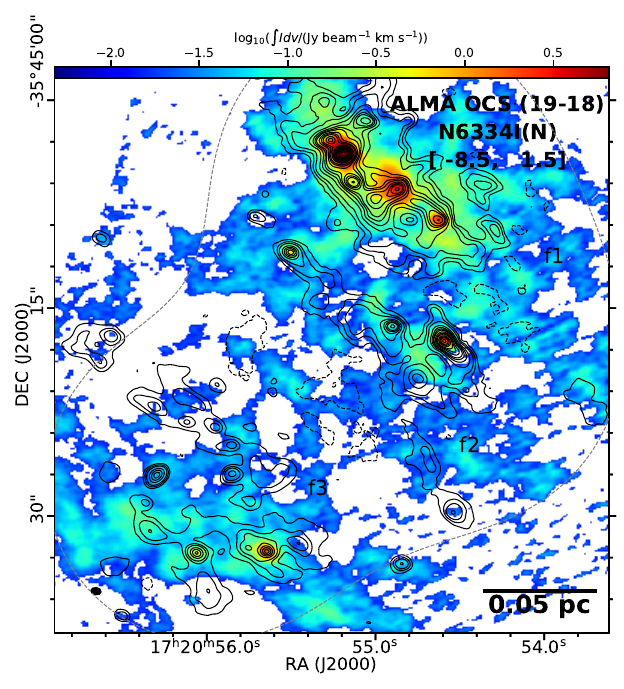}{0.33\textwidth}{(a)}
 \fig{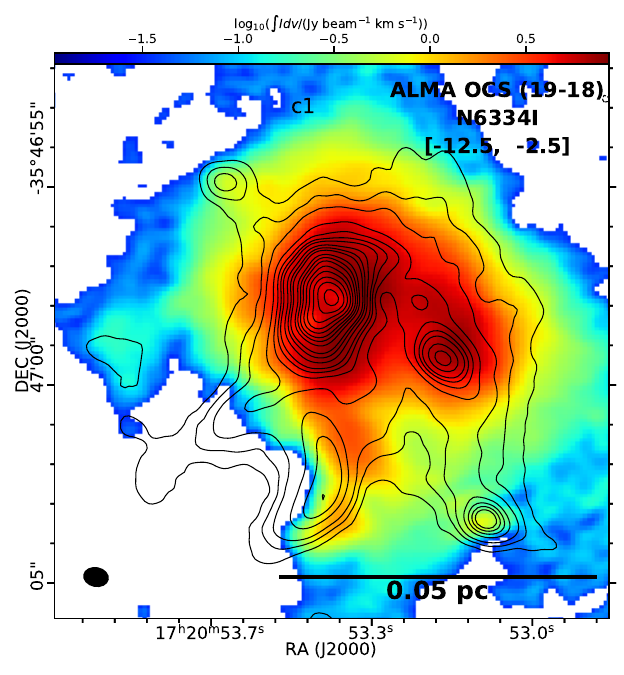}{0.33\textwidth}{(b)}
\fig{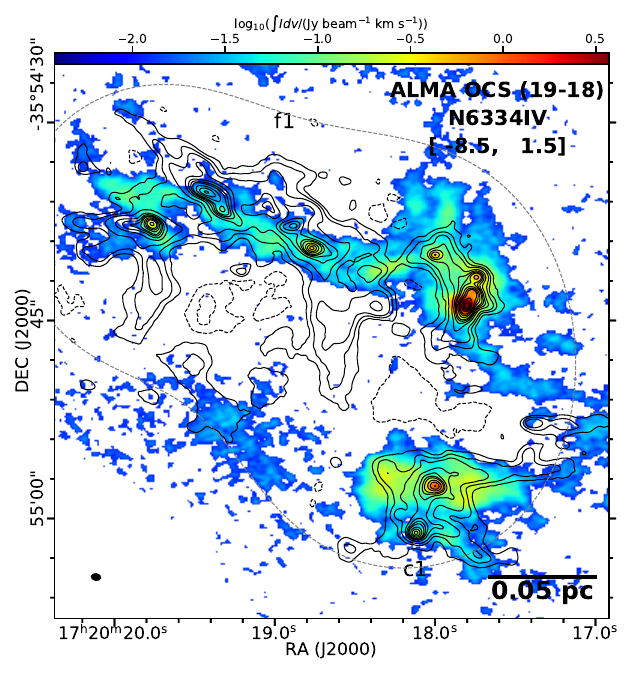}{0.33\textwidth}{(c)}
 }
 \gridline{\fig{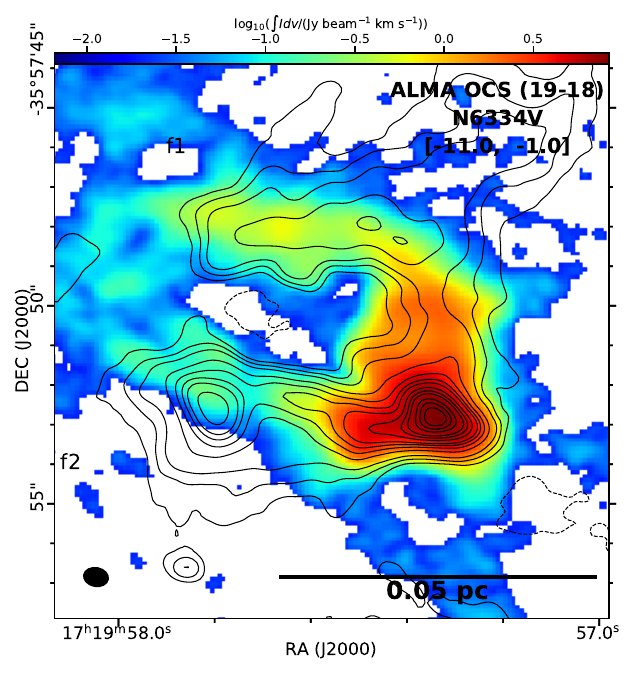}{0.33\textwidth}{(d)}
 \fig{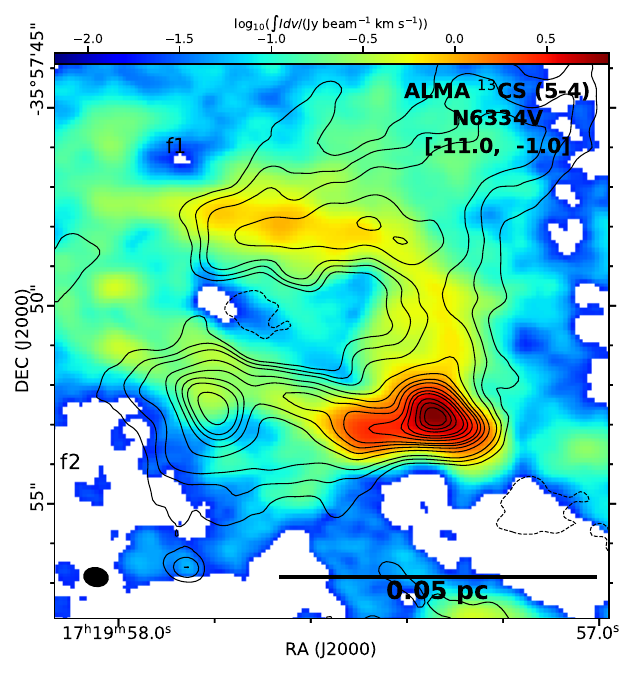}{0.33\textwidth}{(e)}
 }
\gridline{\fig{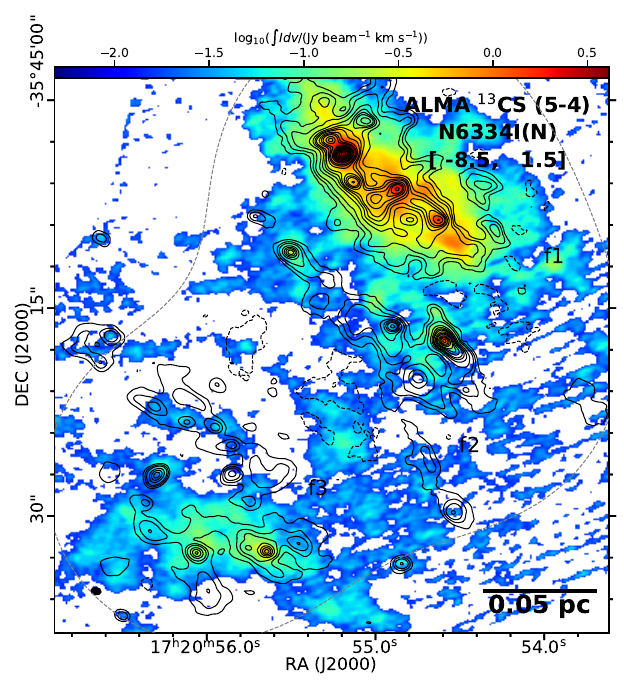}{0.33\textwidth}{(f)}
 \fig{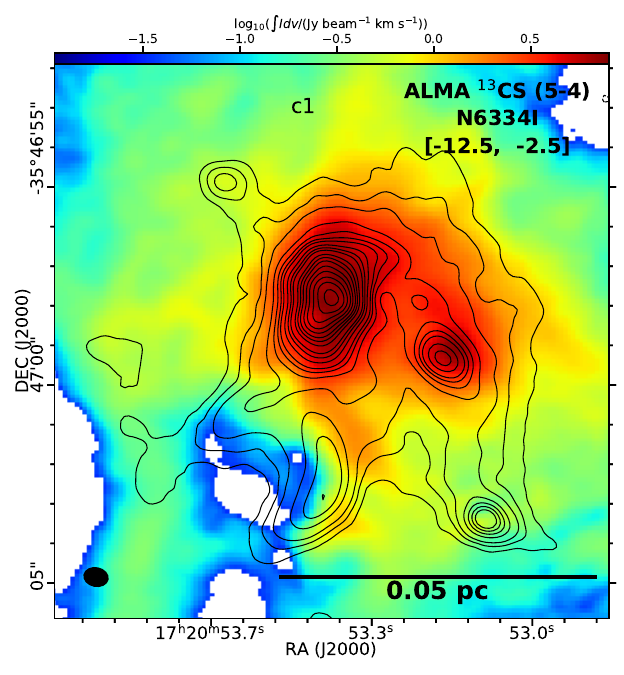}{0.33\textwidth}{(g)}
 \fig{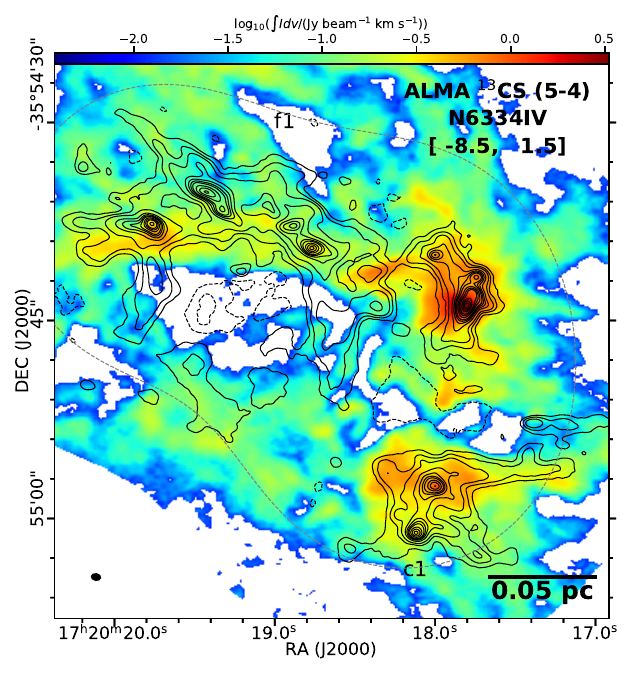}{0.33\textwidth}{(h)}
 }
\caption{Moment 0 maps (colorscale) of ALMA OCS ((a)-(d)) and $^{13}$CS ((e)-(h)) observations. The black contour levels correspond to the ALMA dust continuum map. Contour levels are ($\pm$3, 6, 10, 20, 30, 40, 50, 70, 90, 110, 150, 180, 210, 250, 290, 340, 390, 450) $\times \sigma_{I}$.  Grey dashed contours indicate to the FWHM field of view of the ALMA observations. \label{fig:N6334_alma_line_m0}}
\end{figure*}

Figures \ref{fig:N6334_large_line_m0} and \ref{fig:N6334_alma_line_m0} present the integrated intensity (moment 0) maps of the NANTEN2  $^{12}$CO (1-0), JCMT $^{13}$CO (3-2), and ALMA OCS and $^{13}$CS data. The integrated intensity at position $\boldsymbol{x}$ is calculated with $\Sigma_i^{N_{\mathrm{ch}}} I_i(\boldsymbol{x}) \Delta v_{\mathrm{ch}}$. The propagated uncertainty of the integrated intensity is given by $\sqrt{N_{\mathrm{ch}}} \sigma_{ch} \Delta v_{\mathrm{ch}}$ \citep[e.g.,][]{2002ApJ...572..238C, 2019RNAAS...3...74T}.  Following \citet{2022A&A...660A..56A}, the large-scale NANTEN2  $^{12}$CO (1-0) and JCMT $^{13}$CO (3-2) lines are integrated from -12 to 4 km s$^{-1}$. The ALMA OCS and $^{13}$CS are integrated within slightly different velocity ranges as indicated in Figure \ref{fig:N6334_alma_line_m0}. In general, the integrated line emissions agree with the dust continuum emission near the emission peaks, but show some differences in extended regions.

\section{Uncertainties}\label{sec:uncer}
\subsection{Uncertainty of the gradient orientation}
 The column density gradient is calculated with \citep{2016AA...586A.138P}
\begin{equation}\label{eq:gN}
\nabla N = (G_x \circledast N)\vu*{i} + (G_y \circledast N)\vu*{j} = g_x \vu*{i} + g_y \vu*{j},
\end{equation}
where $G_x$ and $G_y$ are the x- and y-derivatives of the Sobel kernel. The orientation of $\nabla N$ is given by $\theta_{\mathrm{NG}} = \arctan(-g_x, g_y)$. The uncertainty of the column density gradient is given by \citet{2016AA...586A.138P}:
\begin{equation}\label{eq:dgN}
\nabla \delta N = (G_x \circledast \delta N)\vu*{i} + (G_y \circledast \delta N)\vu*{j} = \delta_{gx} \vu*{i} + \delta_{gy} \vu*{j},
\end{equation}
where $\delta N$ is the uncertainty of the column density. The uncertainty of $\theta_{\mathrm{NG}}$ is given by \citet{2016AA...586A.138P}:
\begin{equation}\label{eq:dthetaNG}
\delta \theta_{\mathrm{NG}} = \frac{1}{ g_x^2 + g_y^2 } \sqrt{g_y^2 \sigma_{gx}^2 + g_x^2 \sigma_{gy}^2} ,
\end{equation}
where $\sigma_{gx}$ and $\sigma_{gy}$ are the RMS of $\delta_{gx}$ and $\delta_{gy}$. In our case, we calculate $\sigma_{gx}$ and $\sigma_{gy}$ within the $3\times3$ box. The velocity centroid gradient and its uncertainty can be calculated similarly. 

\subsection{Uncertainty of the alignment measure parameter $AM$}
The uncertainty of $AM = \langle \cos (2\phi) \rangle$ comes from the standard error on the mean and the propagation of the observational uncertainty.

For a statistically independent sample of $n'$ observations toward $f$ (in our case $f = \cos (2\phi)$), we have 
\begin{equation} 
STD(f) = \sqrt{RMS(f)^2-\langle f \rangle^2},
\end{equation}
where $STD(f)$ is the standard deviation of $f$. The relation between the statistical standard error of $\langle f \rangle$ (i.e., $\delta \langle f \rangle_{\mathrm{stat}}$) and the standard deviation of $f$ is 
\begin{equation} 
\delta \langle f \rangle_{\mathrm{stat}} = \frac{1}{n'}STD(f).
\end{equation}
Thus, the uncertainty of $AM$ from the statistical error on the mean is given by 
\begin{equation} 
\delta AM_{\mathrm{stat}} = \delta \langle f \rangle_{\mathrm{stat}} = \sqrt{(\langle (\cos (2\phi))^2 \rangle - AM^2)/n'}.
\end{equation}

On the other hand, the propagated observational uncertainty of $f=\cos (2\phi)$ is 
\begin{equation} 
\delta f_{\mathrm{obs}} \sim \vert 2\sin(2\phi) \delta \phi \vert.
\end{equation}
For $\langle f \rangle$, the propagated observational uncertainty is 
\begin{equation} 
\delta \langle f \rangle_{\mathrm{obs}} = RMS(\delta f_{\mathrm{obs}}).
\end{equation}
Thus, the propagated observational uncertainty of $AM$ is given by
\begin{equation} 
\delta AM_{\mathrm{obs}} = \delta \langle f \rangle_{\mathrm{obs}} = \sqrt{(\Sigma_i^{n'} (2\sin (2\phi_i) \delta \phi_i)^2)/n'}.
\end{equation}
%(i.e., $\delta \langle f \rangle_{\mathrm{obs}}$)

Finally, the combined uncertainty of $AM$ is given by 
\begin{equation} 
\delta AM = \sqrt{\delta AM_{\mathrm{stat}}^2+\delta AM_{\mathrm{obs}}^2}.
\end{equation}

\section{Temperature and column density}\label{sec:Tgas}

\begin{figure*}[!htbp]
 \gridline{\fig{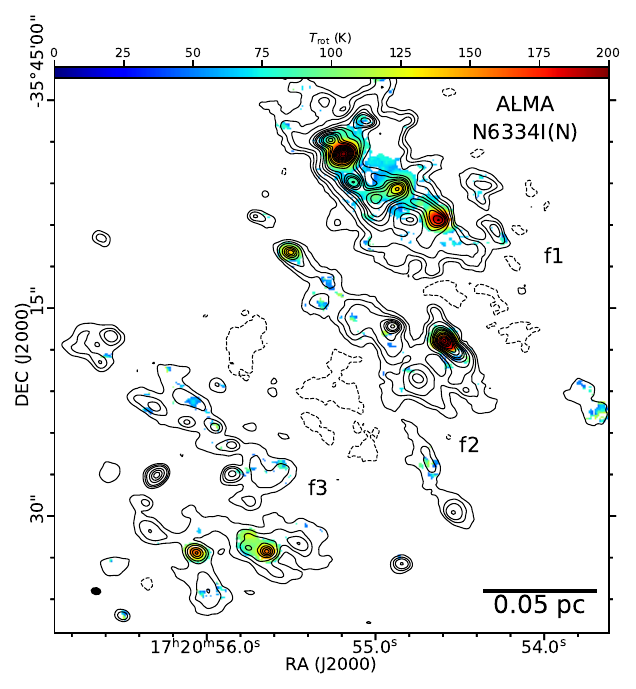}{0.48\textwidth}{}
 \fig{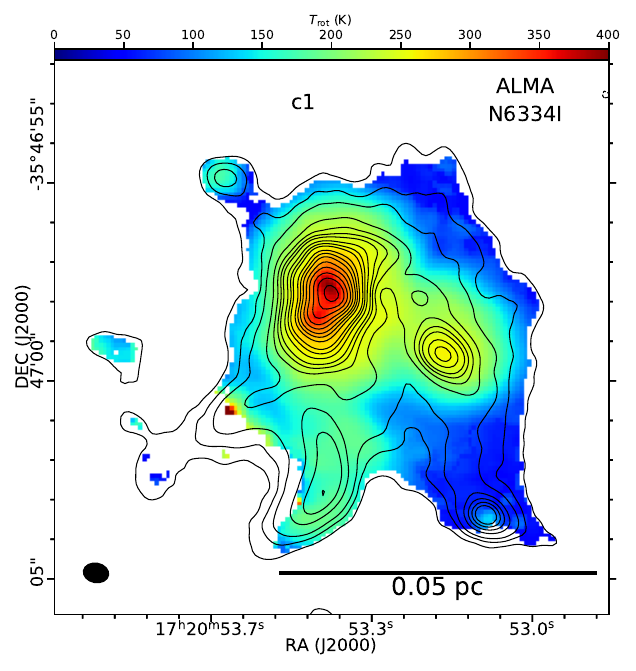}{0.48\textwidth}{}
 }
  \gridline{\fig{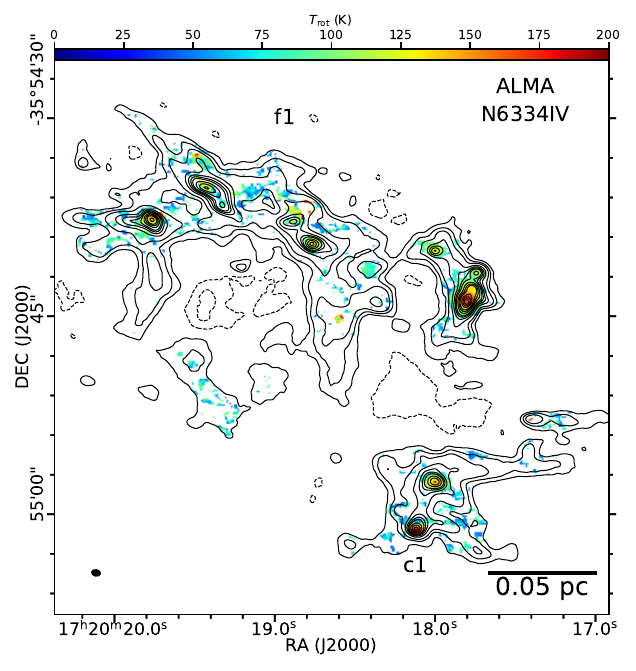}{0.48\textwidth}{}
 \fig{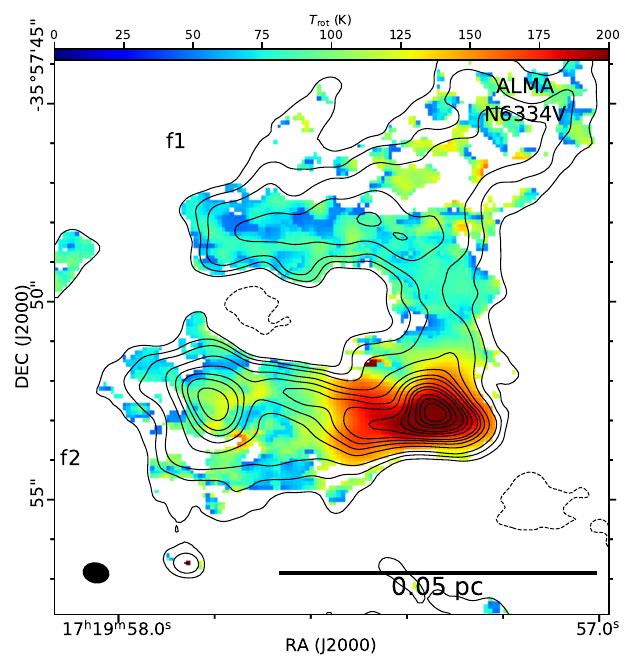}{0.48\textwidth}{}
 }
\caption{Gas temperature maps (colorscale) derived from rotation diagram analysis of ALMA multi-transition CH$_3$OH observations. The contour levels correspond to the ALMA dust continuum map. Contour levels are ($\pm$3, 6, 10, 20, 30, 40, 50, 70, 90, 110, 150, 180, 210, 250, 290, 340, 390, 450) $\times \sigma_{I}$.\label{fig:N6334_Trot}}
\end{figure*}

We use multi-transition CH$_3$OH lines from ALMA observations to derive the physical conditions near the young stellar objects. Table \ref{tab:ch3oh} lists the information of these CH$_3$OH lines from the CDMS\footnote{https://cdms.astro.uni-koeln.de/} catalog. We perform a simple rotation diagram analysis \citep{1999ApJ...517..209G} with the CH$_3$OH lines to estimate the gas temperature under the assumptions of local thermal equilibrium and optically thin. The upper state level population of CH$_3$OH is given by 
\begin{equation}
N_{\mathrm{u}} = \frac{N_{\mathrm{CH_3OH}}}{Z} g_\mathrm{u} e^{-E_\mathrm{u}/kT_\mathrm{rot}},
\end{equation}
where $N_\mathrm{u}$ is the column density of the upper state, $N_{\mathrm{CH_3OH}}$ is the total column density of CH$_3$OH, $g_\mathrm{u}$ is the statistical weight of the upper state, $E_\mathrm{u}$ is the upper energy level, $k$ is the Boltzmann constant, $T_\mathrm{rot}$ is the rotation temperature, and $Z$ is the partition function. We fit the rotation diagram of the 4 transitions of CH$_3$OH to derive the rotation temperature of each pixel. If the transition with the highest $E_\mathrm{up}$ (i.e., $\sim$508 K) is not detected, we only fit the other 3 transitions. If the transition with $E_\mathrm{up}\sim190$ K is not detected, we do not fit the rotation diagram. Figure \ref{fig:N6334_Trot} shows the rotation temperature maps of the 4 clumps. A general trend is that the $T_\mathrm{rot}$ decreases from hundreds of Kelvins near dust emission peaks to less than 100 K in extended regions. The peak temperatures in N6334I(N), I, IV, and V are $\sim$220, 400, 250, and 220 K, respectively, suggesting a ubiquity of hot cores in the massive clumps in NGC 6334.  

\begin{deluxetable}{ccccc}[t!]
\tablecaption{Summary of CH$_3$OH lines \label{tab:ch3oh}}
\tablecolumns{5}
\tablewidth{0pt}
\tablehead{
\colhead{Frequency} &
\colhead{Transition} &
\colhead{$g_{\mathrm{u}}$ \tablenotemark{a} } &
\colhead{$E_{\mathrm{u}}$ \tablenotemark{b}} & 
\colhead{$A_{\mathrm{ul}}$ \tablenotemark{c}} \\
\colhead{(GHz)} & \colhead{} & \colhead{} &
\colhead{(K)} & 
\colhead{((10$^{-5}$s$^{-1}$))} 
}
\startdata
216.945521 &  5$_1$-4$_2$E & 44 & 55.87116 & 1.21 \\
217.886504 & 20$_1$-20$_0$E & 164 & 508.37554 & 3.38 \\
218.440063 & 4$_2$-3$_1$E& 36 & 45.45944 & 4.69 \\
232.945797 & 10$_{-3}$-11$_{-2}$E & 84 & 190.36958 & 2.13 \\
\enddata
\tablenotetext{a}{Statistical weight of the upper state.}
\tablenotetext{b}{Upper energy level.}
\tablenotetext{c}{Einstein A coefficient.}

\end{deluxetable}

\begin{figure*}[!htbp]
 \gridline{\fig{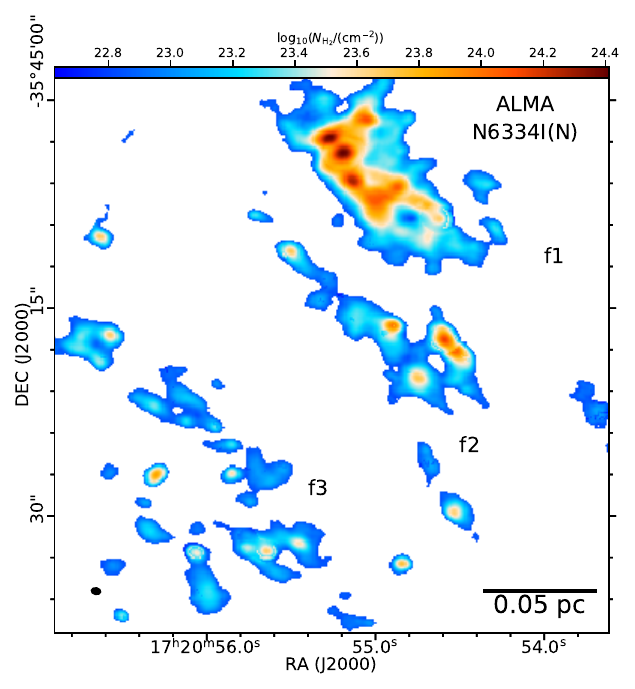}{0.48\textwidth}{}
 \fig{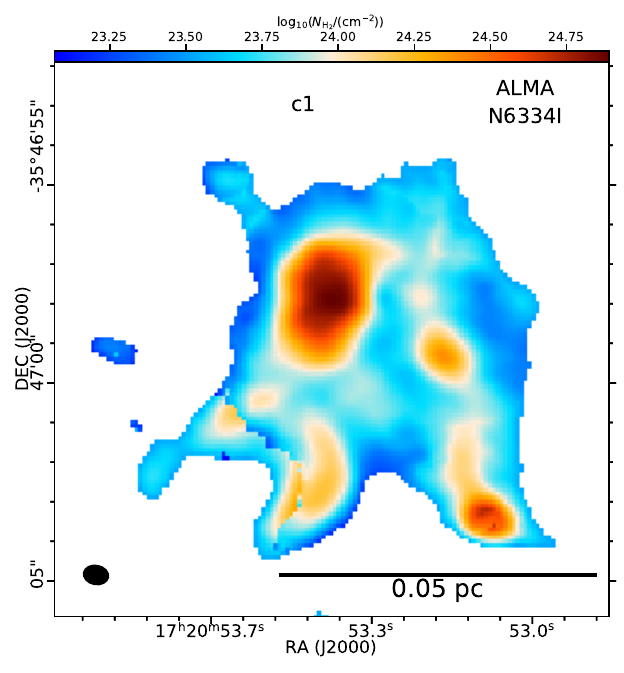}{0.48\textwidth}{}
 }
  \gridline{\fig{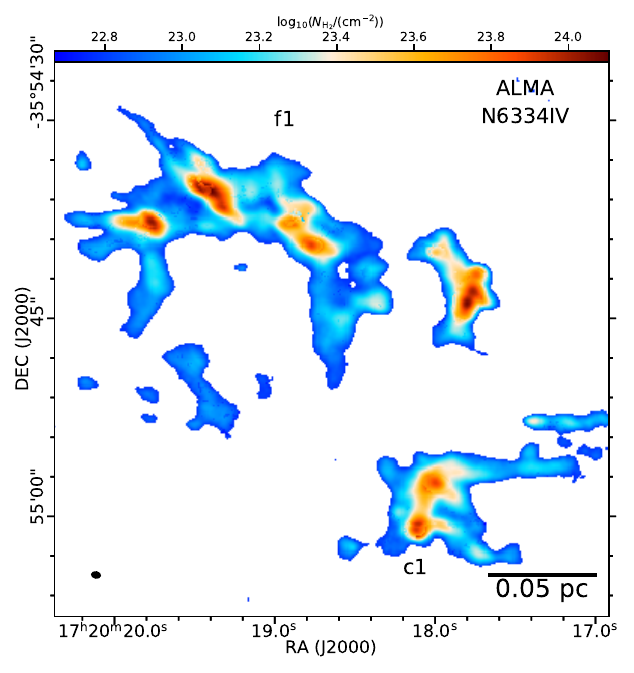}{0.48\textwidth}{}
 \fig{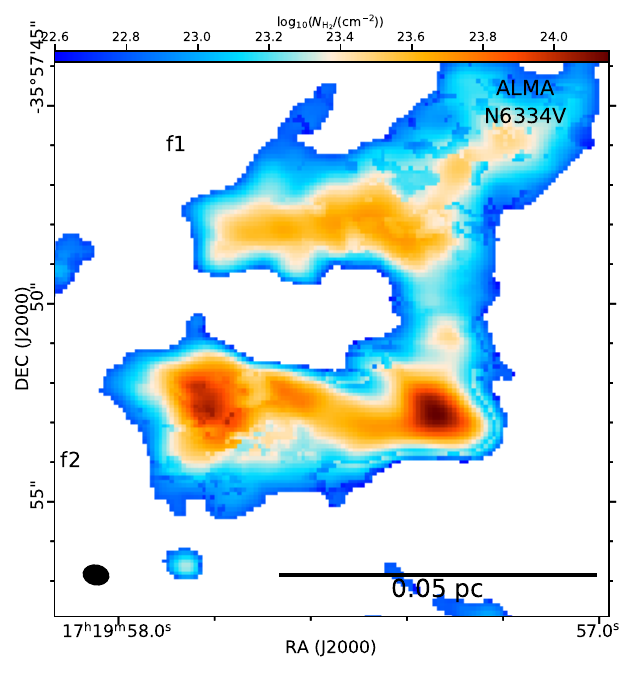}{0.48\textwidth}{}
 }
\caption{Column density maps (colorscale) derived from ALMA dust emission observations.\label{fig:N6334_cd}}
\end{figure*}

Assuming optically thin dust emission, the dust mass can be estimated as 
\begin{equation}
M_{\mathrm{dust}} = \frac{F_{\mathrm{\nu}} d^2}{ \kappa_{\nu} B_{\nu} (T)}, \label{eq:M}
\end{equation}
where $F_{\mathrm{\nu}}$ is the flux density at frequency $\nu$, $d$ is the distance, $\kappa_{\nu} = (\nu / 1 \mathrm{THz})^{\beta}$ is the dust opacity \citep{1983QJRAS..24..267H} in m$^2$ kg$^{-1}$, and $B_{\nu} (T)$ is the Planck function at temperature $T$. Previous multi-wavelength dust emission observations toward massive star-forming regions have found dust emissivity indexes ($\beta$) of $\sim$1.5 \citep[e.g.,][]{2007A&A...466.1065B, 2007ApJ...654L..87C}. Adopting $\beta=1.5$, the $\kappa_{\nu}$ is estimated to be 0.10 m$^2$ kg$^{-1}$ at $\nu\sim220$ GHz. We adopt the rotation temperature derived from the rotation diagram analysis as the dust temperature. For regions without $T_\mathrm{rot}$ estimation, we adopt $T=80$ K, which is approximately the most common temperature in extended regions (see Figure \ref{fig:N6334_Trot}). Adopting a gas-to-dust ratio of $\Lambda = 100$ \citep{1972ApJ...172..491S}, the gas mass is estimated with $M_{\mathrm{gas}} = \Lambda M_{\mathrm{dust}}$. The gas column density is then estimated with   
\begin{equation}
N_{\mathrm{H_2}} = \frac{M_{\mathrm{gas}}}{\mu_{\mathrm{H_2}} m_{\mathrm{H}} A},
\label{eq:N}
\end{equation}
where $\mu_{\mathrm{H_2}} = 2.8$ is the mean molecular weight per hydrogen molecule \citep{2008A&A...487..993K}, $m_{\mathrm{H}}$ is the atomic mass of hydrogen, and $A$ is the area. Figure \ref{fig:N6334_cd} shows the column density maps of the 4 clumps. 

For the JCMT observations from BISTRO, we estimate the gas mass and column density from the Stokes-I map of dust emission with Equations \ref{eq:M} and \ref{eq:N} but adopt a constant temperature $T=20$ K \citep{2021A&A...647A..78A} and $\kappa_{\nu}=$0.21 m$^2$ kg$^{-1}$ (at $\nu\sim353$ GHz). \citet{2021A&A...647A..78A} found that the JCMT observations of NGC 6334 filters out the large-scale emission on the order of $N_{\mathrm{H_2}} = 3 \times 10^{22}$ cm$^{-2}$. 

For the Planck observations, we scale the dust optical depth ($\tau_{353}$) map to atomic hydrogen column density ($N_{\mathrm{H}}$) map with the relation \citep{2014A&A...571A..11P}
\begin{equation}
\tau_{353}/N_{\mathrm{H}} = 1.2 \times 10^{-26} \mathrm{cm}^{2}.
\end{equation}
The variation of the $\tau_{353}/N_{\mathrm{H}}$ ratio can be more than a factor of 2 from diffuse to dense ISM, but the statistics of relative orientation does not critically depend on this calibration \citep{2016AA...586A.138P}. We convert $N_{\mathrm{H}}$ to $N_{\mathrm{H_2}}$ with the relation $N_{\mathrm{H}} = 2N_{\mathrm{H_2}}$.

\section{Relative orientation between magnetic field and local gravity}\label{sec:B_LG_map}

\begin{figure}[!htbp]
 \gridline{\fig{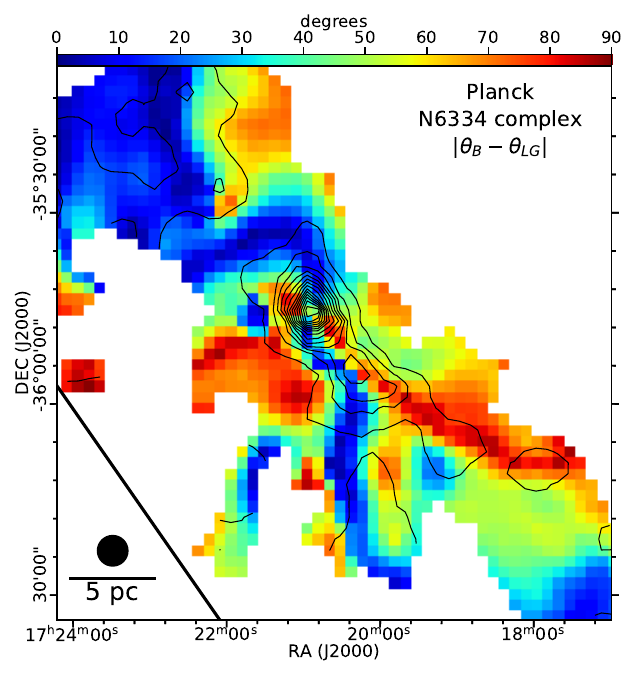}{0.48\textwidth}{(a)}
 \fig{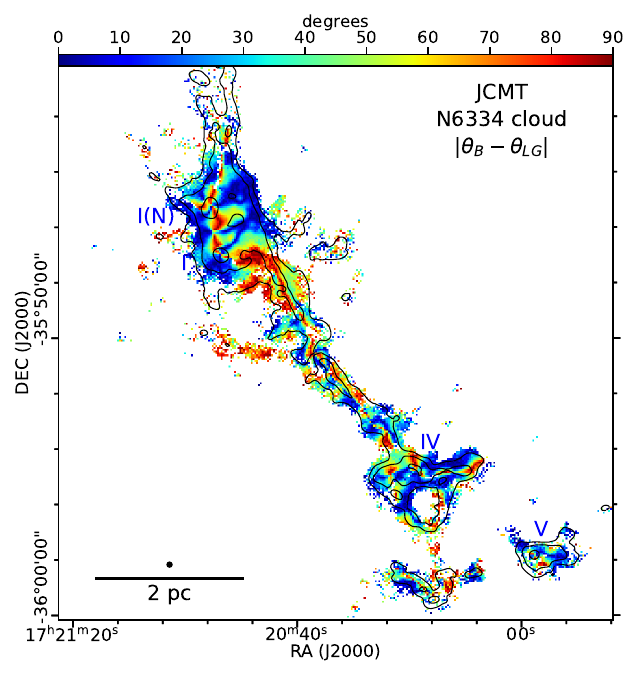}{0.48\textwidth}{(b)}
 }
\caption{(a) Relative orientation between magnetic fields and local gravity from Planck observations. The contour levels correspond to the Planck $\tau_{353}$ map. Contour starts at 0.0004 and continues with an interval of 0.0004. Only data points with NANTEN2 integrated  $^{12}$CO (1-0) intensity greater than 25 K km s$^{-1}$ are shown. (b) Relative orientation between magnetic fields and local gravity from JCMT observations. Values of $\phi_{B}^{LG}$ at positions with SNR($PI$)$>$3 are shown. The contour levels correspond to the JCMT 850 $\mu$m dust continuum map. Contour starts at 2 Jy beam$^{-1}$ and continues with an interval of 4 Jy beam$^{-1}$. \label{fig:N6334_large_B_LG}}
\end{figure}

\begin{figure*}[!htbp]
 \gridline{\fig{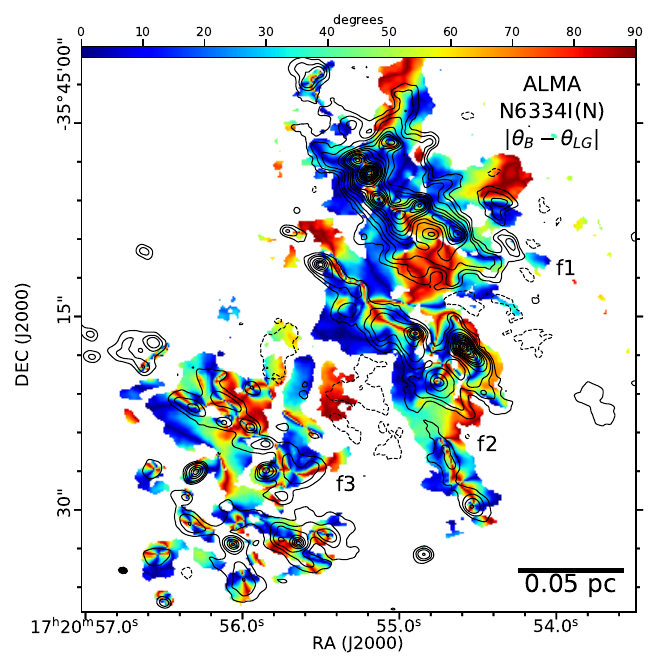}{0.48\textwidth}{}
 \fig{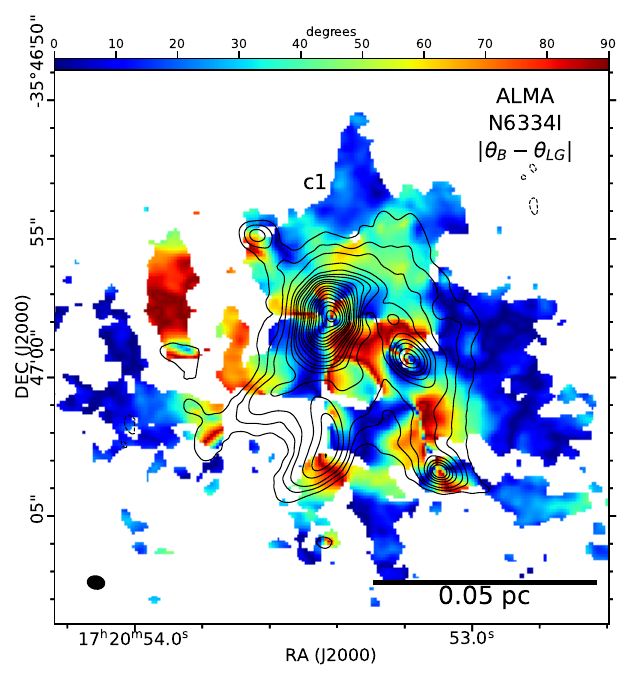}{0.48\textwidth}{}
 }
  \gridline{\fig{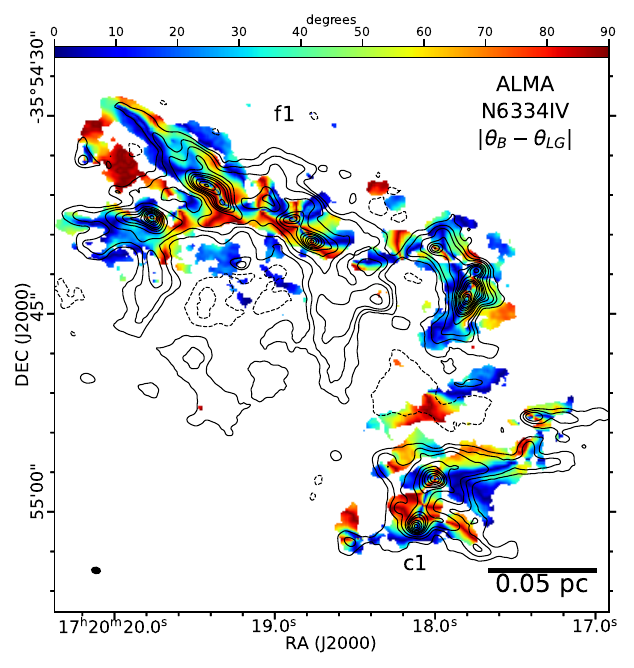}{0.48\textwidth}{}
 \fig{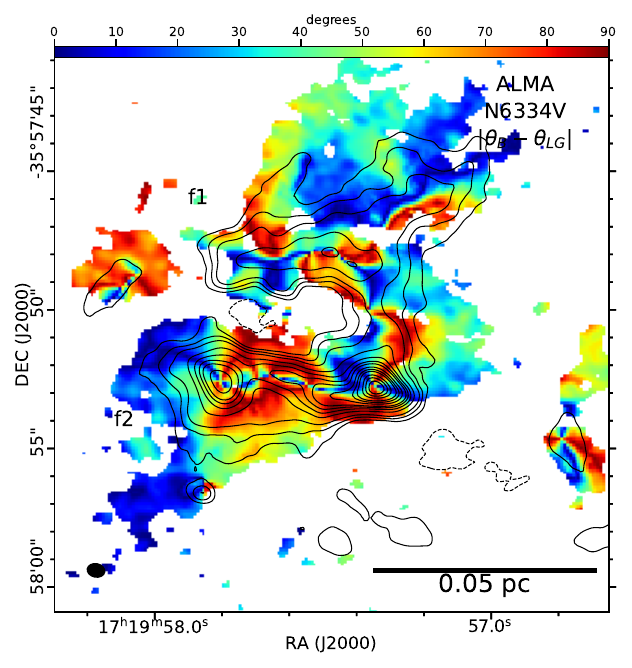}{0.48\textwidth}{}
 }
\caption{Relative orientation between magnetic fields and local gravity from ALMA observations. Values of $\phi_{B}^{LG}$ at SNR($PI$)$>$2 are shown. The contour levels correspond to the ALMA dust continuum map. Contour levels are ($\pm$3, 6, 10, 20, 30, 40, 50, 70, 90, 110, 150, 180, 210, 250, 290, 340, 390, 450) $\times \sigma_{I}$. \label{fig:N6334_alma_B_LG}}
\end{figure*}

The angle $\phi_{B}^{LG} = \vert \theta_{\mathrm{B}} - \theta_{\mathrm{LG}} \vert$ characterises the relative orientation between magnetic fields and local gravity. Figures \ref{fig:N6334_large_B_LG} and \ref{fig:N6334_alma_B_LG} show the $\phi_{B}^{LG}$ maps from Planck, JCMT, and ALMA observations. The spatial distribution of $\phi_{B}^{LG}$ is not random. At different scales, a common pattern is the tangential fan-like distribution of $\phi_{B}^{LG}$ with low and high values appear alternately near the emission peaks, which may suggest the gravitational infall/collapse can occur locally through the magnetic channels with small $\phi_{B}^{LG}$ values \citep{2018ApJ...855...39K}. The $\phi_{B}^{LG}$ distribution is clear in some regions. e.g., small and large $\phi_{B}^{LG}$ values are found in the main part and ends of N6334I(N)-f1, respectively, which agrees with the scenario of a magnetized gravitational collapse and clearly suggests the magnetic field structure is shaped by gravity in the main part and is being distorted by gravity in the ends. However, the $\phi_{B}^{LG}$ distribution is complex in most regions. Although \citet{2012ApJ...747...79K} and \citet{2018ApJ...855...39K} have suggested that small and large $\phi_{B}^{LG}$ (or $\omega$ in their work) values indicate weak and strong magnetic resistance against gravity, respectively, more detailed analytical explanations of different spatial $\phi_{B}^{LG}$ distributions are yet to be established. The local variation for the spatial distribution of other relative orientations ($\phi_{NG}^{LG}$, $\phi_{B}^{NG}$, $\phi_{VG}^{B}$, $\phi_{VG}^{NG}$, and $\phi_{VG}^{LG}$) are less clear than that of $\phi_{B}^{LG}$, thus we do not shown them in this paper.

%\nocite{*}

\bibliography{N6334ro}{}
\bibliographystyle{aasjournal}


\end{CJK*}
\end{document}